\documentclass[preprint,12pt,prd]{revtex4}
\usepackage{latexsym}
\usepackage{amsthm}
\usepackage{graphicx}
\usepackage{epsfig}
\usepackage{bbm}
\usepackage{amsmath}
\usepackage{amssymb}

\newcommand{\bef}{\begin{figure}[htbp]\begin{center}}
\newcommand{\eef}{\end{center}\end{figure}}
\newcommand{\bet}{\begin{table}[htbp]\begin{center}}
\newcommand{\eet}{\end{center}\end{table}}
\newcommand{\MET}{\not\!\!\!E_T} 

\begin{document}

\begin{flushright}
SLAC-PUB-13425\qquad \\
SU-ITP-08/24\qquad\ \\
\end{flushright}
\vspace{.0in}

\title{Simplified Models for a First Characterization of New Physics at the LHC}

\author{Johan Alwall} \email{alwall@slac.stanford.edu}
\author{Philip C. Schuster} \email{schuster@slac.stanford.edu}
\affiliation{Theory Group, Stanford Linear Accelerator Center}

\author{Natalia Toro} \email{ntoro@stanford.edu}
\affiliation{Stanford Institute for Theoretical Physics, Stanford University}

\begin{abstract}
Low-energy SUSY and several other theories that address the hierarchy problem predict pair-production at
the LHC of particles with Standard Model quantum numbers that decay to
jets, missing energy, and possibly leptons.  If an excess of such
events is seen in LHC data, a theoretical framework in which to
describe it will be essential to constraining the structure of the new
physics.  We propose a basis of four deliberately simplified
models, each specified by only 2-3 masses and 4-5 branching ratios,
for use in a first characterization of data.  Fits of these simplified
models to the data furnish a quantitative presentation of the jet
structure, electroweak decays, and heavy-flavor content of the data,
independent of detector effects.  These fits, together with plots
comparing their predictions to distributions in data, can be used as
targets for describing the data within any full theoretical model.
\end{abstract}

\maketitle

\tableofcontents
\section{Introduction}
The LHC experiments are the largest and most complex in human history,
with great potential to shed light on fundamental physics.  As the
experimental collaborations prepare to search for evidence of new
physics at the TeV scale, particle physicists must also prepare
for the next step: \emph{finding a framework in which
to describe the data.}

The Standard Model served this role through the entire history of
hadron colliders, from the discoveries of the $Z$, $W$, and top quarks
through percent- and sub-percent-level measurement of their properties
with Tevatron Run II data.  But there are many proposed extensions of
the Standard Model; many have qualitatively similar phenomenology,
which depends dramatically on a large number of free parameters.
Within the Minimal Supersymmetric Standard Model (MSSM), for example,
each signature that is commonly searched for can be produced in
multiple ways.  When a signal is seen, it will not be immediately
clear what particles are producing it, what their dominant decay
modes are, or what other species are simultaneously produced.  
For this reason, it is useful to step back from the detailed
predictions of any one model or region of parameter space, and
characterize these basic properties first in a manner that allows
comparison to \emph{any} model.

In this paper, we propose a specific approach to characterizing the
first robust evidence for new physics seen at the LHC.  We 
present four ``simplified models'', each with a small set of
unambiguous parameters, based on the phenomenology typical
of SUSY but stripped of much of the complexity possible in the full
parameter space of supersymmetry. Despite their small size, these simplified models will give a good coarse-level description of SUSY-like physics, especially appropriate in the low luminosity limit. We discuss and illustrate by
example how the parameters of the simplified models can be
constrained, and how deviations from the simplified models  
in the data can be used to further characterize the underlying physics.
We also discuss how to use these models as
a basis for comparison of data 	with theoretical models such as the MSSM.

These simplified models are a useful first description for any
``SUSY-like'' new-physics signal in jets+MET+X. By this we mean that
the new physics has a discrete spectrum of narrow resonances, that the new
particles are odd under some exact parity and are ``partners'' of a
Standard Model particle (with the same Standard Model gauge and flavor
quantum numbers), and that the lightest parity-odd particle, which is
necessarily stable, is neutral (and hence a dark matter candidate).
These theories include not only the $R$-parity conserving MSSM (see,
e.g., \cite{Martin:1997ns}), but also UED models with conserved KK
parity \cite{Appelquist:2000nn}, Little Higgs with $T$ parity
\cite{Cheng:2003ju}, and Randall-Sundrum models with custodial SU(2) and discrete symmetries \cite{Carena:2006bn}. 

The simplified models are expected to reproduce kinematics and multiplicities of
observed particles remarkably well in a wide variety of SUSY-like new
physics models --- even when the spectrum of unstable particles in the
full model is far more complex than the simplified model permits.  
The simplified model fits can then be used as a representation of
the data, and can be compared to any full model
by simulating both in a simple detector simulator. This last process of comparison can be done by phenomenologists outside the LHC collaborations.

The paper is organized as follows: In the rest of this introduction,
we will motivate the approach of using ``simplified models'' to
characterize data, and our particular choice of simplified models.  We
first consider alternative characterizations, and why we are led to
the counterintuitive choice of trying to match data with models that
we know to be incomplete (Section
\ref{sec:alternativeCharacterization}). We then discuss which
features we wish our approach to well describe, and define the
simplified models (\ref{sec:simplifiedModelMotivation}).

In Section \ref{sec:SusyLikePheno} we review basic features of
SUSY-like phenomenology (\ref{sec:susyLikeTopologies}).  This leads to
a set of questions we can ask about any robust excess of new-physics
events that is seen in jets and missing energy searches (\ref{sec:PhysicsQuestions}).  This section can be
skipped by the reader familiar with SUSY phenomenology.  We give a
detailed description of the
four simplified models, and introduce variables that can be used to
constrain their parameters, in Section \ref{sec:fourModels}.

We present the first of two examples in Section
\ref{sec:example1}. In this example, the simplified models
describe the data very well, which allows us to use the results directly
as a basis for model-building.

In many other cases, the structure of new physics breaks one or
more of the assumptions in the simplified models.  In this case, the
``best fit'' within the simplified models must be interpreted
carefully.  We discuss such subtleties in Section
\ref{sec:beyondSimple}.  They also play an important role in our
second example (Section \ref{sec:example2}).  In this case the
simplified models reproduce some features of the data, but not others,
and we focus on how the simplified models can be used to test
particular hypotheses for new physics.

\subsection{Motivation for Simplified Models}
\label{sec:alternativeCharacterization} 
If evidence for SUSY-like new physics is seen at the LHC, it
will be presented and characterized in several ways.  Both CMS and
ATLAS are expected to present kinematic distributions, comparisons of
data to SUSY benchmarks, and parameter fits within small parameter
spaces such as the CMSSM and possibly larger ones.  Why add another
characterization to this list?  Moreover, why characterize data in
terms of deliberately incomplete models, when full models can be
simulated quite accurately?

To address this question, we begin by summarizing the importance of
comparing distributions in data to models, rather than presenting
distributions alone.  
Some observables, such as the locations of kinematic features, can
readily be read off plots.  The analysis for producing such plots --- jet
energy scale correction, etc. --- has become standard at CDF and D0.
Properties that do not lead to sharp features are harder to
determine. For example, counting events with different numbers of $b$
quarks from the frequency of $b$-tags in data requires inverting
differential tagging efficiency and mis-identification functions that
depend on the kinematics of every jet in an event; to the authors'
knowledge this degree of ``unfolding'' based on data alone (not to be
confused with measuring the rate of a process with a fixed number and
distribution of $b$-jets, e.g. $\sigma(W\,b\bar b)$), has not been
done in the past.

We encounter the same kind of difficulty in answering also much simpler
questions: is a search that finds twice as many muon as electron
events above expected backgrounds evidence for new physics that
couples differently to electrons and muons?  The answer depends on the
differences in isolation requirements and detection efficiencies for
the two flavors.

Both problems illustrate a key reason to study the consistency of data
with models of new physics --- the detector's response to a model is
subject to uncertainties, but is at least well-defined.  A model that
describes the data well is an invariant characterization of that data,
independent of the detector.  Finding such a characterization is the
primary reason that comparing data to models, even at an early stage,
is useful.    

There is clearly a delicate balance between studying large and small
parameter spaces: a small parameter space can be studied more
efficiently and more thoroughly, but is less likely to contain a point
that describes the data well than a larger space.  Small subspaces
such as mSUGRA (with four real parameters) are obtained by imposing
relations between masses with little physical basis --- for instance,
a nearly fixed ratio of gaugino masses.  The most dramatic changes in
phenomenology typically arise from a \emph{re-ordering} of the
spectrum, which the mSUGRA mass relations prohibit.  On the other
hand, to cover all LHC phenomenology possible within the MSSM, one
must scan over $\approx 15-20$ Lagrangian parameters.

The compromise --- enlarging a parameter space until it can explain
the data, but no further --- is a natural one.  It is likely that an
MSSM point consistent with early new-physics data will be found in
this manner, and it is a very useful result; what is unreasonable to
expect is a thorough scan of the MSSM parameter space, in which each
point is compared directly to data.  Of course, this is even less
probable where it concerns models beyond the MSSM.  Therefore, we
would like not only \emph{a} consistent model, but an understanding of
what structure is \emph{required} for consistency (some subtleties of
model discrimination at the LHC have been studied in
\cite{ArkaniHamed:2005px} and \cite{Hubisz:2008gg}).

In answering this question, the simplified models, which have the
simplest spectra compatible with SUSY-like structure, are an ideal
starting point because they are --- both practically and morally ---
minimal.  They have few parameters not because of relations but
because they contain only 2-4 new particles.  Deviations from the
phenomenology of the simplified models can be taken as evidence for a
larger set of particles playing a role in new physics, and they are a
natural starting point for building more accurate models.

Besides being a stepping stone to finding more accurate models, the
simplified models work as baseline models to present in their
own right.  As we will illustrate in Sections
\ref{sec:example1}-\ref{sec:example2}, they can describe many features
of the data in a manner that is useful to further model-building, even
when they do not reproduce all observables.  This description
motivates specific consistent models, which in turn suggest particular
experimental tests to distinguish among them.

Finally, the simplified models suggest that imposing parameter relations is not
the only way of reducing the SUSY parameter space, and may not be the
most useful.  As noted earlier, mass relations are particularly
restrictive because they prevent interchanges of particle masses that
qualitatively change phenomenology.  By design, the simplified models
have a small set of parameters whose variations have large effects on
observables.  These are the most important parameters to focus on in a
first characterization of evidence for new physics.  If technical
obstacles permit detailed study of only a few-parameter space of
models, the simplified models may be the most efficient alternative.

\subsection{Introducing the Simplified Models}\label{sec:simplifiedModelMotivation}

Two of the problems that motivate searching for TeV-scale physics ---
and that motivate a ``SUSY-like'' structure of Standard Model particle 
partners, among which the
lightest is stable --- are dark matter and the hierarchy problem. As
we try to study the structure of physics at the LHC, it is useful to
keep both in mind.  

The lightest parity-odd particle (LSP) is stable, and a leading
candidate for the dark matter in the universe.  Being stable and
invisible, it cannot be probed directly at a hadron collider, but its
couplings can affect the decay chains of other, heavier particles.
Within any model, such as the MSSM, the LSP can account for dark
matter in some regions of parameter space but not others, and in some
regions the standard cosmology is inconsistent with direct dark matter
detection experiments. The decay modes of color-singlet particles
offer one probe of what regions of parameter space we could be living
in; these typically result in emission of weakly interacting Standard
Model particles --- $W$, $Z$, and Higgs bosons, and/or pairs of
flavor-correlated leptons or lepton/neutrino pairs.  Characterizing
the relative rates of these decays is thus a first step in relating
TeV-scale physics discovered at the LHC to cosmology.

Natural solutions to the hierarchy problem also require relatively
light partners of the top quark, and by association the bottom
quark. The lack of discovery of such partners at LEP and the Tevatron
already makes all known solutions to the hierarchy problem look a
little fine-tuned, creating what has been called the ``little
hierarchy problem''.  Confirming the presence of these partners
confirms that the new physics solves the hierarchy problem;
determining where they appear in relation to other new states sheds
light on how natural the solution is.  If these partners are
produced, either directly or through decays of a gluon-partner, there will be
an excess of $b$ and/or $t$-rich events.  Another feature that can
give rise to extra $b$ or $t$ quarks is a light Higgs partner --- this
gives some hint at the structure of electroweak symmetry breaking.

With this in mind, there are three \emph{initial} questions that can tell
us a great deal about the structure of TeV-scale physics, and touch on
questions of fundamental interest like dark matter and the electroweak
hierarchy:

\begin{description}
\item[1. What colored particle(s) dominate production?]  
\item[2. What color-singlet decay channels are present, and in what fractions?] 
\item[3. How $b$-rich are events?] 
\end{description}

The four simplified models proposed in this paper are designed to
answer these questions.  They are compact --- 2-3 masses and 3-4
branching ratios and cross sections for each model --- and retain a
structure motivated 
by solutions of the hierarchy problem, but pared down to a parameter
space that can be easily studied.  This makes them ideal for
experimental analyses with limited statistics and an
excellent starting point for developing more refined theoretical
frameworks to test against data. These simplified models are illustrated
in Figure \ref{fig:ModelPicture}; we will specify them fully in
Section \ref{sec:fourModels}.

\bef
\parbox{0.8in}{Lep(Q)}\parbox{5in}{\includegraphics[width=5in]{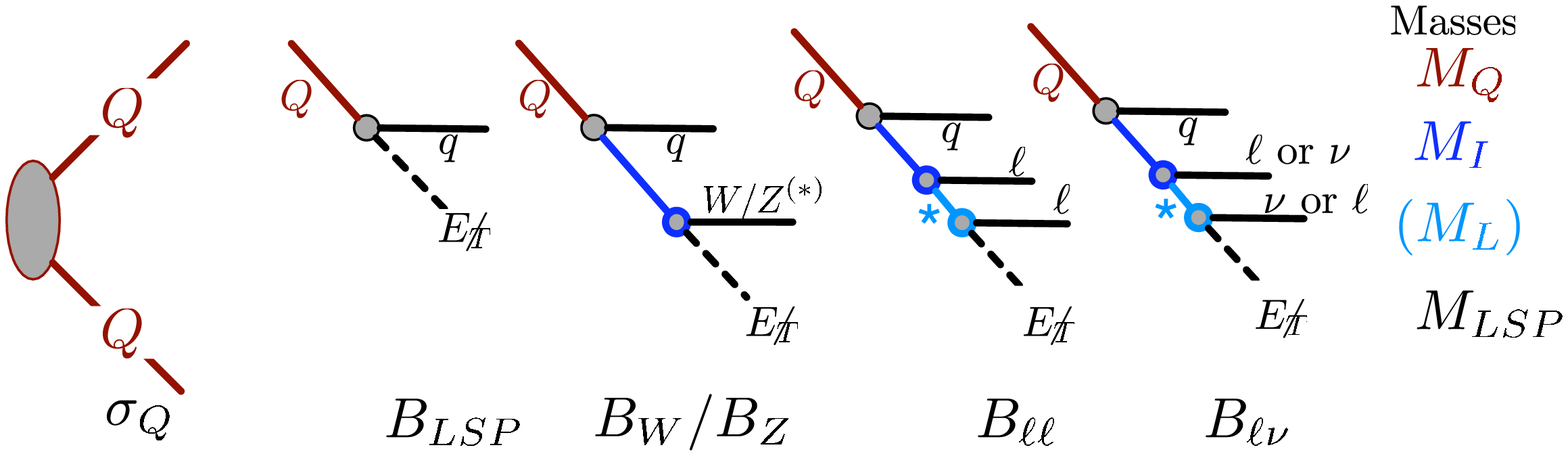}}\\
\includegraphics[width=5in]{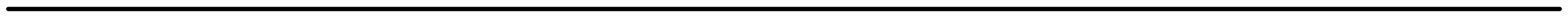}
\parbox{0.8in}{Lep(G)}\parbox{5in}{\includegraphics[width=5in]{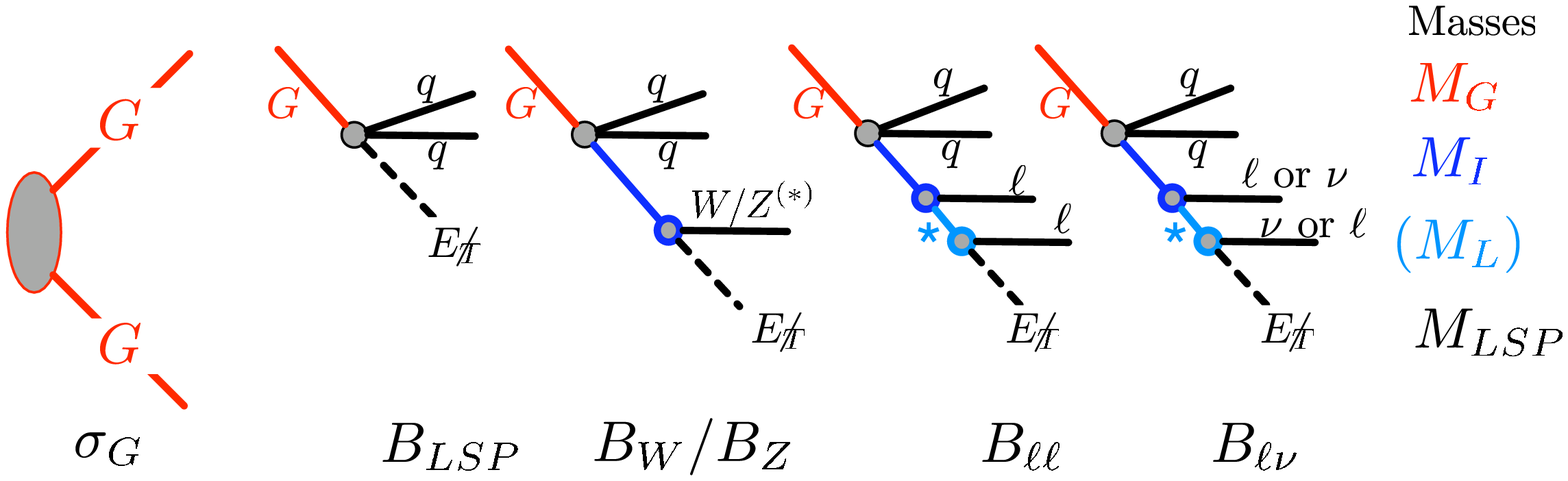}}\\
\includegraphics[width=5in]{graffles/line.eps}
\parbox{0.8in}{Btag(Q)}\parbox{5in}{\includegraphics[width=5in]{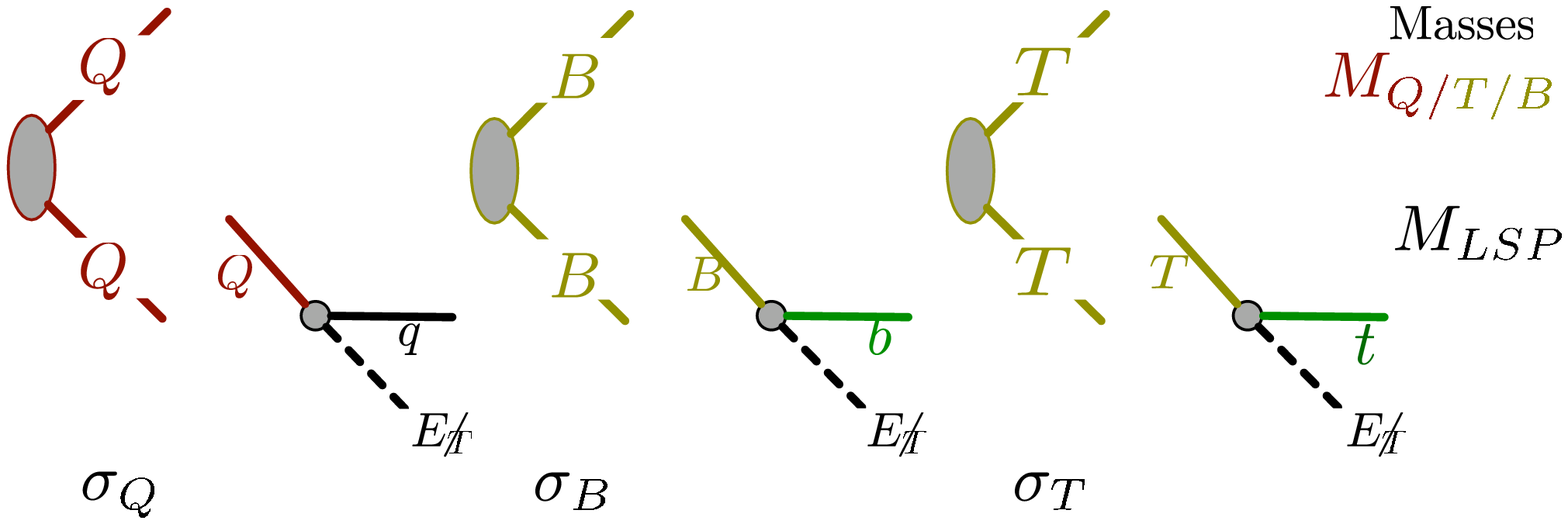}}\\
\includegraphics[width=5in]{graffles/line.eps}
\parbox{0.8in}{Btag(G)}\parbox{5in}{\includegraphics[width=5in]{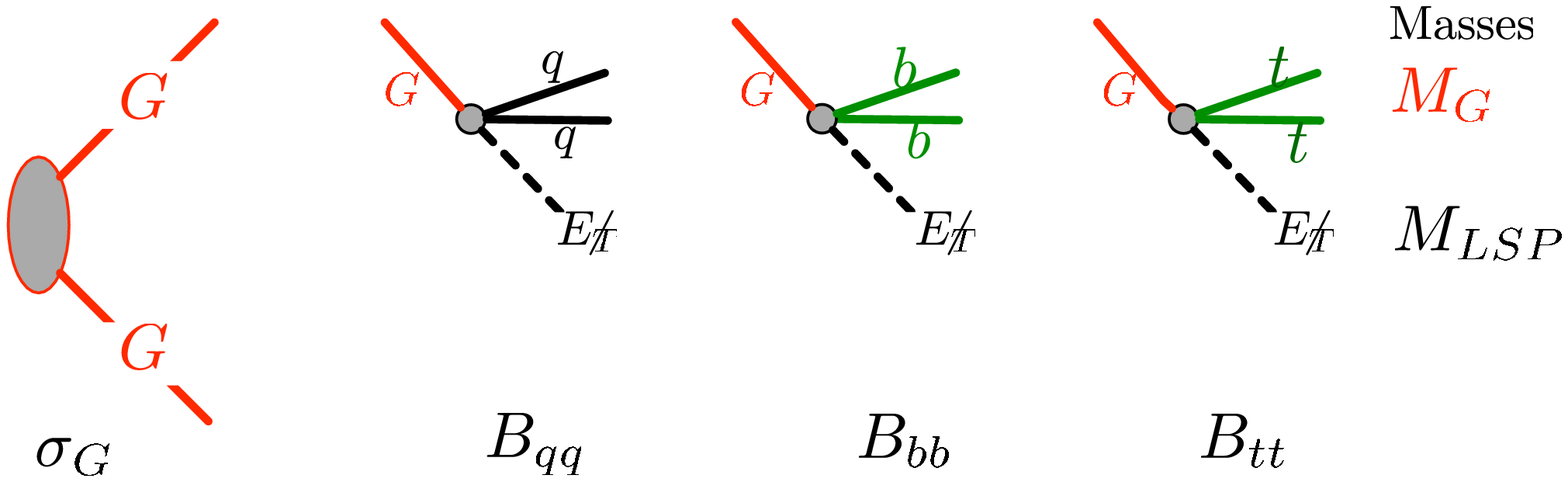}}\\
\caption{Particle and parameter content of the simplified models. From
top to bottom: The two leptonic decay models, originating from production of
either a quark-partner or gluon-partner; the two b-tag models,
originating from either a quark-partner or a gluon-partner.  Please
see text and Section \ref{sec:fourModels} for further discussion.
\label{fig:ModelPicture}}
\eef

Each of the four simplified models includes direct production of only
one type of strongly interacting species, either a quark or a gluon partner.
The leptonic models Lep(Q) and Lep(G) are designed to parameterize the
color-singlet particles produced in decays, question 2 above.  The
quark or gluon partner decays to one or two light quarks, plus either
an LSP or an intermediate state that decays to the LSP by emitting a
$Z$ or $W$ boson, or a $\ell\ell$ or $\ell\nu$ pair.  We will see that
these two models typically provide a good description of new physics even when
it contains multiple cascades, or multiple initial states.  Associated
production of quark and gluon partners when they have similar masses
can be viewed as an interpolation between Lep(Q) and Lep(G).

The $b$-tag models Btag(Q) and Btag(G) are designed to parameterize
heavy-flavor production (question 3), as well as 1.  By comparing data
to these two models, one can quantitatively describe the rates of heavy
flavor processes, and establish whether data is consistent with quark
flavor universality, or whether the third generation is enhanced or
suppressed. In the gluon partner model, Btag(G), the gluon partner
decays to light quark pairs, pairs of $b$-quarks or pairs of
$t$-quarks. In the quark partner model, Btag(Q), there are instead
three different pair-produced species: light-flavor quark partners,
$b$-quark partners and $t$-quark partners, which decay to their
respective partner quarks.

Despite their simplicity, these four models can describe the
kinematics and numbers of reconstructed physics objects remarkably
well (including jets and either leptons or $b$-tags, though not
necessarily both leptons and $b$-tags in the same model).  Therefore,
one may conclude that they also reproduce the properties of idealized
physics objects, and the agreement is not sensitively dependent on
details of the simulations.  When this agreement is observed, the
best-fit simplified models furnish a clear and simple representation
of the data, which any physicist can compare to a full model by
simulating both.  This procedure should be valid, even when the
simplified model-to-full model comparison is done with a simple
parameterized detector simulator that has not been tuned to data.  This
application underscores that, though model-independent, the simplified
models are most effectively used in conjunction with full models.  Their
virtue is that this characterization can be easily used with \emph{any}
model, because no correlations are imposed between different
parameters.


\section{The Phenomenology of SUSY-like BSM Physics}\label{sec:SusyLikePheno}

In this section, we elaborate on what ``SUSY-like'' physics means,
defining and discussing its important phenomenological features. In
order to clearly understand how our simplified models are motivated
theoretically, it will be useful to first consider the structure of
the MSSM, highlighting its features at the level of quantum numbers
and typical decay patterns. In the process, we will discuss non-MSSM
(though still ``SUSY-like'') physics using a more topology-based
language.

Operationally, ``SUSY-like" includes theories with new particles that
carry Standard Model quantum numbers (partner particles) and a parity
(under which partner particles are odd) that makes the lightest such
partner particle stable. This in turn means that LHC processes are
initiated by pair production. We begin by summarizing the particle content of different SUSY-like models.
Within the MSSM, the particle content is fixed --- two
Higgs doublets complete the matter fields, and every Standard Model
particle has partner with the opposite spin, but the same charges
under $SU(3)\times SU(2) \times U(1)$.  We expect all of these
partners to have TeV-scale masses, and flavor-conserving decays. 

Universal Extra Dimensions (UED) has an infinite tower of KK modes for
each Standard Model state; if the theory is compactified on an
interval, these KK towers alternate between parity-even and -odd
states.  UED is often considered as a ``foil'' for supersymmetry
because the first set of KK modes, like SUSY partners, are parity-odd
\cite{Appelquist:2000nn}.  Little Higgs models have a smaller slate of
partner particles --- minimally, for the top and the $SU(2)\times
U(1)$ gauge bosons \cite{Cheng:2003ju}.  Like UED, they also have
parity-even new states.  In this note, we focus on the phenomenology
of the first KK level, or the T-parity-odd states in Little Higgs
models, whose phenomenology is quite similar to that of supersymmetric
partners.

The unique challenge of describing SUSY-like phenomenology comes from
the sensitivity of both particle production and decay to particle
masses.  Typically, production is dominated by strongly interacting
particles, i.e., in SUSY, either squarks or gluinos.  The production
cross-sections vary greatly as a function of mass --- roughly as
$M^{-4}$ or faster as shown in Figure
\ref{fig:productionXSecByParticle}.  Thus, depending on the squark and
gluino masses, either gluino or squark production can have a much
larger cross-section than the other, and the dominant production mode
will change accordingly.

\bef
\includegraphics[height=1.9in]{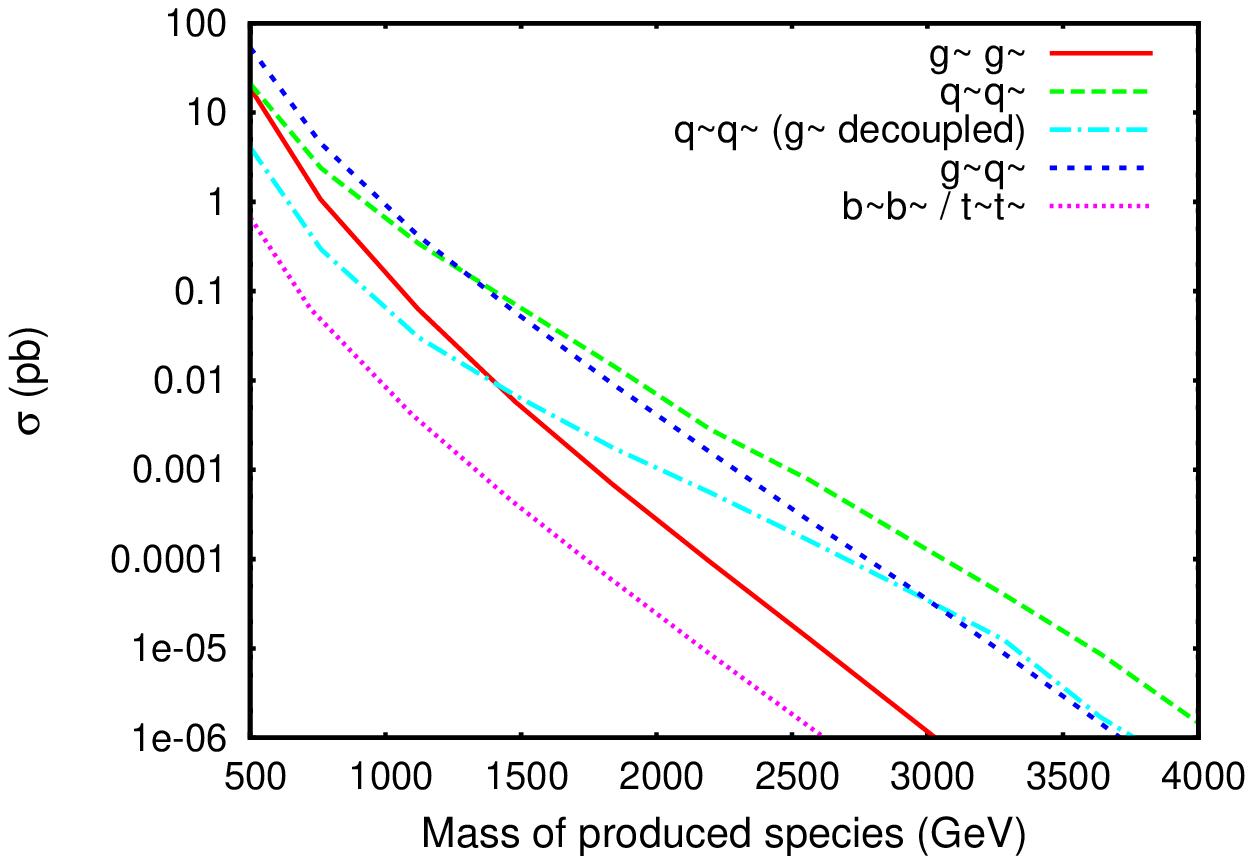}
\includegraphics[height=1.9in]{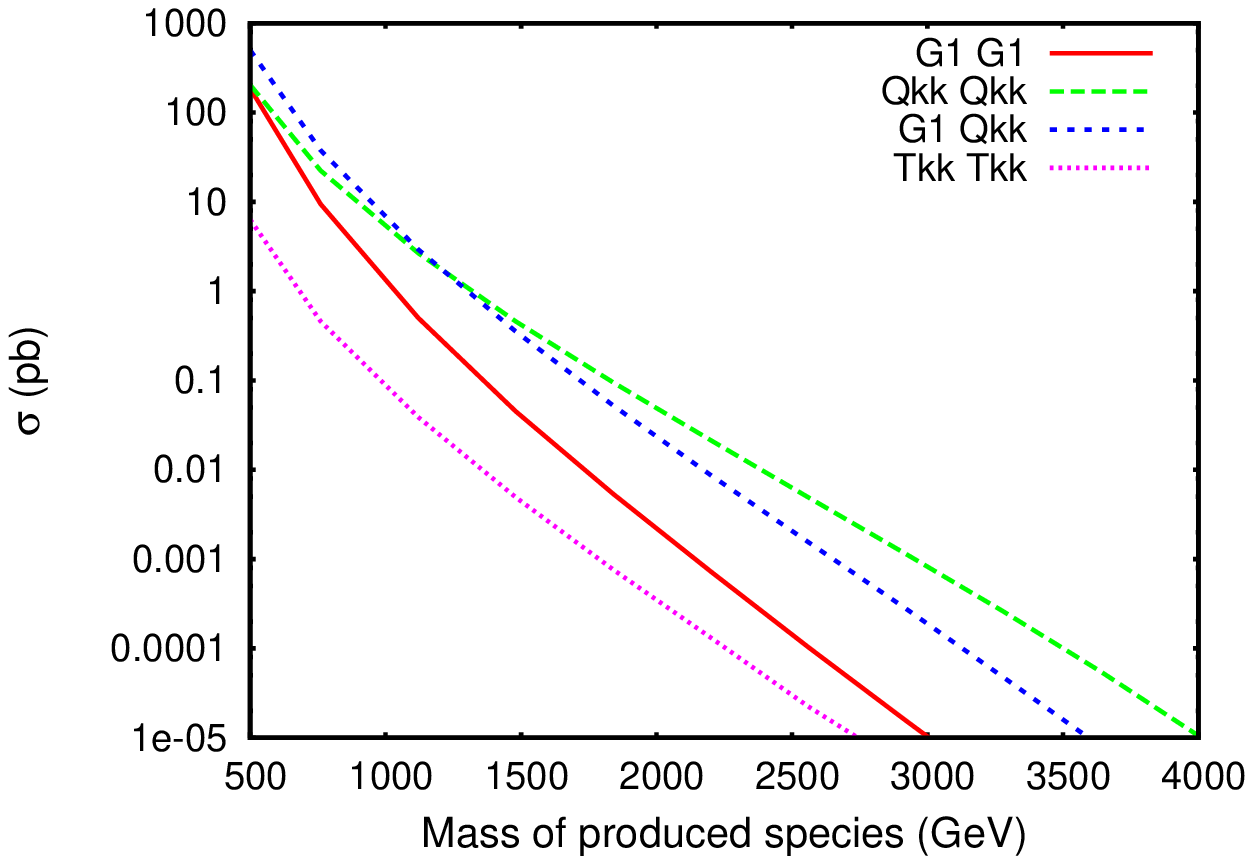}
\caption{Left: Production cross-sections for gluino pair production,
  gluino-squark associated production, and squark-pair production, for
  light and heavy flavors of squarks.  The mass on the $x$-axis is the
  mass of the produced particles. We have assumed equal masses for
  squarks and gluinos unless otherwise indicated. Right: The
  corresponding cross sections for UED production, with all KK particle
  masses taken equal. The MSSM generations are made in Pythia 6.4
  \cite{Sjostrand:2006za} and the UED generations using
  MadGraph/MadEvent 4 \cite{Alwall:2007st,UEDMadGraph}.
\label{fig:productionXSecByParticle}}
\eef

Similarly, a given particle's decays depends sensitively on masses:
if, as we assume here, there is a conserved parity under which
``partner'' particles are odd (or even if the parity is only
approximate), every partner decays to a Standard Model particle and a
partner.  The couplings controlling these decays range from the
hypercharge coupling $\alpha' \approx 0.01$ to $y_t \approx 1$ --- the
partial widths of different 2-body decay modes span 2-4 orders of
magnitude (some also depend on mixing angles), so of all the
kinematically allowed modes, those modes with the largest coupling
constants dominate; however, re-arrangement of masses can forbid
would-be dominant decay modes, so that small-coupling modes dominate.
The range of partial widths is wider still for three-body decays.

The task of the rest of this section will be to map out the common
topologies and decays that partner particle production can give rise
to. We will focus on the MSSM, as it provides a complete set of
partner particles, though other SUSY-like theories have very similar
processes. Having done this, we'll propose a set of questions that can
guide the process of identifying what qualitative patterns of
production and decay might exist in data. The goal of our simplified
models will be to neatly encapsulate these patterns in a minimal way.

\subsection{Typical Topologies for SUSY-Like New-Physics Production
  and Decay}\label{sec:susyLikeTopologies}

We will find it useful to organize the
production and decay modes according to the couplings of the
particles: the production modes we focus on here are those with QCD
couplings, which typically dominate at the LHC, so we will start from
strongly interacting particles.  Because partners are odd
under a conserved parity, they decay through ``cascades'', with one
parity-odd particle and one or more parity-even particles produced at
each stage of the decay. If possible, $SU(3)$-charged states decay
into other $SU(3)$-charged states. once an
$SU(3)$-neutral state is reached, it is unlikely for decay to another
$SU(3)$-charged state to occur.  

As a result, the R-parity conserving MSSM-like decay chains we are
interested in here can be effectively divided into one $SU(3)$-charged
segment and one $SU(3)$-neutral segment, in which decays are typically
dominated by the electroweak interactions.  Identifying the possible
\emph{topologies} for each stage allows us to ignore some ambiguities
that will arise between, e.g., left vs. right-handed squarks and
interchanges of the gaugino states (the ``flipper'' degeneracies of
\cite{ArkaniHamed:2005px}), which are in general challenging to discriminate.  This
approach will suggest the simplifications we adopt in Section
\ref{sec:fourModels}.

In this paper we focus on phenomenology that can be described in the
R-parity conserving MSSM with a heavy gravitino (this also includes,
e.g., little Higgs and UED models with parities that have species with
the same gauge and global quantum numbers but different spins), but
does not describe models with light gravitinos or R-parity
violation \cite{Martin:1997ns}.  However, the classification we describe has a
natural extension to these models, provided the additional
interactions have small couplings to MSSM-like states, so that their
main effect is on the decays of the LSP. In this case, we should add a
3rd ``small-coupling'' stage to every decay, after the ``strong'' and
``electroweak'' stages.

\subsubsection{Production and Initial Decay of Colored Particles}
Because they carry $SU(3)$ charges, squarks and/or gluinos will
probably be the most abundantly produced new physics particles at the
LHC.  Depending on their masses, one or both will be readily
produced. The associated mode ($q\,g\rightarrow \tilde q\, \tilde g$)
can also be competitive.  Depending on their decay chains, the
production of same-sign squarks from quark pdf's (e.g. $u\,u
\rightarrow \tilde u\,\tilde u$) can be distinctive.  Note that the
associated and same-sign modes rely on quarks of the same flavor as
the squarks that are produced --- thus, they do not effectively
produce third-generation squarks.  Some estimate of the production
cross-sections is given in Figure \ref{fig:productionXSecByParticle}
(note that the cross-section depends not only on the masses of the
produced particles, but on the masses of t-channel exchanged states as
well).

The colored parts of decay chains, which typically end in a
chargino or neutralino, are often quite simple:

\paragraph{Gluinos} can at tree level only decay through the $\tilde q q \tilde g$
interaction.  When there are squarks lighter than the gluino, it will
decay to these squarks (with relative rates determined by phase
space), and not via off-shell states to three body; when the squarks
are heavier, the gluino will decay through off-shell squarks to two
quarks and a chargino or neutralino. In the latter case the branching
ratios depend on the identities of neutralinos that are kinematically
available, and the masses of the squarks.  A simple parameterization is
especially useful in that case. Note that, in either case, the decay
is to two quarks of the same generation.

\paragraph{Squarks} can decay to a quark and a gaugino.  When the gluino
is lighter than a squark, decays to the gluino and a quark of the same
flavor is often a preferred mode because it has the largest
coupling. Other possible decays (and the only ones if the gluino is
heavy) are to quark of the same flavor plus a wino,
bino, or higgsino. When all of these are kinematically accessible,
third-generation squarks will often favor the higgsino, left-handed
squarks of the first two generation will favor the wino (third
generation may be split between higgsino and wino), and right-handed
squarks of the first two generations will go to the bino. Mixing among
the gauginos can be important if they are light (~300 GeV or less) and
nearby in mass to one another, and this can slightly change these decay
guidelines.

A squark always decays to an odd number of quarks, and a gluino to
an even number.  This can be useful, but there are kinematic regions
where one of the emitted quarks is very soft (for example, a gluino
decay to a nearly degenerate squark).  In this case, it is possible to
think of the gluino production as an additional contribution to
the squark production cross-section, with an additional (softer) jet.

Note that three-body decay to a state with lepton number is in both cases
suppressed, and is present only if none of the gauge or Higgs
partners are available. 

\subsubsection{Electroweak Decay Chains}
If the final decay product of the $SU(3)$ decay is the LSP, then there
are no additional emissions; but when it is not the LSP, it
decays down through one or more cascades from one electroweak-ino to
another.  These can be mediated by: 
\begin{itemize}
\item Electroweak interactions of leptons, and/or Yukawa couplings of
  $\tau$ leptons --- these lead to decays mediated by on- or off-shell
  sleptons, producing two leptons that may be charged or neutral.
  Different fractions of $\ell^+\ell^-$, $\ell^\pm \nu$, and $\nu\nu$
  can result from different hierarchies of masses and depending on
  whether the slepton is left- or right-handed: for example, decays
  through a right-handed slepton are often dominated by $\ell^+
  \ell^-$, while left-handed sleptons and sneutrinos mediate all three
  modes, often dominated by $\ell^\pm \nu$.  The $\ell^+\ell^-$ events
  are easily recognizable as an excess in opposite-sign leptons of the
  same flavor (``OSSF'') whose invariant mass does \emph{not}
  reconstruct the $Z$ mass.  Moreover, these have a well-known
  characteristic invariant mass distribution, either an ``edge'' if
  the slepton is on-shell or an ``endpoint'' if it is off-shell.
\item Higgsino-Higgs-gaugino interactions (SUSY partners of Higgs
  gauge couplings) allow decay through emission of a Higgs boson or
  (through the longitudinal mode) of the $W$ and $Z$ gauge bosons.
  Off-shell $h$, $W$, and $Z$ instead mediate a 3-body decay if the
  electroweak-ino mass splittings are too small for on-shell decays.
\item $SU(2)$ gauge self-couplings allow emission of $W$'s, and
  in chargino-chargino decays, emission of $Z$ bosons or (suppressed
  by $\sin^2 \theta_W$) photons.   Again, the gauge bosons may mediate
  three-body decays if on-shell emissions are kinematically forbidden.
\end{itemize}
Note that all of these decays can occur irrespective of whether
the initial and final electroweak-inos are bino-like, wino-like, or
higgsino-like.  The branching fractions are certainly sensitive to
these changes, but the identification of partners is not
\emph{sufficient} to constrain these branching ratios.  One reason is
that decays through mixing can be competitive or even dominant, and
the mixing matrix is extremely difficult to measure (knowing them
requires measuring the masses and phases of \emph{all} entries in the
mass matrix, which most likely cannot be done at the LHC).  The masses
of sleptons and $\tan\beta$ also play a significant role in
determining the decay patterns.

A pragmatic treatment is to disregard the $SU(2)$ quantum numbers of
the electroweak-inos, which are insufficient to determine branching
fractions, and instead treat the unknown fractions as free
parameters.  We should then allow for direct decays of the
$SU(3)$-interacting sector into the LSP, and the cascades of
heavier electroweak-inos to lighter ones as shown in Figure
\ref{fig:electroweak}.  If the bino, wino, and higgsino are all light
(or just the wino and higgsino, with significant splitting
between charginos and neutralinos), there can be \emph{multiple}
cascades, each stage with different branching fractions.

\bef
\includegraphics[height=2.0in]{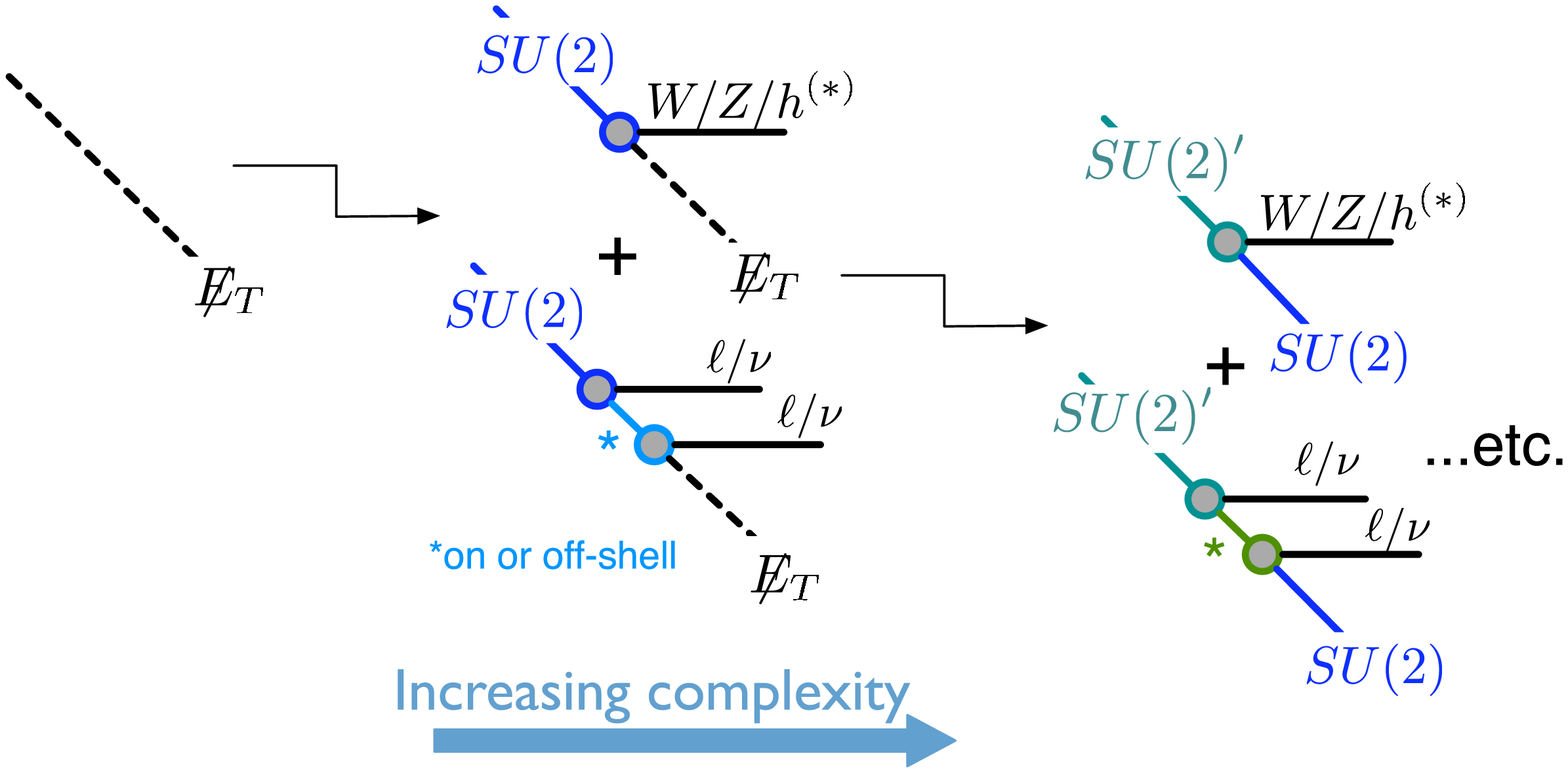}
\caption{Topologies typical of electroweak decays. \label{fig:electroweak}}
\eef

We have been deliberately unclear in Figure \ref{fig:electroweak}, in
the case of charged emissions (e.g. $W$), about whether the decay is
from a neutral state to a charged state, or a charged state to a
neutral one.  We do so because they might be difficult to distinguish
In the former case, the light charged state must decay
again to the LSP. However, if the state is a charged wino or higgsino
in the same $SU(2)$ multiplet as the LSP, their mass splitting can be
quite small ($\approx 1$ GeV when $|M_2 \pm \mu| \gtrsim M_Z$).  The
charged states decay to the nearly-degenerate LSP by emitting soft
leptons or pions, which are difficult to measure, so the charge of
this final state is lost.

We have assumed so far that decays of $SU(3)$-neutral states to
$SU(3)$-charged states do not occur.  Though they are impossible in
CMSSM-like mass spectra, with winos and binos much lighter than the
squarks, spectra for which these decays are permitted are logically
consistent (for example, with a spectrum ordering $m(\tilde q_R) >
m(\tilde B) > m(\tilde q_L) > m(\widetilde W)$). Depending on mixing, the
cascade $\tilde q_R \rightarrow \tilde B + q \rightarrow \tilde q_L +
\bar q \,q \rightarrow \widetilde W^0 q\, \bar q \,q$ can be possible.
These additional jets are however typically significantly softer than
the prompt jets from the directly produced $SU(3)$ state.

\subsection{Eight Questions for the LHC}\label{sec:PhysicsQuestions}
The phenomenology of SUSY-like models is quite rich, and the language
presented above provides a useful framework in which to ask questions
about data and build evidence for the answers.  Many of the most
interesting questions we would like to answer about the structure of
new-physics production and decay processes are overly ambitious for a
\emph{first} characterization of data --- they require a firm
foundation as a starting point.  The first goal is to build this
foundation, by determining which of the processes discussed above are
in play:
\begin{description}
\item[1.] \textbf{Is production dominated by events with 2 hard
  partons (quark partners) or 4 (gluon partners)?}  As mentioned
  above, the physical interpretation is not always so simple --- if
  the gluon partners are only slightly heavier than the quark
  partners, gluon partner production may dominate, but kinematically
  these events may look more like quark partner events.  This may still be
  distinguishable from true quark partner pair production, either by the
  kinematics of softer jets or by the fraction of same-sign dilepton
  events.  But in a first pass, these two alternatives are the ones to
  consider.
\item[2.]\textbf{What $SU(2)$ modes are present, and in what fractions?} 
  If $\ell^+ \ell^-$ pairs are seen, characterizing their kinematics
  (in particular, on- vs. off-shell sleptons and the implications of
  the edge or endpoint in their invariant mass for the mass
  spectrum). Decays to $Z$, and to $\ell^+ \ell^-$ pairs off the $Z$
  pole, are rather distinctive in dilepton invariant mass; more care
  is required in distinguishing $W$ bosons from $\ell\nu$ pairs from
  sleptons.
\item[3.] \textbf{How $b$-rich are events?}  The minimal interpretation of
  $b$-rich events depends on whether we are dominated by quark or
  gluon partner production.  If quark partners are light, the third-generation
  quark partners may well be even lighter --- though they
  typically have smaller production cross-sections for the same mass,
  these modes are certainly worth looking for.  On the other hand,
  three-body decays of a gluon partners can be approximately universal among
  flavors, or can be dominated by decays to the third
  generation --- measuring these rates provides useful information.
\end{description}

These questions are well-posed within the simplified models we
consider in this paper.  But of course, they are only the beginning
--- to characterize the phenomenology, we must also consider
correlations.  In this paper, we will instead try to study these
questions \emph{qualitatively} using deviations of observed
distributions from those predicted by our simplified models.  Among these
questions are:

\begin{description}
\item[4. Is there evidence for tops] (as opposed to independent production of $W$'s and $b$ quarks)?
\item[5. Is there evidence for double cascades] from multiple electroweak-ino multiplets?
\item[6. Are there differences in $SU(2)$ decays] between quark and
  gluon partners, between different quark partner species (e.g.\
  left-and right-handed), or between heavy- and light-flavor decays?
\item[7.] \textbf{Is there evidence for competition between gluon and quark partner production modes?}
\item[8.] \textbf{Are different tests of the features above consistent with one
  another?} (for example: gluon partner pair production is expected to
  produce both a large number of jets and, if single-lepton
  decays are present, same-sign and opposite-sign dileptons in equal numbers).
\end{description}

Answering these questions gives insight into many of the most
important properties of new physics that we can hope to establish at
the LHC.

We should note that we are here deliberately leaving out questions
concerning certain aspects of the new physics. Two omissions in
particular deserve mention: spin determination and precision
measurement of the new-physics spectrum. These questions have been
extensively covered in the literature
\cite{Barr:2004ze,Smillie:2005ar,Alves:2006df,Athanasiou:2006ef,Bachacou:1999zb}.
While these can, and should, be pursued in parallel with the questions
we have emphasized, they often require correct assumptions about the
mass hierarchy of partners and their decay topologies. Therefore the
approach we are suggesting can be seen as a preparatory step for such
further studies.

\section{Four Simplified Models}\label{sec:fourModels}

In this section, we present our proposal for how to characterize
early excesses, keeping in mind the questions outlined above. We
propose to use four simplified models, based on a number of
well-justified approximations. Two models are aimed at capturing
cascade decays that give rise to Standard Model leptons or gauge
bosons, while two are aimed at describing the heavy flavor structure. We stress
that each model is designed mainly to answer a targeted set of
questions, not to necessarily provide a globally good description of
data.

In these simplified models, we will assume that new-physics particle
production is dominated by the pair-production of one new particle ---
either a gluon partner or a quark partner (where ``a quark partner''
indicates that we assume all prominently produced quark partners to
have at least nearly degenerate masses, and that their properties can
be characterized by averaging their decay branching ratios). This is a
good approximation if mass scales for the gluon and quark partners the
are widely split. If not, the resulting distributions will be a
baseline of comparison for estimating the fractions of different
production modes.  The key reason for this assumption, however, is
pragmatic: the clearest characterization of the data is found by
comparing it to \emph{simple} models, and this should come before any
attempt to study more complex (and potentially more accurate) ones.

Our simplified models will only involve production of heavy particles,
and we therefore expect that an On-Shell Effective Theory (OSET)
approximation scheme is accurate \cite{ArkaniHamed:2007fw}. 
In this scheme, matrix elements for
production and decay can be approximated as constants (or with simple
leading order corrections) and decays are described by pure phase
space. With these approximations, the parameters of our simplified models
will always take the form of cross sections for production, branching
ratios for decays, and masses of on-shell particles (see
Appendix~\ref{app:Implementations}). We want to stress though, that
this is only one way to simulate the processes, and while the
simplified models can easily be constructed using a Lagrangian and
implemented in a Matrix Element generator, the difference in the
simulation is in practice negligible at the accuracy we target here.

\subsection{Definition of the Simplified Models}
\subsubsection{Two Models for Leptonic Decays and Rates}
\label{ssec:LeptonModels}
We propose two models as a framework for studying electroweak cascade
branching ratios in early data.  The two models have identical decay
structure, and differ only in that one is quark partner-initiated and
the other gluon partner-initiated, which gives different jet structure
of the decays. Each contains three mass scales:
the primary produced particle $Q$ or $G$ (quark partner or
gluon partner), an intermediate electroweak state $I$, and the
lightest stable particle (LSP). 
The primary produced particle can decay either directly to the LSP, or to
the intermediate state which then decays down by one of several channels:
\begin{itemize}
\item A $Z$ boson (or off-shell $Z^*$ with $Z$ branching ratios if 
  $M_I - M_{LSP} < M_Z$).
\item A $W$ boson (or off-shell $W^*$ with $W$ branching ratios if 
  $M_I - M_{LSP} < M_W$).
\item An $\ell^+ \ell^-$ pair, decaying through three-body phase space,
  unless there is kinematic evidence for an edge in the opposite
  sign-same flavor invariant mass, in which case it is replaced by a
  decay through an on-shell lepton partner.
\item An $\ell \nu$ pair, again decaying through three-body phase space,
  unless there is kinematic evidence for an edge in $\ell^+
  \ell^-$ events, in which case the same on-shell lepton partner mass is
  used.
\end{itemize}

As we will argue, simplified models of this form are very effective
for characterizing cascade decays involving Standard Model gauge
bosons and leptons, even if when the underlying physics has a more
complicated structure (multiple cascades, or multiple produced species
with different decays).

The only ``flat direction'' in this choice of parameter space is the
distinction between $W$ and $\ell \nu$ decays. While this distinction
is difficult to constrain (see sec.\ \ref{ssec:LeptonSignatures}
below), it is important to include to understand systematic effects on
the fits (for example, the lepton or $Z$ fractions, due to differences
in signal efficiency, and the jet structure of the fits).

For quark partners, there is an ambiguity in the decay to the charged
modes, $W$ and $\ell \nu$, between charge asymmetric and charge
symmetric production. The former occurs when the intermediate color
singlet state is charged, in which case up-type and anti-down type
quark partners decays only to the positive intermediate state, and
down-type and anti-up type quark partners decay only to the negative
state. At the LHC, this means that $\ell^+{\ell'}^+$ production will
dominate over $\ell^-{\ell'}^-$ final states. If the intermediate
state is instead neutral while the LSP is in an SU(2) multiplet, as is
the case, e.g., in anomaly mediation, the decay is flavor independent
and charge symmetric (the remaining charge is then shed as soft pions
or leptons when the charged SU(2) partner of the LSP decays to the
neutral LSP). In order to avoid modeling different cross sections
between $QQ$ and $Q\bar Q$ production, we choose the latter,
charge-symmetric, decay mode for our simplified models. If data
displays a difference between $\ell^+\ell^+$ and $\ell^-\ell^-$
production, this assumption can be modified as needed.

\bet
\begin{tabular}{|c|c|c|}
\hline
Model& Particle content, SU(3)$\times$EM, mass & Rate parameters\\
\hline
Lep(Q) &
\begin{minipage}{1.8in}
\begin{tabular}{rcl}
$Q$ & ($\mathbf{\overline3}\times\frac23$ & / $M_Q$)\\
$I$ & ($\mathbf{\overline1}\times 0$ & / $M_I$)\\
$[$ $L$ & ($\mathbf{\overline1}\times 0,\pm1$ & / $M_L$) $]$\\
${LSP}^\pm$ & ($\mathbf{\overline1}\times \pm1$ & / $M_{LSP}+\epsilon$)\\
${LSP}^0$ & ($\mathbf{\overline1}\times 0$ & / $M_{LSP}$)
\end{tabular}
\end{minipage}
&
\begin{minipage}{3in}
\begin{tabular}{l}
$\sigma_Q=\sigma(gg\to QQ)$\\
$B_W=B(Q\to q\,I)B(I\to W^\pm{LSP}^\mp)$\\
$B_Z=B(Q\to q\,I)B(I\to Z\,{LSP})$\\
$B_{\ell\nu}=B(Q\to q\,I)B(I\to \ell^\pm\nu\,{LSP}^\mp)$\\
$B_{\ell\ell}=B(Q\to q\,I)B(I\to \ell^\pm\ell^\mp\,{LSP})$\\
$B_{LSP}=B(Q\to q\,{LSP})$
\end{tabular}
\end{minipage}
\\
\hline
Lep(G) &
\begin{minipage}{1.8in}
\begin{tabular}{rcl}
$G$ & ($\mathbf{\overline8}\times\frac23$ & / $M_G$)\\
$I$ & ($\mathbf{\overline1}\times 0$ & / $M_I$)\\
$[$ $L$ & ($\mathbf{\overline1}\times 0,\pm1$ & / $M_L$) $]$\\
${LSP}^\pm$ & ($\mathbf{\overline1}\times \pm1$ & / $M_{LSP}+\epsilon$)\\
${LSP}^0$ & ($\mathbf{\overline1}\times 0$ & / $M_{LSP}$)
\end{tabular}
\end{minipage}
&
\begin{minipage}{3in}
\begin{tabular}{l}
$\sigma_G = \sigma(gg\to GG)$\\
$B_W=B(G\to q\bar q\,I)B(I\to W^\pm{LSP}^\mp)$\\
$B_Z=B(G\to q\bar q\,I)B(I\to Z\,{LSP})$\\
$B_{\ell\nu}=B(G\to q\bar q\,I)B(I\to \ell^\pm\nu\,{LSP}^\mp)$\\
$B_{\ell\ell}=B(G\to q\bar q\,I)B(I\to \ell^\pm\ell^\mp\,{LSP})$\\
$B_{LSP}=B(G\to q\bar q\,{LSP})$
\end{tabular}
\end{minipage}
\\
\hline
\end{tabular}
\caption{The particle content and parameters of the Leptonic Decay Models
Lep(Q) and Lep(G). The models differ in the number of quarks emitted
in the primary decay; one quark for the quark partner and two quarks
for the gluon partner. Lep(Q) is based on the assumption of one
pair-produced active quark partner state (or several degenerate
states), while Lep(G) describes pair-produced gluon partners. The
quark or gluon partners decay directly to the LSP, and, through an
intermediate color singlet state $I$ to $Z$+LSP, $W$+LSP,
$\ell^+\ell^-$+LSP or $\ell^\pm\nu$+LSP. In the latter two decays, an
on-shell lepton partner with mass $M_L$ can be added between the
intermediate state and the LSP, if there is evidence for this in the
lepton kinematics data.
\label{tab:Leps}}
\eet

The leptonic decay models are illustrated in Table~\ref{tab:Leps}, and
in the upper panes of Fig.~\ref{fig:ModelPicture}. The leptonic decay
model for quark partner production, or Lep(Q) for short, has a total
of 3 (or 4) mass parameters, $M_Q$, $M_I$, $M_{LSP}$ (and $M_L$ if
there is an edge in the dilepton mass spectrum), the most readily
constrainable of which are the mass differences. There are 5 branching
ratios (i.e.~4 unconstrained parameters), $B_Z$, $B_W$,
$B_{\ell\ell}$, $B_{\ell\nu}$ and $B_{LSP}$. Finally, there is 1
overall production cross section $\sigma_{Q}$. Throughout most of
this paper, we will just use the 7 (3 mass, 3 branching ratio, 1 cross section) parameter Lep(Q).
  
The leptonic decay model for gluon partner production, Lep(G), is
identical to Lep(Q), except for the initial decay which is to two
quarks. We use off-shell gluon partner decays, for two reasons. First,
a series of two-body decays requires the quark partner to be lighter
than the gluon partner; if it is not much lighter then the first decay
will produce a rather soft quark, and will not look so different from
direct quark partner production; while if it is significantly lighter,
then direct quark partner production will dominate. Second, we want
the two models to act as extremes for the jet structure of the decays,
in order to ``fence in'' the underlying model; the greatest difference
between quark partner and gluon partner decay is achieved with the
gluon partner decaying to two jets of similar energy.  Also Lep(G) has
a total of 3-4 mass parameters ($M_G$, $M_I$, $M_{LSP}$, and if needed
$M_L$), 5 branching ratios $B_Z$, $B_W$, $B_{\ell\ell}$, $B_{\ell\nu}$
and $B_{LSP}$ (giving 4 unconstrained parameters), and 1 cross section
$\sigma_{G}$.

Why do we suggest two models, 
rather than simply fit the number of quarks coming off the decay of
the QCD particles to data? The reason is that fitting to the jet
structure is in general quite difficult. Not only are jets
complicated experimental object, 
jets are also abundantly produced in the underlying event and
pileup at a hadron collider. While these effects can be subtracted,
there are also many different sources for jets from the hard
interaction itself: Initial state radiation jets, which depend on the
masses of the produced particles; jets from the decay of the heavy QCD
particles to color singlet states, with characteristics dependant on the
different mass splittings present; and finally jets from the decay
chains, in particular due to the presence of electroweak bosons and
Higgs. This means that in order to get a fit right to the number of
jets from the initial decay of the QCD particles, one must first model
all the other aspects correctly. We therefore
keep the description of the jet structure qualitative, and
use different extreme choices to get a measure of which scenarios are
more or less compatible with the data.

\subsubsection{Two Models for B-tags and Rates}\label{ssec:HFModels}

The b-tag models are constructed with the primary intention to
quantify the heavy flavor quark fraction of the data, ignoring the lepton
structure (which is studied using the leptonic decay models). This
means that they are considerably simpler than the two leptonic models;
in particular the presence of intermediate cascade decays is
ignored. Differences in jet structure and kinematics are studied by
varying the fraction of b quarks vs.\ top quarks in the decays.

The reason for this division between lepton and heavy flavor
properties, is that there is no a priory reason to expect the same
decays of light and heavy flavor quark partners. Any model that
attempts to fit them both simultaneously therefore risks getting many
parameters, thereby reintroducing possible flat parameter directions
as well as unconstrained choices for the model structure. See
sec.~\ref{sec:leptonBCorrel} for a discussion of particular cases when
combined fits might be feasible and useful in a second pass.

\bet
\begin{tabular}{|c|c|c|}
\hline
Model& Particle content, SU(3)$\times$EM, mass & Rate parameters\\
\hline
Btag(Q) &
\begin{minipage}{1.8in}
\begin{tabular}{rcl}
$Q$ & ($\mathbf{\overline3}\times\frac23$ & / $M_Q$)\\
$B$ & ($\mathbf{\overline3}\times\frac23$ & / $M_Q$)\\
$T$ & ($\mathbf{\overline3}\times-\frac13$ & / $M_Q$)\\
${LSP}^0$ & ($\mathbf{\overline1}\times 0$ & / $M_{LSP}$)
\end{tabular}
\end{minipage}
&
\begin{minipage}{3in}
\begin{tabular}{l}
$\sigma_Q=\sigma(gg\to Q\bar Q)$, $Q\to q\,LSP$\\
$\sigma_B=\sigma(gg\to B\bar B)$, $B\to b\,LSP$\\
$\sigma_T=\sigma(gg\to T\bar T)$, $T\to t\,LSP$\\
\end{tabular}
\end{minipage}
\\
\hline
Btag(G) &
\begin{minipage}{1.8in}
\begin{tabular}{rcl}
$G$ & ($\mathbf{\overline8}\times\frac23$ & / $M_G$)\\
${LSP}^0$ & ($\mathbf{\overline1}\times 0$ & / $M_{LSP}$)
\end{tabular}
\end{minipage}
&
\begin{minipage}{3in}
\begin{tabular}{l}
$\sigma_G=\sigma(gg\to GG)$\\
$B_{qq}=B(G\to q\bar q\,LSP)$\\
$B_{bb}=B(G\to b\bar b\,LSP)$\\
$B_{tt}=B(G\to t\bar t\,LSP)$\\
\end{tabular}
\end{minipage}
\\
\hline
\end{tabular}
\caption{The particle content and parameters of the b-tag models. The
b-tag model for Quark partner production or Btag(Q) 
includes flavor-conserving pair production of
light-flavor quark partners (modeled using only one light quark
partner state), bottom quark partners and top quark partners, all with
the mass $M_Q$. They each decay directly to the LSP, emitting a
light quark, bottom quark and top quark respectively in the
decay. In the b-tag model for Gluon partners, Btag(G), gluon partners
of mass $M_G$ are pair produced. They decay to the LSP in three modes,
emitting two light quarks, two bottom quarks and two top quarks
respectively.}
\label{tab:Btags}
\eet

The b-tag models are shown in Table \ref{tab:Btags}, and the lower
panes of Fig.~\ref{fig:ModelPicture}. The b-tag model for quark
partner production, Btag(Q), has a total
of 2 mass parameters, $M_Q$ and $M_{LSP}$ and 3 cross sections,
$\sigma_{Q}$, $\sigma_{B}$ and $\sigma_{T}$. In the b-tag model for gluon partner production, or Btag(G), we assume
that the gluon partners do not carry flavor, and so we use only one
primary production mode with multiple decays. In its simplest form,
Btag(G) has only a single light flavor mode along with a $b\bar{b}$
mode. It is often useful to include $t\bar{t}$ modes as well,
especially if there is evidence for $W$ bosons from the
leptonic fits, or heavy flavor-lepton correlations. Throughout most of
this paper, Btag(G) will include a $t\bar t$ mode. In all, Btag(G) has
2 mass parameters, $M_G$ and $M_{LSP}$, 3 branching ratios,
$B_{qq}$, $B_{bb}$ and $B_{tt}$, and 1 production cross section
$\sigma_{G}$.

\subsection{Observables for Constraining Simplified Model Parameters}\label{sec:fittingSMs}

In this section, we present a general discussion about how to fit the
simplified models to experimental data. These methods will be used,
and elaborated on, in Sections~\ref{sec:example1} and \ref{sec:example2}.

The observables we discuss are very standard, and indeed among the
simplest used in the literature.  The minimality of the simplified
models will, however, allow us to use these very simple observables to
fully constrain their parameters in a transparent manner.

\subsubsection{Mass Signatures}\label{ssec:MassSignatures}
Scalar $\sum p_T$-type observables $H_{T,X} = \sum_{i \in X} |p_T(i)|$
are common mass estimators for SUSY-like topologies.  There are many
conventions for the set $X$ of objects included in the sum; we will
include up to four jets with the highest $p_T$'s, all leptons, and the
missing energy.  This ``effective mass'' is sensitive to the mass
difference $M_1-M_{LSP}$ between the produced particle and the
lightest stable particle (times a prefactor in the range $\approx 1.5
- 1.8$).  The peak location depends on both the production matrix
element and the decay chain undergone by the heavy particles.  This
will be seen in our examples, where the favored mass estimate differs
from one simplified model to another.

Intermediate mass scales can be constrained by lepton kinematics,
particularly by the dilepton invariant mass if a prominent dilepton
mode exists.  The dilepton invariant mass
distribution will have either an edge discontinuity if the decay
proceeds through an on-shell lepton partner $L$, or endpoint if it is
three-body, at 
\begin{equation}
M_{edge} = \frac{\sqrt{(M_2^2 - M_L^2)\, (M_L^2 -
    M_{LSP}^2)}}{M_L},\qquad\mbox{or }M_{end} = M_2 - M_{LSP}.
\end{equation}
The distinction, which can be difficult to discern at low statistics,
is whether the distribution in $m_{\ell\ell}$ is discontinuous at
$m_{edge}$, or falls continuously to zero at $m_{endpoint}$.  Lepton
$p_T$ distributions provide a second constraint on kinematics,
necessary to fix two out of three masses that play a role in the
on-shell decay ($M_2$, $M_L$, and $M_{LSP}$).

The constraints above always leave one mass unconstrained.  Absolute
mass scales can be determined from endpoints in observables such as
$M_{T2}$ \cite{Barr:2007hy,Cho:2007dh}, or simultaneously using
constraints from several decay chains (e.g.~\cite{Nojiri:2007pq}).
Initial-state QCD radiation may also be useful in measuring masses, or
at least provide a strong cross-check \cite{SUSYJets}.  These
techniques however rely on knowledge about the decay chains of the
produced particles.

For the early stage of analysis we consider, when
decay chains are unknown, we take a more pragmatic approach --- it is
typically sufficient to present a fit at one mass for the LSP, and the
lowest mass consistent with data is a good benchmark for this purpose.
In addition, it is useful to vary the masses coarsely over the widest
consistent range.  For this purpose, rough upper and lower bounds on
the mass can be obtained from the new-physics production
cross-section; these are particularly useful for gluon partners, whose
the cross section is dominantly determined by QCD couplings, the mass and
spin of the produced particles, and parton luminocities.

\subsubsection{Signatures for Leptonic Model Rates}\label{ssec:LeptonSignatures}

With the masses in the models fixed, the canonical lepton counts are
in principle sufficient for constraining the branching fractions of
the different decay modes in the leptonic models (with the exception
of the $\ell\nu$ vs.\ $W$ boson fraction): counting dilepton ``$Z$
candidate'' events ($e^+e^-$ or $\mu^+\mu^-$ pairs with dilepton
invariant mass in a window around $M_Z$), events with opposite-sign,
same-flavor pairs of leptons that do not reconstruct near the $Z$
mass, and single-lepton events, can be used to constrain the frequency
of events with one $Z$, lepton partner dilepton cascades, and
single-lepton cascades (from lepton partners or $W$),
respectively. Single-lepton decays are also constrained by the
frequency of events with opposite-flavor or same-sign dileptons, and
(to a lesser degree, due to smaller event samples) the different decay
fractions are constrained by 3- and 4-lepton events. The production
cross section and branching fraction for direct decay to the LSP can
be estimated from the total number of events passing the cuts.

In some cases, lepton kinematics may permit an empirical distinction
between $W$ and lepton partner-mediated $\ell\nu$ modes, but they
are often quite kinematically similar.  In this case, it is difficult
to fix these two modes independently, and the strongest handle will be
the jet structure of different types of events. In the interest of minimality, it is reasonable to expect presence of
$W$ but not $\ell\nu$ (from lepton partners) if evidence is seen for
$Z$'s, and $\ell\nu$ but not $W$ if evidence is seen for
$\ell^+\ell^-$ from lepton partners (unless there are $W$'s from top
decays). Another effect to which one may be sensitive, is an
anti-correlation between jet multiplicity and lepton counts, in the
case of a $W$, which should be missing in the case of pure
$\ell\nu$. In general, we suggest fitting to extremes (no $\ell\nu$/no
direct decay
and no $W$ respectively) as well as a free fit, to study the
systematic effects related to this ambiguity. Since different
jet cuts are typically used on the different lepton number signal
regions, the relation of $W$ and $\ell\nu$ events might affect the
fits of other parameters, such as $Z$ vs.\ $\ell\ell$, or the total
cross section vs.\ $B_{LSP}$, in particular for Lep(Q) fits (see
the examples in sec.~\ref{sec:example1} and \ref{sec:example2}).

\subsubsection{Signatures for B-tag Model Fractions}\label{ssec:HFSignatures}

The most important function of the b-tag models is to
parameterize the heavy flavor fraction of the events, why the most
important discriminator here is the frequency of events with different
number of $b$-tags. 

If a sample has a large fraction of events with leptons, this is good
evidence for the presence of leptonic or electroweak cascades, which
are \emph{not} present in the two $b$-tag models.  In this case, it is
most reasonable to constrain heavy-flavor decay modes in a
lepton-inclusive event selection, using the proportion of top quark
events to study systematics, similarly to the comparisons between
$\ell\nu$ and $W$ decays for the leptonic models. 

If, on the other hand, a sample has fewer leptons (and, in particular,
if there is no evidence for $Z$ or $\ell^+\ell^-$ decay modes), it is
quite interesting to see whether these can be explained entirely by
$b$ and $t$ production processes, with no electroweak cascades. In
that case, the top quark fraction can be fitted using the $b$ tag
fractions in different lepton number signal regions.

In general, we suggest that the approximately ``flat direction''
corresponding to including only $b$ quarks or $b$ and $t$ quarks in
the decays, should be investigated in a similar manner that the
difference between $W$ and $\ell\nu$ decays can be studied in the
Lep(Q/G) models. By doing the fits with no $b$ and no $t$ decays,
respectively, as well as allowing the ratio between them to float
freely, it is possible to estimate the systematic uncertainty of the
fits due to differences in jet and lepton structure, as well as
investigate to which extent the leptons in the data can be described
by top quarks only or if there are indications for cascade decays
involving leptons.

\subsection{Using the Simplified Model Fits}

For a first characterization of the data, the results of fits of the
simplified models can already by themselves answer many questions, and
in particular contribute to the questions we laid out in
sec.~\ref{sec:PhysicsQuestions}. An example of a case when this is
particularly true will be given in sec.~\ref{sec:example1}
below. However, this is not the only use of the simplified
models. Another purpose is to give people outside an experimental
collaboration a ``target'' to guide them in the attempts to explain
the data, defined independently of the detector. A simplified model that is consistent with data (in some limited and
well-defined sense), after the experimental collaborations' best
accounting for Standard Model backgrounds as well as effects like jet
energy scale, $b$-tag efficiencies, and electron reconstruction, is a
target that physicists outside the collaboration can try to match,
either analytically or with their own simulations.

A simplistic map from any model onto the simplified model space can be
defined by averaging over the decay modes of different states,
weighted by production cross-sections.  The map is not
one-to-one, but rather reflects the wide variety of models that may be
consistent with data, until specifically optimized discriminating
variables have been studied.  Nonetheless, when a simplified model
agrees very well with data (as in the example of Section
\ref{sec:example1}), it is reasonable to look for full models of new
physics that do have one production mode with the branching ratios
found in the simplified models.  

A more robust procedure for precise characterization is to generate
Monte Carlo for the simplified models and compare it to other models;
it is reasonable to expect that, where the simplified model is
consistent with data and (in some simulation environment) with a
proposed model, the model is also a reasonable hypothesis for the
data.  This procedure is is illustrated in detail in Example 2, see
Sec.~\ref{sec:example2} below. For this to be possible, it is of vital
importance that also a set of diagnostics plots are published by the
experiments, with the simplified model fits indicated. It should be
noted that, when the simplified model does not reproduce all
kinematics, the comparison in a different simulator may introduce
systematic effects; optimizing the observables used in the fit to
reduce dependence on detector modeling merits further study.

Once one or several models have been found which in this sense
reproduce the data, theorists can focus on finding further predictions
and discriminating observables due to the models, which can then be
further analyzed by the experiments. The simplified models, together
with comparison plots, would be an excellent starting point for this work.

\section{Example 1: Simple New Physics}\label{sec:example1}
In most of their parameter space, complete models of new physics have
more complex structure than the four deliberately simplified models we
have suggested.  Nevertheless, we will show in this and the next two
sections  how comparing data to the simplified models
provides information about new physics beyond what one can conclude from
published data alone.

The model we consider in this section is a limit of the MSSM that is
well described by the simplified models (we will consider the opposite
case --- a model with far more structure than the simplified models
--- in Section \ref{sec:example2}).  As our purpose is to treat the
SUSY model as an \emph{unknown} signal, we defer a summary of its
physics to the end of Section \ref{sec:ex1_mssm}.

We have separated the discussion into two parts: first, in Section
\ref{sec:ex1_comparison}, the ``experimental'' task of constraining
parameters of the simplified models and comparing them to data, and
second, in Section \ref{sec:ex1_mssm}, the ``theoretical'' exercise of
drawing conclusions about model parameters from these results.  We
will interpret the fits of simplified models to the data in the context
of the MSSM, but there is nothing SUSY-specific about the exercise and
it could be repeated for any of the ``SUSY-like'' models.  We
emphasize that the ``theorist'' need not have access to the raw data
or to an accurate detector simulator, but only to experimental results
of the kind presented in \ref{sec:ex1_comparison}.

\subsection{Comparison of New-Physics Signal With Simplified
  Models}\label{sec:ex1_comparison} We have idealized the experimental
problem in several ways.  We work in six ``signal regions'': five with
exclusively 0, 1, 2, 3, or 4 or more leptons ($e$ or $\mu$), and a
sixth lepton-inclusive region used only in the heavy flavor studies.
Each region has different requirements on jet $p_T$'s, $H_T$, and
missing energy (specified in Section \ref{app:signalRegions}); these
cuts have been chosen to mimic event selections for SUSY searches
proposed by ATLAS or CMS in TDRs and notes.  We have not included
backgrounds in this study, but by design they are expected to be small
and controllable in these signal regions, with the main effect of
slightly increasing the uncertainty of the simplified model fits
(examples of comparison plots including backgrounds will be given in
Sec.~\ref{sec:6Interpretation}. We have represented the
detector by the parameterized simulation program PGS \cite{PGS}.  We
expect the LHC experiments would use the full set of tools they have
available --- modeling of backgrounds validated on control regions,
detailed detector simulation, and corrections applied to signal Monte
Carlo where necessary --- to make the fits that we propose.  Likewise,
the variables we use in this discussion are only representative; the
most discriminating variables that can be reliably modeled should be
used.  The SUSY pseudo-data and all simplified models were generated using
Pythia 6.404 \cite{Sjostrand:2006za}, with the On-Shell Effective
Theory implementation of simplified models as described in Appendix
\ref{app:Implementations}.

In the discussion that follows, we will use four kinds of observables
(also described in Section \ref{sec:fittingSMs}) to find
consistent simplified models:
\begin{description}
\item[Lepton Counts] (the number of events in each signal region, the
  breakdown by sign and flavor in 2- and 3-lepton events, and
  identification of pairs reconstructing a $Z$) constrain the
  total cross-section and branching fractions in each leptonic model,
  but \emph{do not distinguish} $W$'s from an admixture
  of $\ell\nu$ and direct LSP decay modes.  Lepton counts can also be
  useful in comparing bottom to top quark modes in the b-tag models.
\item[Jet multiplicity and kinematics] constrain the ambiguous
  direction within either Lep(G) or Lep(Q) --- distinguishing $W$'s
  and $\ell\nu$ contributions --- as it is sensitive to the hadronic
  $W$ fraction.  The results are quite different between Lep(G) and
  Lep(Q), because they have different hadronic decays of the initial
  state.  Jet-counting should be interpreted with care, especially if
  the qualitative differences between data and a simplified model vary
  depending on jet definitions.  Correlations between lepton and jet
  multiplicities are more robust, but require more statistics.
\item[Overall kinematic distributions] such as $H_T$ and $\not\!\!E_T$,
  as well as lepton kinematics can be used to constrain the masses of
  particles in each of the four simplified models.
\item[$b$-tag Multiplicity] is used to
  determine the $b$ branching ratios in the Btag(Q/G) models, and
  \emph{tagged jet kinematics} is a useful diagnostic.  
\end{description}

We will start by looking at the Lep(G) model, as it turns
out to give the best fit to the data, and dissect it in stages.  
We summarize the ranges
of parameters we will consider in Section \ref{sec:ex1_G-LCM_summary};
sections \ref{sec:ex1_G-LCM_rates}-\ref{sec:ex1_G-LCM_mass} each focus
on constraints on Lep(G) parameters coming from a different set of
observables.  The observables used to constrain Lep(Q) are quite
similar, and the Btag(G) model is significantly simpler than the
leptonic models, so we treat these two more briefly in Sections
\ref{sec:ex1_Q-LCM} and \ref{sec:ex1_HFM2} respectively, focusing on
the notable features of these fits rather than the methodology.

In these studies, we wish to determine what regions of parameter space
are most consistent with data and what inconsistencies cannot be
removed by varying parameters of the simplified models --- we are not
simply interested in a single ``best-fit'' point in parameter space.
The simplified models are small enough that the more ambitious goal is
attainable.  We minimize a $\chi^2$ defined over lepton count
distributions, and quote best-fit cross-section and branching ratios
(subject to a constraint that removes the $W/\ell\nu$ ambiguity).  We
treat overall kinematics and jet counting more qualitatively --- rather than
fitting to these distributions, we illustrate how they are affected in
different extreme parameter choices, and between the Lep(Q) and Lep(G)
models. We use this hybrid approach because quoting a ``best fit''
with error bars is a clear presentation of the leptonic constraints,
but is misleading in the latter two cases, where true uncertainties in simplified model parameters 
are likely to be dominated by detector and modeling systematics.

In plots in this and following sections, we include a lower pane
showing the fraction of the simplified model curve divided by the
pseudo-data, with the data error bars represented as grey bands, in order to
facilitate comparison of models with data.

\subsubsection{Comparison of data to the Lep(G) Model: A First
  Look}\label{sec:ex1_G-LCM_summary} We will illustrate the role of each
discriminating variable by studying parameter points that are nearly
consistent with data, excepting a single discrepant distribution.
Table \ref{tab:ex1_G-LCM_bestFitTable} summarizes the parameter values
of interest.  At each point, we have fixed either $B_W=0$ or
$B_{LSP}=0$ (as motivated in Sec.\ \ref{sec:ex1_G-LCM_WvsLNU}), and
other leptonic branching fractions are optimized using the leptonic
$\chi^2$; we have also fixed $B_Z=0$ because there is no evidence for
a non-zero $Z$ mode.  The table is divided into decay mode
variations (top), consistent kinematic variations (middle), and
inconsistent kinematic variations (bottom).  The first line, ``Model
A'', reproduces all distributions of interest quite well, as do C and
D.  Model A appears as a solid red line in every plot in the next
three sections.  The alternative models will be displayed as dashed or
dotted lines.

\bet
\begin{tabular}{|c|l|c|c|c|c|c|c|c||c|c|c|}
\hline
\multicolumn{12}{|c|}{\textbf{Leptonic Decay Models for Gluon Partners (Lep(G))}}\\
\hline
Label& Description & $M_G$/$M_I$/($M_L^*$)/$M_{LSP}$ & 
          $\sigma_G$ (pb)  & $B_{\ell\ell}$ & $B_{\ell\nu}$ & $B_{LSP}$  &
$B_W$   & $B_Z$  & Leptons & Jets & Kin.  \\
\hline
\multicolumn{12}{|l|}{Decay mode variations with best-fit kinematics}\\
\hline
A &  $B_W=0$ & 600/300/---/100 & 5.4 & 0.15 & 0.43 & 0.42  &
---   & --- &  + &  + & + \\
B &  $B_{LSP}=0$ & 650/300/---/100 & 5.3 & 0.16  & 0.19
& ---   & 0.65 & ---  & + & - & +  \\
\hline \multicolumn{12}{|l|}{Kinematic variations (including on-shell
  kinematics) with no $W$'s, best-fit rates}\\ \hline
C &  $B_{W}=0$ & 750/500/---/300 & 5.3 & 0.15 & 0.47 & 0.38 & --
& -- &  + & + & + \\
D & $B_{W}=0$ (on-shell) & 600/300/200/100 & 5.3 & 0.14 & 0.43 & 0.43 & -- & -- &  + & + & + \\
\hline
E &  $B_{W}=0$ & 700/300/---/100 & 4.7 & 0.16 & 0.41 & 0.43 &
-- & -- &  + & + & - \\
F &  $B_{W}=0$ & 620/400/---/100 & 5.7 & 0.13 & 0.46 & 0.41 &
-- & -- &  + & + & - \\
G & $B_{W}=0$ (on-shell) & 600/400/345/100 & 6.0 & 0.13 &
0.46 & 0.41 & -- & -- &  + & + & -     \\
H & $B_{W}=0$ (on-shell) & 600/300/250/100 & 5.3 & 0.15 &
0.41 & 0.44 & -- & -- &  + & + & ?     \\
I &  $B_{LSP}=0$ & 600/300/---/100 & 5.6 & 0.16 & 0.22 & --
& 0.63 & --- &  + & - & ?     \\
\hline
& ~ stat. error & N/A  & 0.2  & 0.01 & 0.03 & 0.06 & 0.08
& 0.04 &  \multicolumn{3}{|c|}{}\\
\hline
\end{tabular}
\caption{The set of model parameters for the leptonic decay model for
  gluon partners,
  considered in the next three sections.  $M_L$ is only specified for
  models with on-shell sleptons; the cross-section and branching
  fractions quoted are best fits to leptonic data.  Model A is our
  baseline model, which agrees in all distributions and appears as a
  red solid line in every plot (models C and D are also consistent
  with all distributions, but have different spectra).  Model B
  reproduces inclusive kinematics and lepton multiplicities but has
  discrepant jet counts; models E-H reproduce lepton and jet counts in
  the data, but have different inclusive kinematics, while model I has
  the same mass spectrum as A, but inconsistent kinematics (and jet
  counts).  The last three columns summarize which kinds of variables
  (lepton multiplicity, jet multiplicity and inclusive kinematics)
  agree with data (+) and which disagree (-) in each model.  ?'s denote mild disagreement.  
\label{tab:ex1_G-LCM_bestFitTable}}
\eet

\subsubsection{Constrained Branching Fractions from Leptonic Counts}\label{sec:ex1_G-LCM_rates} Aside from $W$'s, each
of the four leptonic decay modes in the Lep(G) model ($\ell\ell$,
$\ell\nu$, $Z$, and direct LSP decays) leads to a very different
leptonic signature.  For example, the only sources of $3\ell$ events are events where one gluon partner
emits $\ell\ell$ and the other $\ell\nu$, and fakes from independently
constrained processes.  Therefore, lepton counts constrain these
branching fractions well if we force the $W$ fraction to
zero. As
evidenced by the mass variations in the bottom half of Table
\ref{tab:ex1_G-LCM_bestFitTable}, significant mass variations do not
affect the results of these fits by more than $\approx 20\%$. 
We can conclude rather robustly that the
total cross-section is $\approx 470$ to 600 pb, the $\ell\ell$
branching fraction is $13-16\%$, and the $Z$ fraction $\lesssim 4\%$.  

One would expect that these models remain consistent if we replace
$\ell\nu$ and direct LSP decays with $W$'s, which decay to one lepton 32\% of the time, and hadronically (which, in lepton
counts, looks like a direct LSP decay) 68\% of the time.  For
instance, models A and B from Table \ref{tab:ex1_G-LCM_bestFitTable}
both reproduce lepton counts as shown in Figure
\ref{fig:ex1_G-LCM_leptonCounts}.

\bef
\includegraphics[height=2.0in]{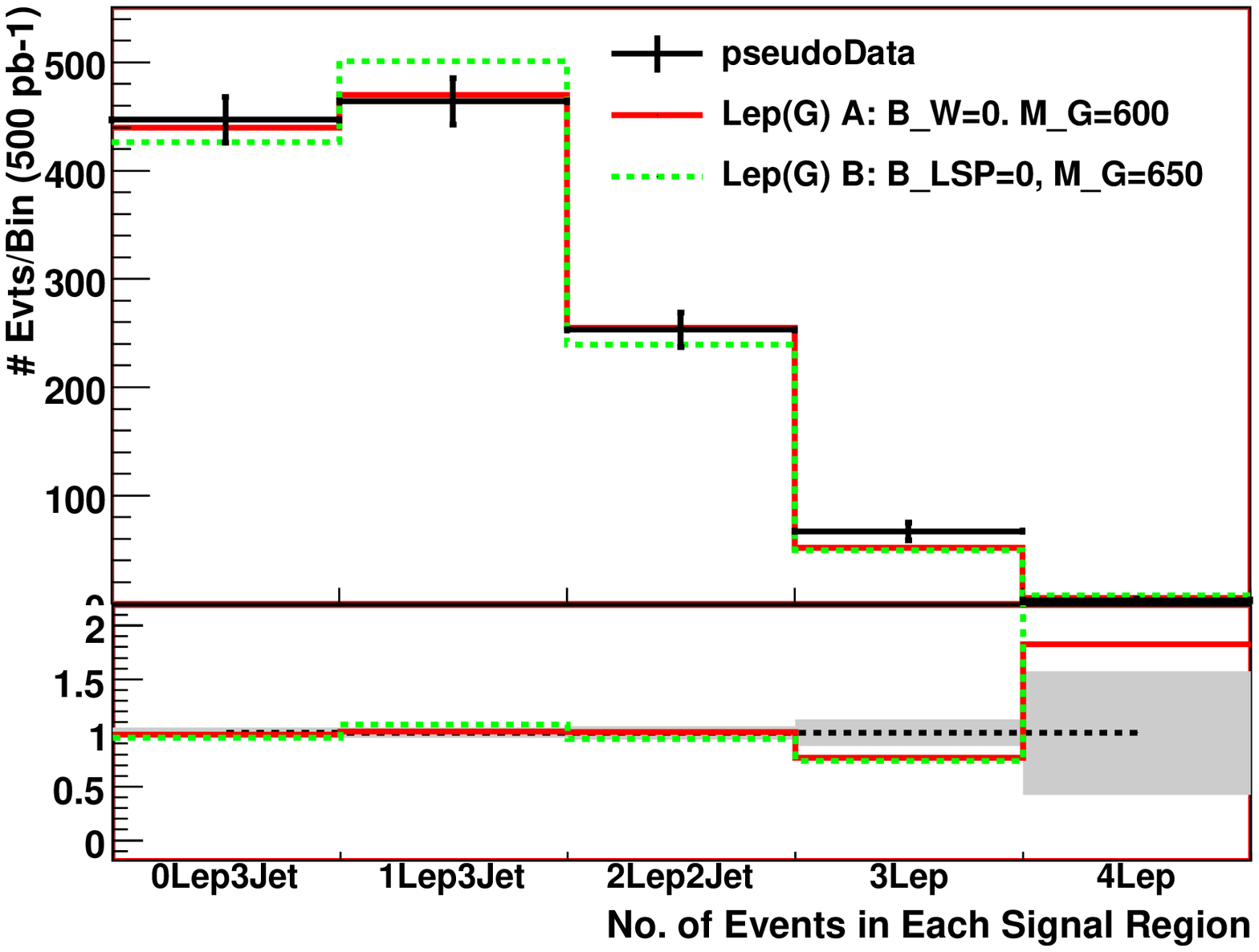}
\includegraphics[height=2.0in]{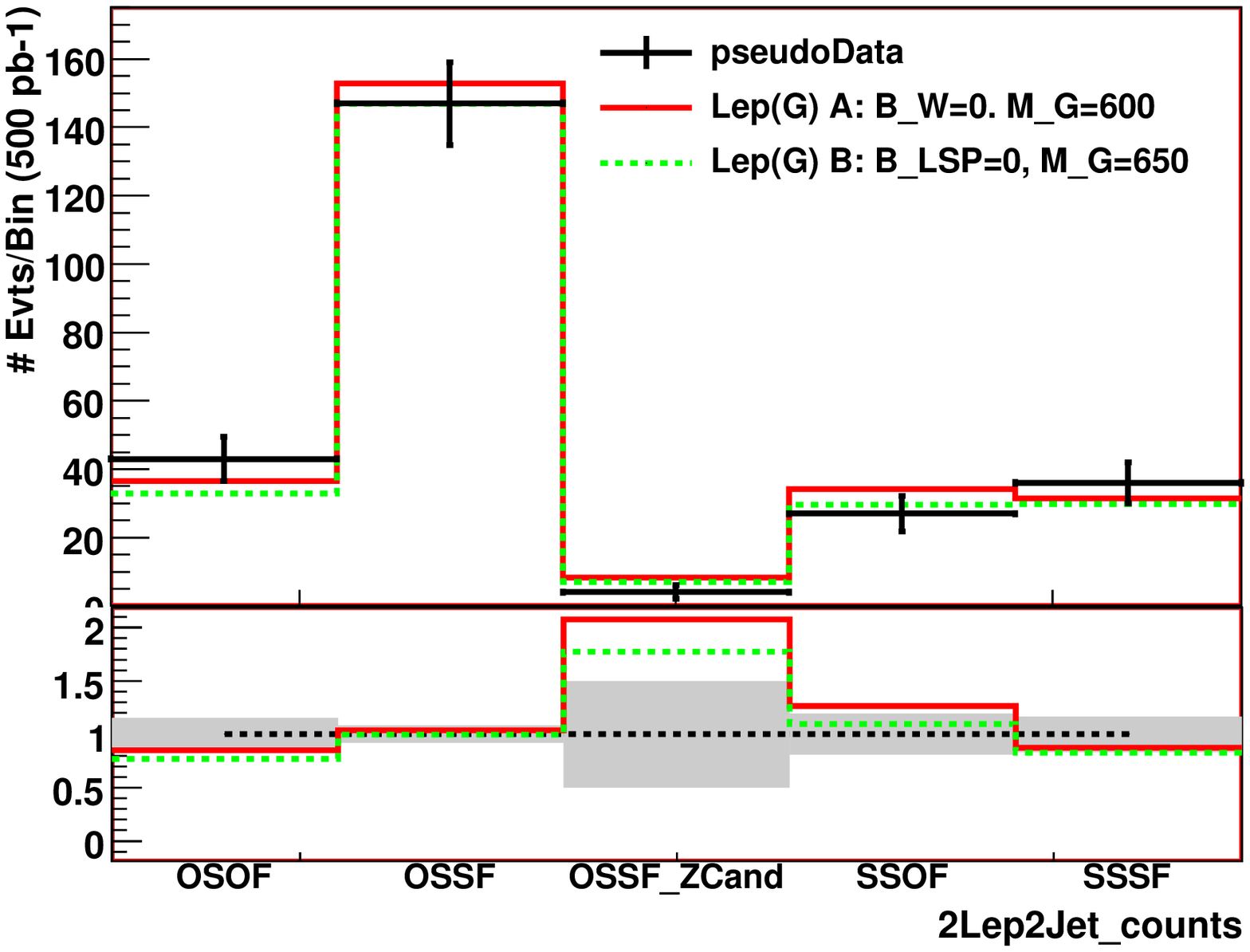}
\caption{Comparisons of lepton count observables between ``data''
  (error bars) and the simplified model Lep(G) with parameter set A
  ($B_W=0$, red solid line) or B ($B_{LSP}=0$, green dashed) from
  Table \ref{tab:ex1_G-LCM_bestFitTable}.  
  \label{fig:ex1_G-LCM_leptonCounts}}
\eef

The ratio of rates of slepton-mediated $\ell\ell$ and $\ell\nu$ events
is a useful constraint on models of new physics, but the $\ell\nu-W$ ambiguity
prevents us from constraining it directly (depending on particle
masses, the two leptonic modes may give rise to different lepton
kinematics, but in this case they are quite similar).  Instead, we
will try to distinguish between \emph{hadronic} $W$'s and direct
decays.

\subsubsection{Jet Kinematics/Counts and the $\ell\nu/W$ Ambiguity}
\label{sec:ex1_G-LCM_WvsLNU} 
From the discussion above, we
expect lepton counts to be nearly invariant under shifts $\delta
Br(W)$ when they are compensated by
\begin{equation}
\delta B_{\ell\nu} \approx - 0.32 \delta B_W \qquad \qquad
\delta B_{LSP} \approx -0.68 \delta B_W.
\end{equation}
By inspecting the ``no-W'' fit parameters in Table
\ref{tab:ex1_G-LCM_bestFitTable}, we see that, if we decrease
$B_W$ while compensating by decreasing $B_{LSP}$ and $B_{\ell\nu}$,
the first to reach zero is $B_{LSP}$.  So the most extreme cases we
can consider while preserving our success in matching lepton counts
with Model A are defined by fixing either $B_W=0$ or $B_{LSP}=0$.
  
The best fits in each of the two extremes are shown in Figure
\ref{fig:ex1_G-LCM_leptonCounts}.  As shown in Figure
\ref{fig:ex1_G-LCM_WvsLNU}, however, the $W$ mode produces too many
jets.  Before drawing strong conclusions from this observation, it is
important to verify that the agreement of the model without $W$'s, and
the disagreement of the model with a high $W$ fraction, are
insensitive to the jet definition.  If, as in this example, the
qualitative conclusion is insensitive to detailed jet definitions, we
can conclude that a gluon-initiated decay chain cannot have a large
fraction of $\ell\nu$ events coming from $W$'s.

\bef
\includegraphics[width=3in]{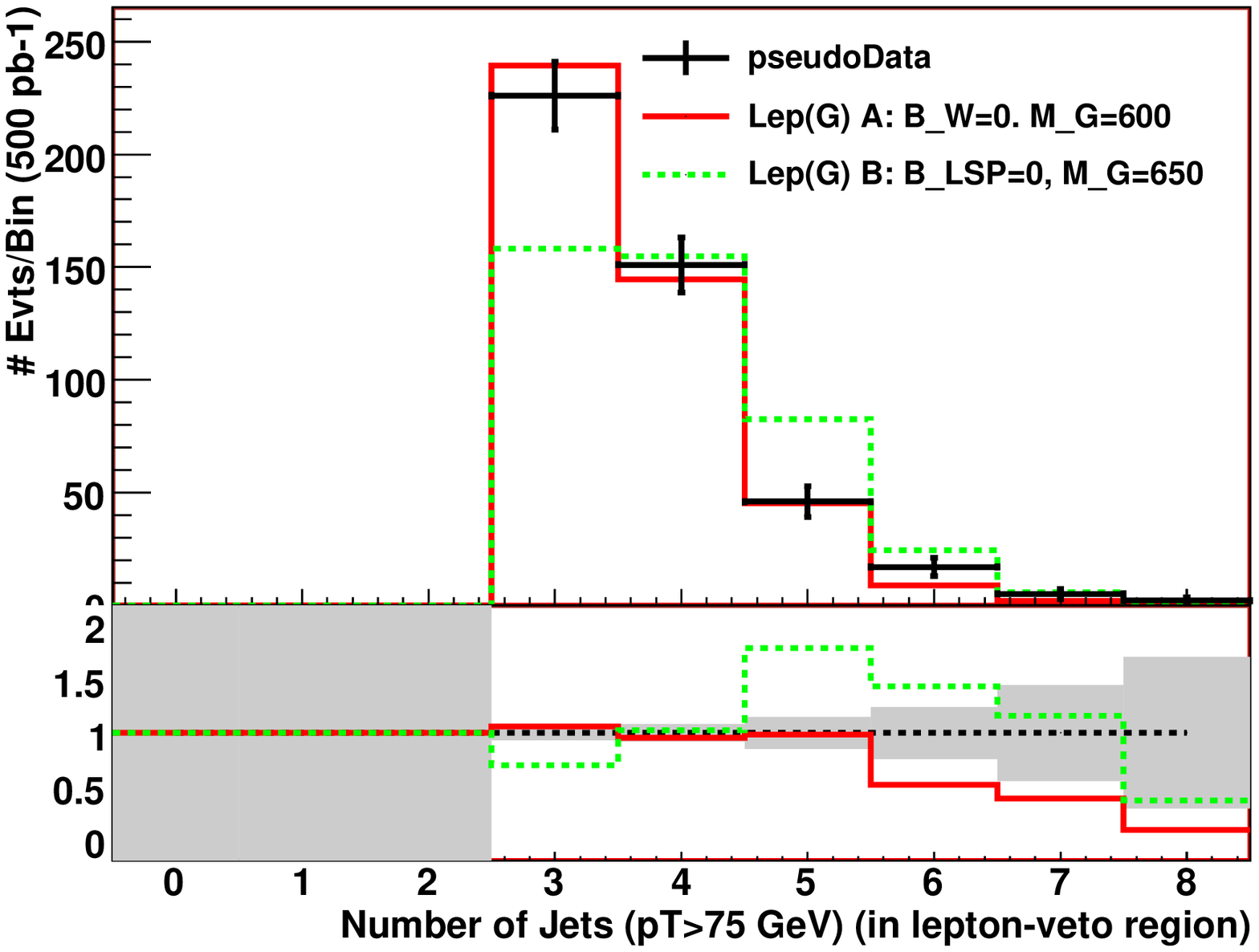}
\includegraphics[width=3in]{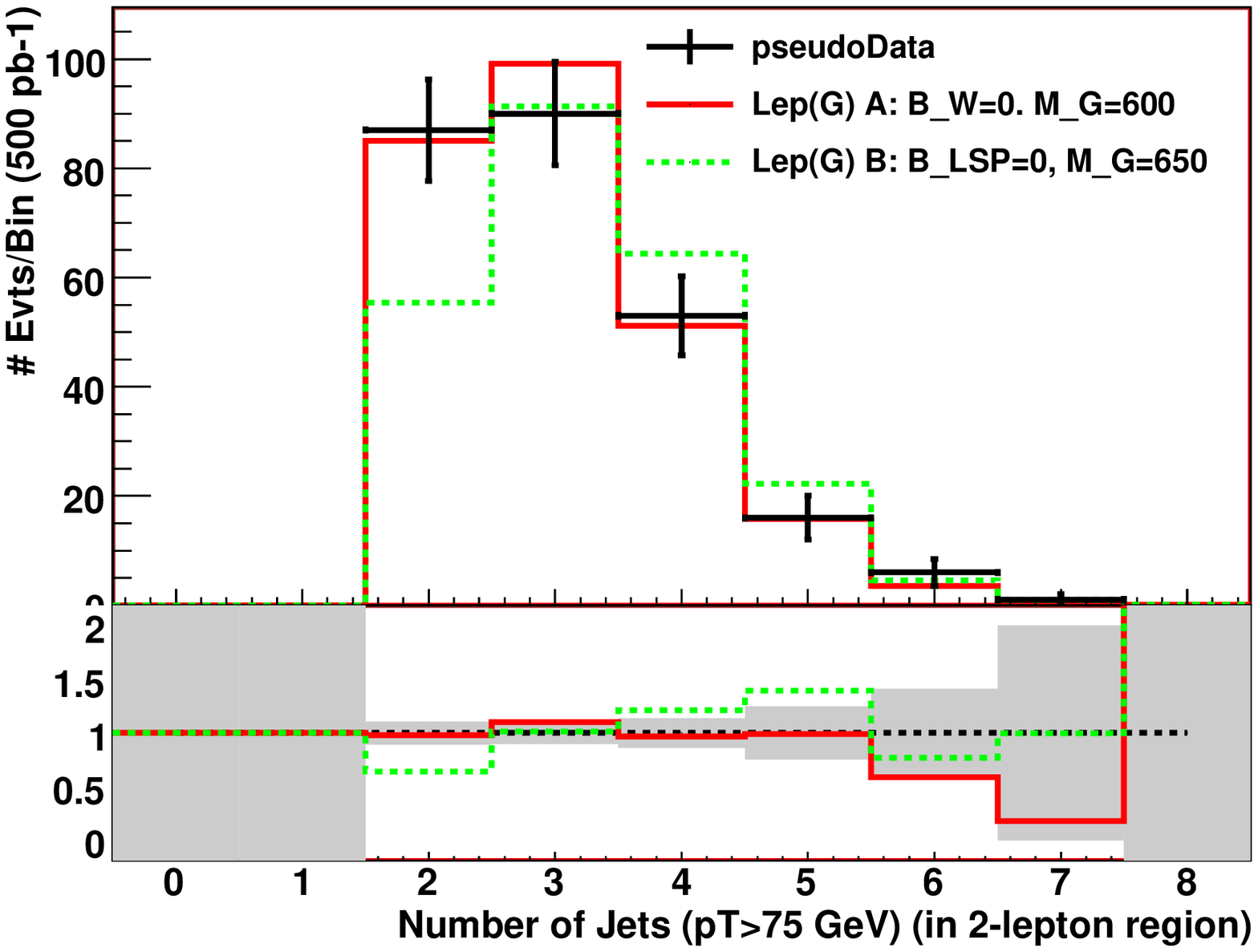}
\caption{Jet counts (in 0- and 2-lepton regions) between ``data''
  (error bars) and the simplified model Lep(G) with parameter set A
  ($B_W=0$, red solid line) or B ($B_{LSP}=0$, green dashed) from
  Table \ref{tab:ex1_G-LCM_bestFitTable}. \label{fig:ex1_G-LCM_WvsLNU}}
\eef

\subsubsection{Mass Variations in Lep(G) Models} \label{sec:ex1_G-LCM_mass} 
Kinematic distributions such as $H_T$ and the di-lepton invariant mass
constrain mass splittings between new-physics particles.  Roughly,
$H_T$ is sensitive to changes in the top-to-bottom mass splitting
($M_G - M_{LSP}$), such as between lines A and E in Table
\ref{tab:ex1_G-LCM_bestFitTable}, while $m_{\ell\ell}$ is sensitive to changes in the
intermediate splitting $M_I-M_{LSP}$ (e.g. lines A and F).
These variations are shown in Figure
\ref{fig:ex1_LCM2_massVarOffShell}.  In each case, we use the best-fit
branching ratios and cross-section, with $B_W=0$ fixed.

As mentioned in Section \ref{ssec:MassSignatures}, one direction in mass space does not significantly affect these
kinematic distributions --- it corresponds approximately (but not
exactly) to shifting the masses of all new particles by the same
amount (e.g. compare lines A and C from Table
\ref{tab:ex1_G-LCM_bestFitTable}, with $M_G=600$ and 750
respectively).  

\bef
\includegraphics[width=3in]{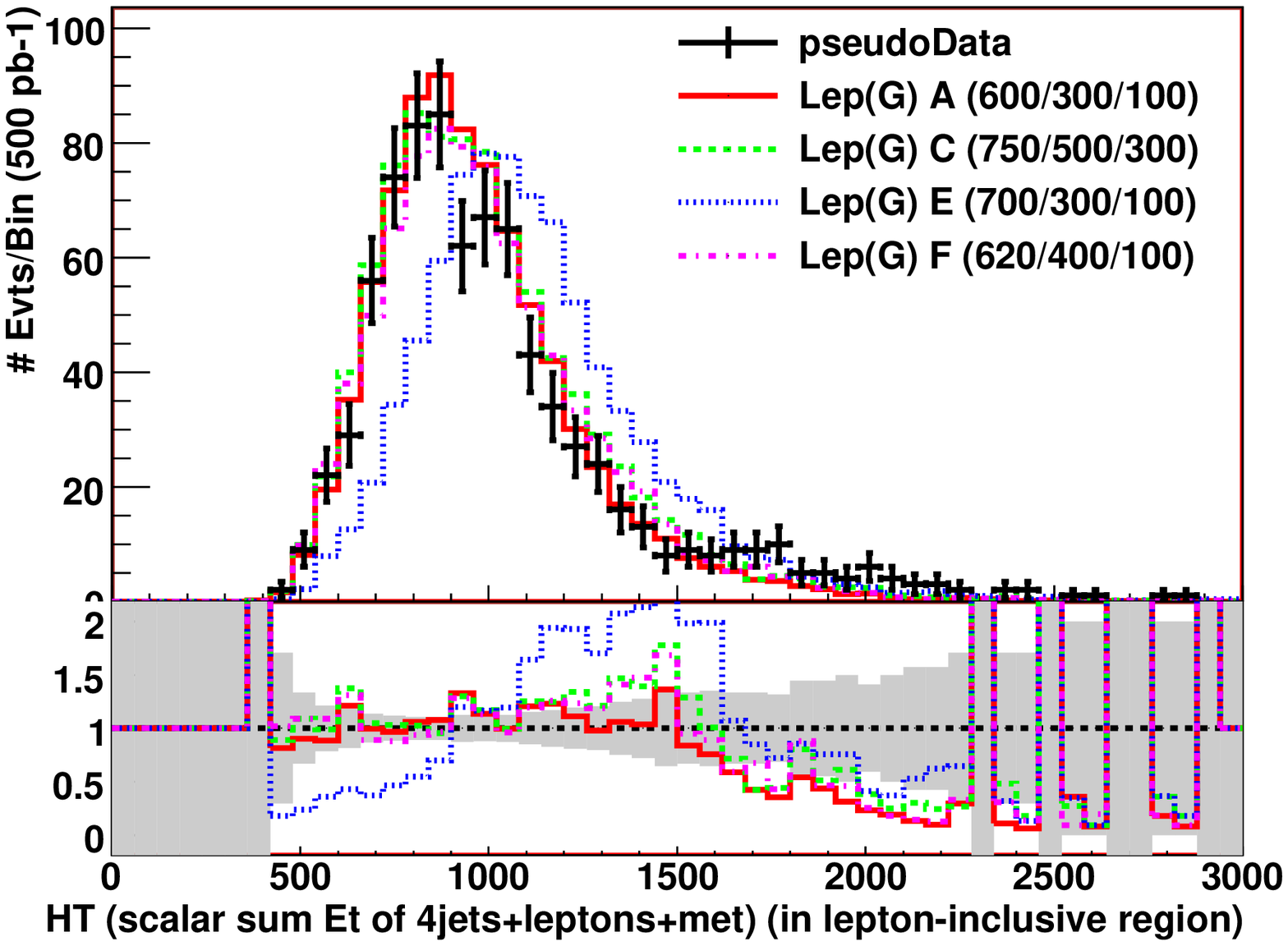}
\includegraphics[width=3in]{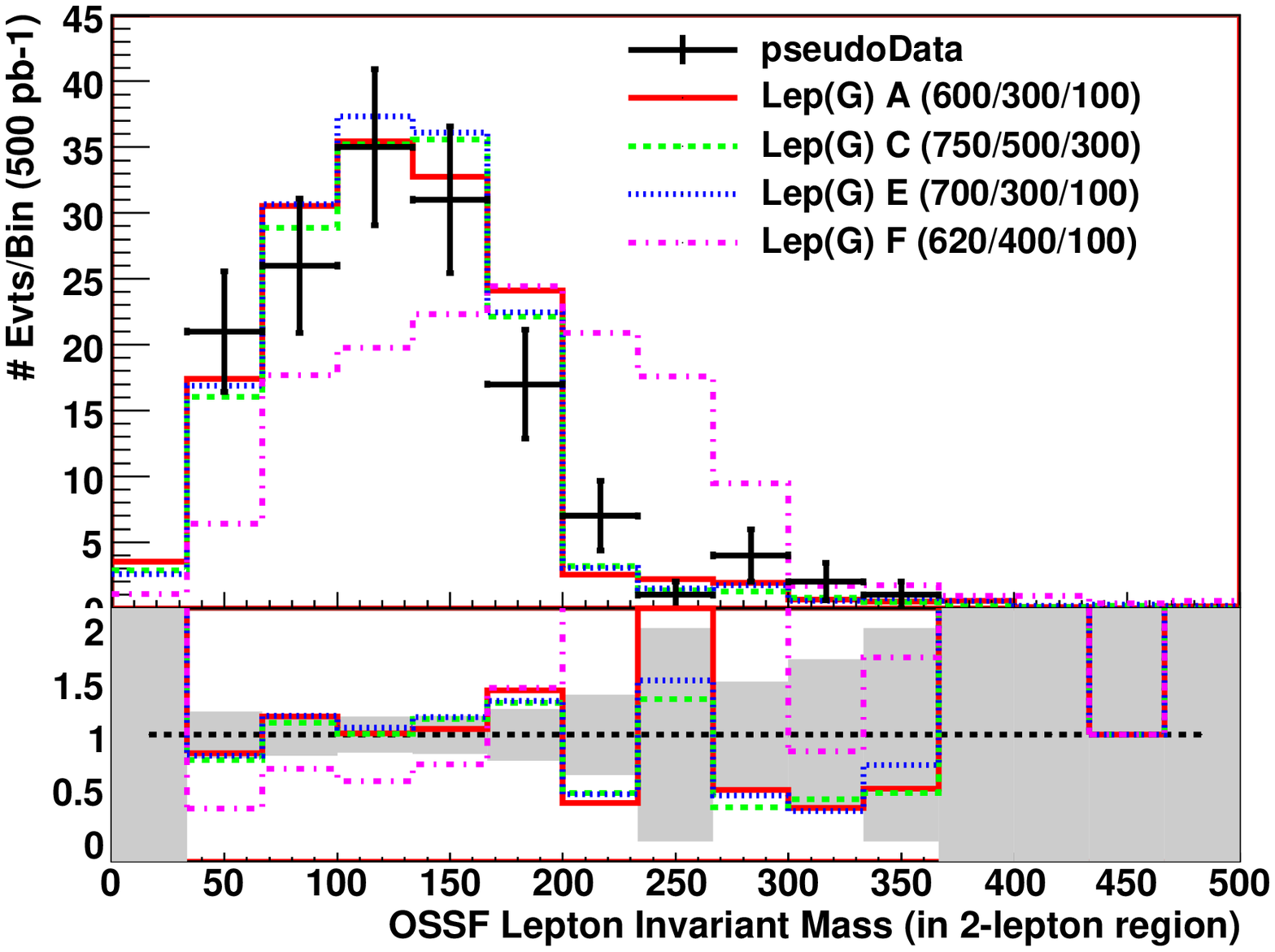}
\includegraphics[width=3in]{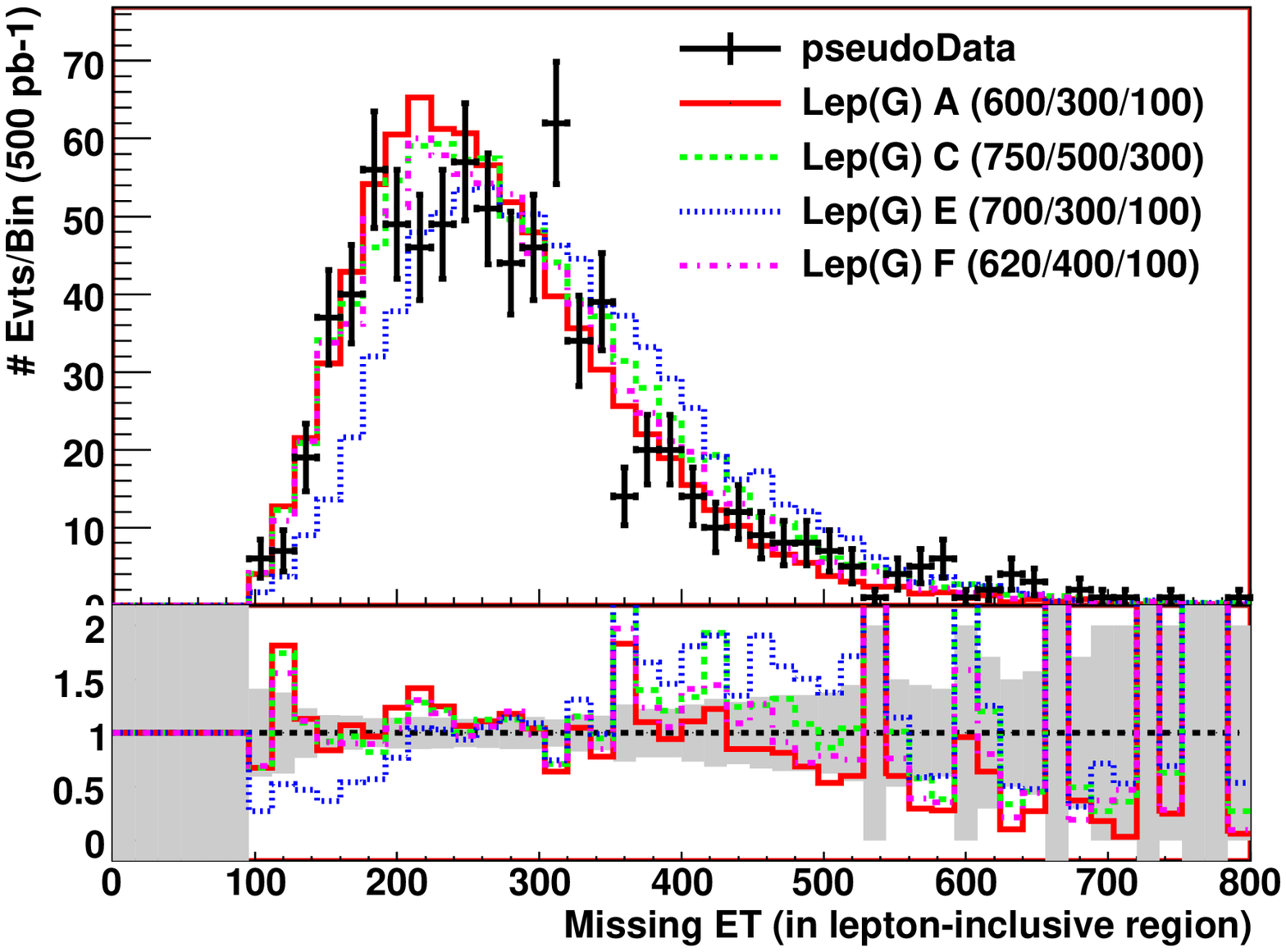}
\includegraphics[width=3in]{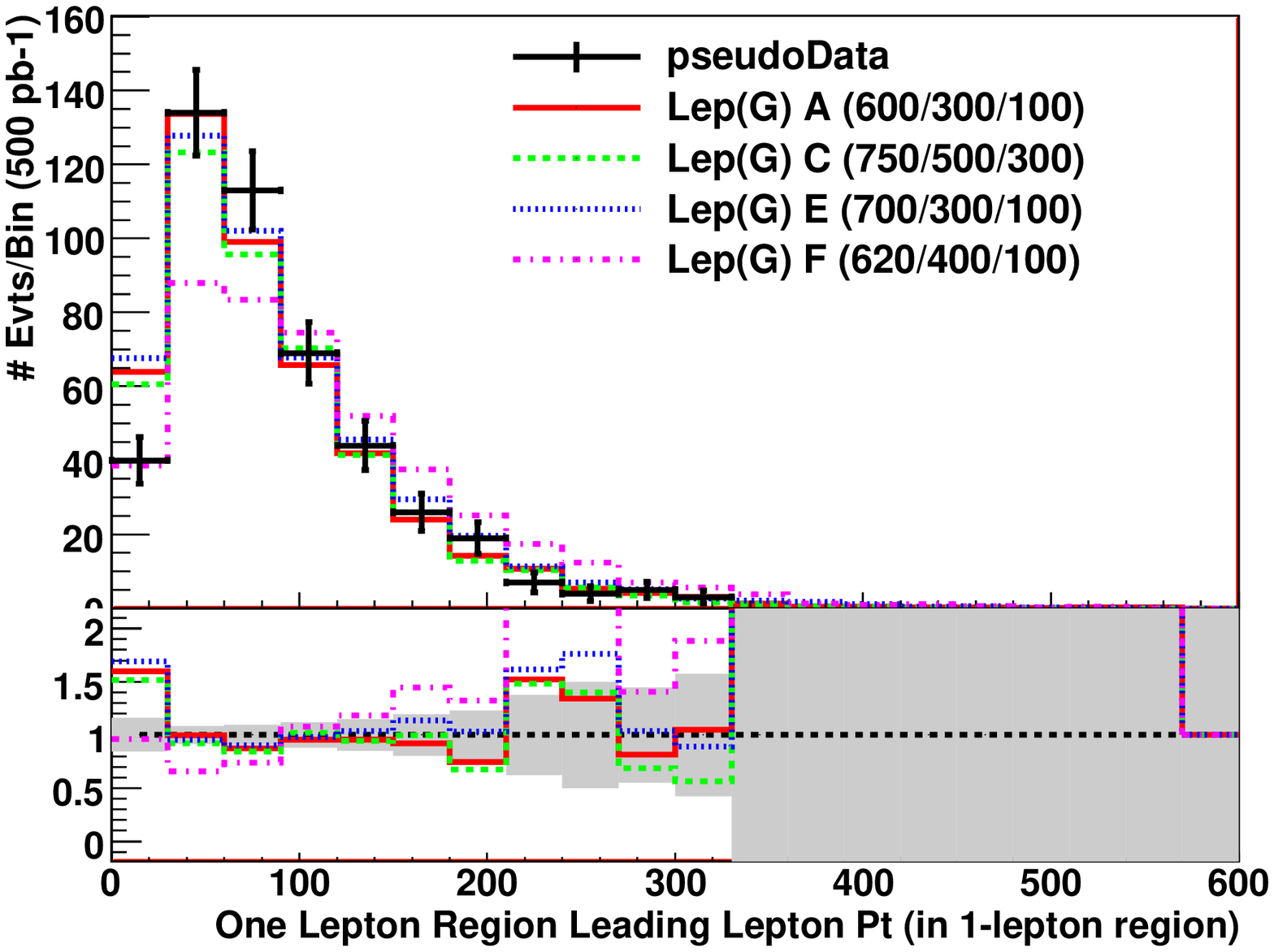}
\caption{Two variables that constrain the kinematics are $H_T$ and the
  $\ell^+\ell^-$ invariant mass in dilepton events (top left and
  right, respectively).  $\MET$ and single-lepton $p_T$ have similar
  sensitivity (bottom left and right).  Over the pseudo-data (black
  error bars), we show Lep(G) models at four parameter values, all
  from Table \ref{tab:ex1_G-LCM_bestFitTable}: the best fit A (red,
  solid), and mass variations C, E, and F as described in the plot
  legends.  Model C (the green dashed line) is globally consistent
  with both the data and model A, although it has a rather different
  mass spectrum; E and F are constrained by $H_T$ and
  $m_{\ell^+\ell^-}$,
  respectively.\label{fig:ex1_LCM2_massVarOffShell}} \eef

Given that the data has a significant $\ell\ell$ branching fraction,
and in fact is consistent with a slepton-mediated $\ell\nu$ branching
fraction as large as 45\%, it is reasonable to ask whether the slepton
could be on-shell (we would typically expect the 3-body decay to have
a much smaller branching fraction).  As shown in Figure
\ref{fig:ex1_LCM2_massVarOnShell}, at an integrated luminosity of
500 $pb^{-1}$ both off- and on-shell kinematics
are
consistent with the data; for example is all kinematics well described by line
D from Table \ref{tab:ex1_G-LCM_bestFitTable}.  We are again sensitive to the
intermediate masses, which must produce a kinematic edge at $\approx
200$ GeV and reproduce lepton $p_T$'s.  These constrain, for example,
lines G and H of Table \ref{tab:ex1_G-LCM_bestFitTable}.

\bef
\includegraphics[width=3in]{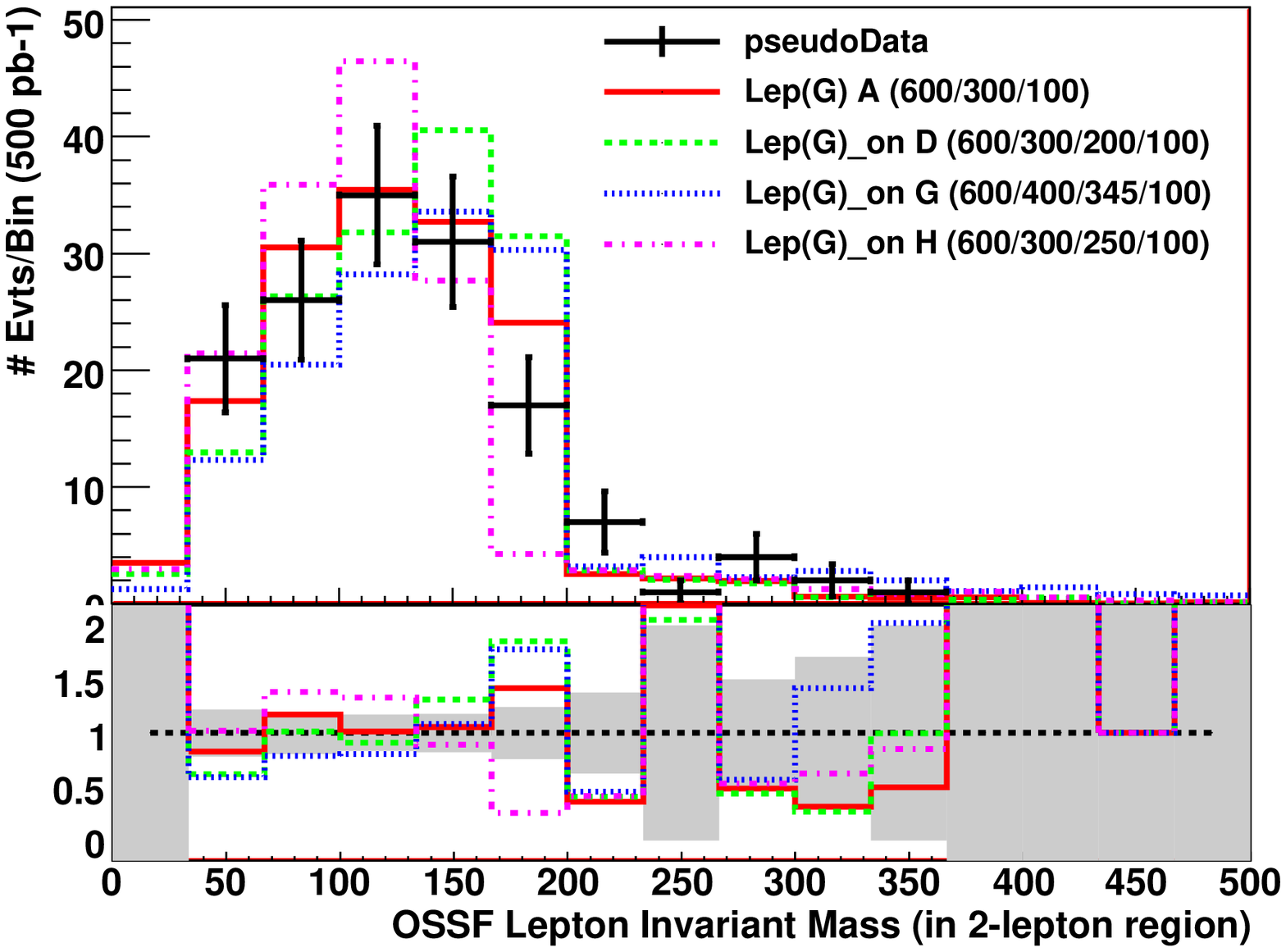}
\includegraphics[width=3in]{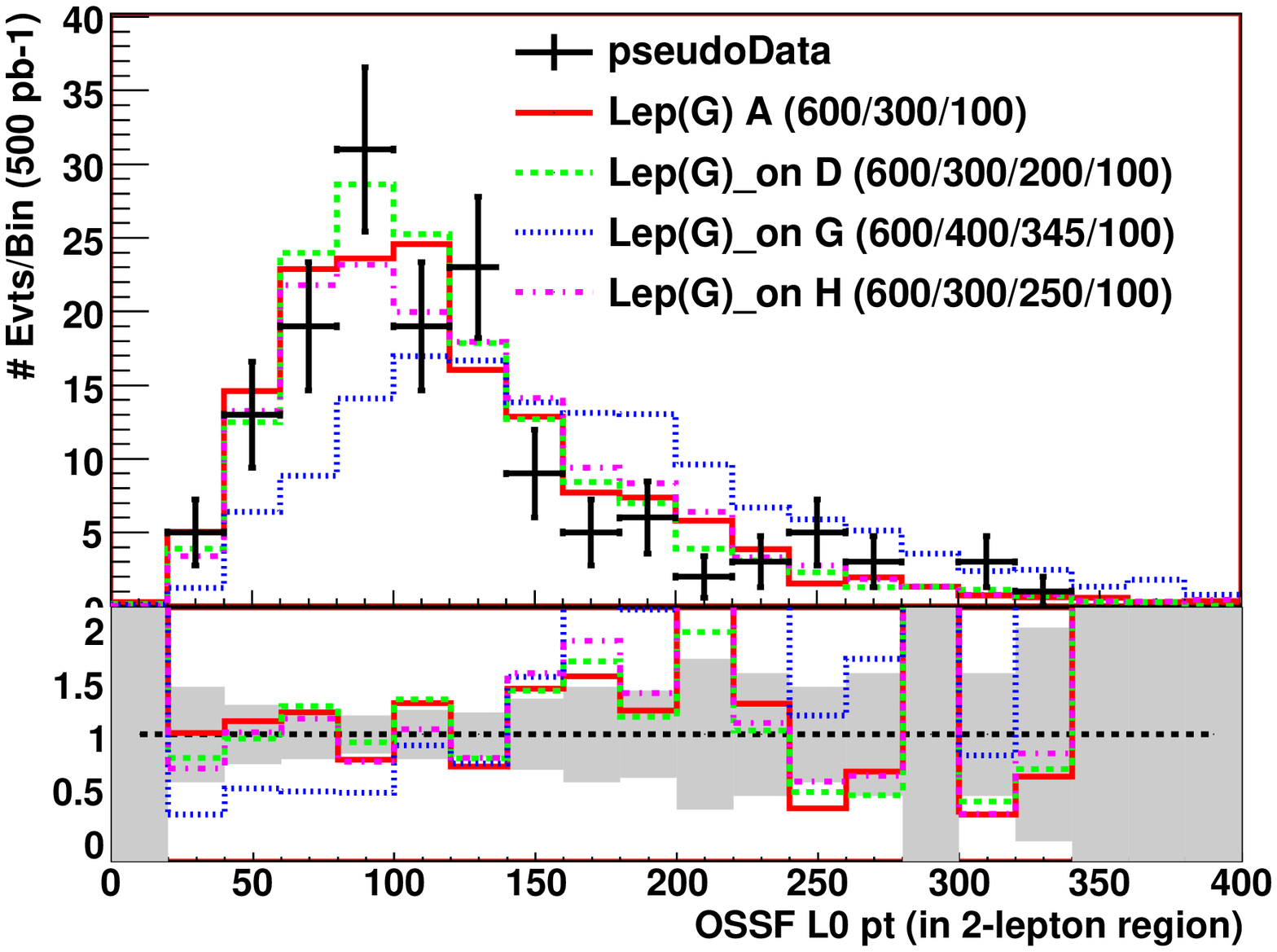}
\caption{Two observables from dilepton events used to constrain
  lepton kinematics and discriminate between on- and off-shell slepton modes: the dilepton invariant mass and the $p_T$ of the
  harder lepton.  We show four models:  one off-shell and one on-shell model that adequately reproduce all kinematics (A --- red solid, and D --- green dashed respectively) , and two inconsistent on-shell variations (line H --- blue dotted, excluded by the $m_{\ell\ell}$ distribution, and line G --- purple dash-dotted excluded by the leading lepton $p_T$). The models are specified fully in Table
  \ref{tab:ex1_G-LCM_bestFitTable}.  \label{fig:ex1_LCM2_massVarOnShell}}
\eef

It is worth noting that kinematic distributions such as $H_T$ depend
on branching fractions of the parent particle.  For instance, if we
substitute $W$'s for some of the $\ell\nu$ fraction of best-fit model
A, as in the second line of Table \ref{tab:ex1_G-LCM_bestFitTable},
the peak of the $H_T$ distribution shifts downward (see Model I, the
green dashed line on the left panel Figure \ref{fig:HT_varyBranching}).  To compensate
for this, we must raise $M_G$, as was done in model B in
Table \ref{tab:ex1_G-LCM_bestFitTable} (see right panel of Figure
\ref{fig:HT_varyBranching}).  A weakly constrained $B_W$
gives rise to a systematic uncertainty in the mass of the gluon
partner \emph{within the Lep(G) model} (more precisely, this is an
uncertainty in the mass difference between the color octet and
invisible LSP).  

\bef
\includegraphics[width=3in]{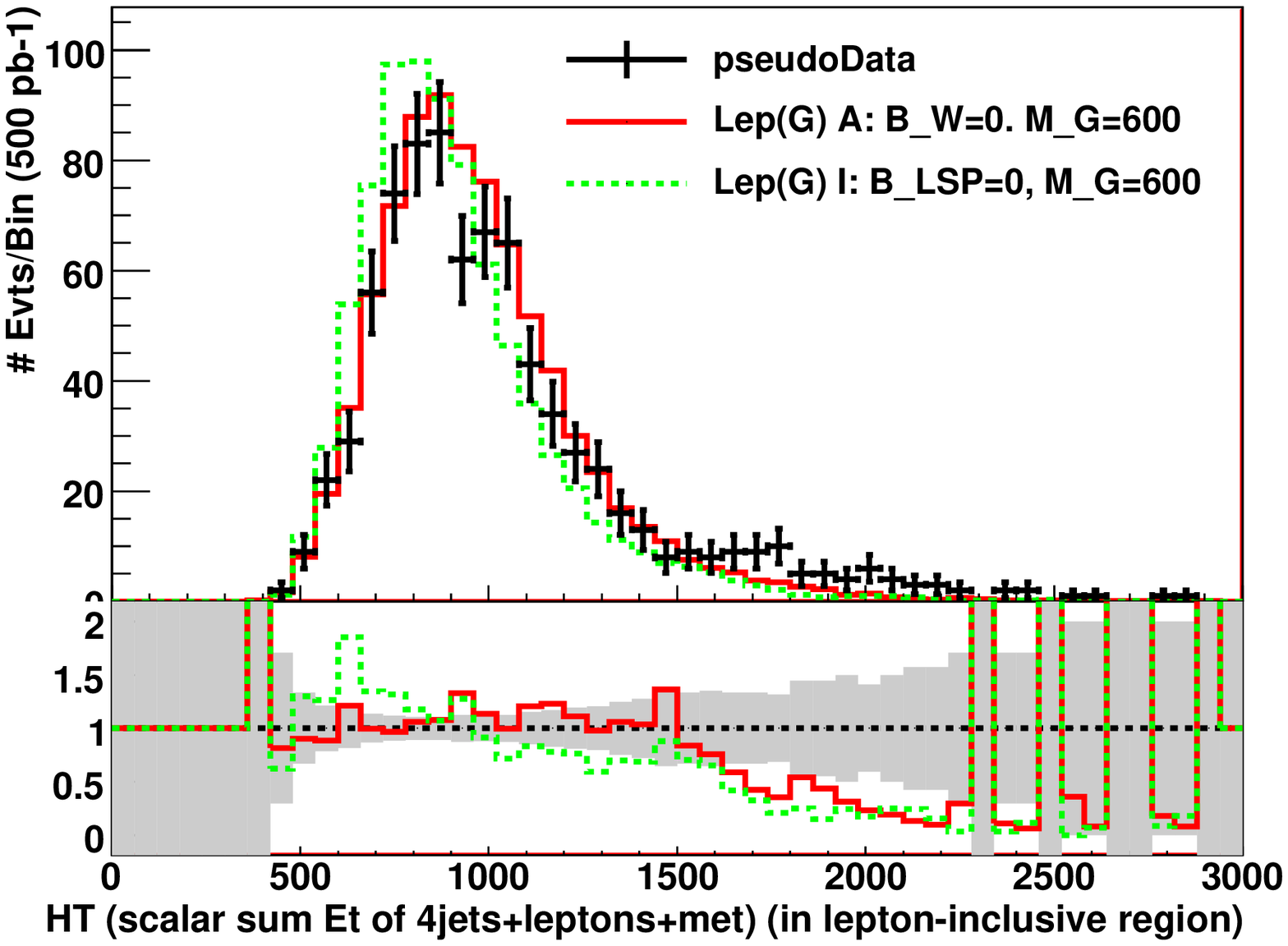}
\includegraphics[width=3in]{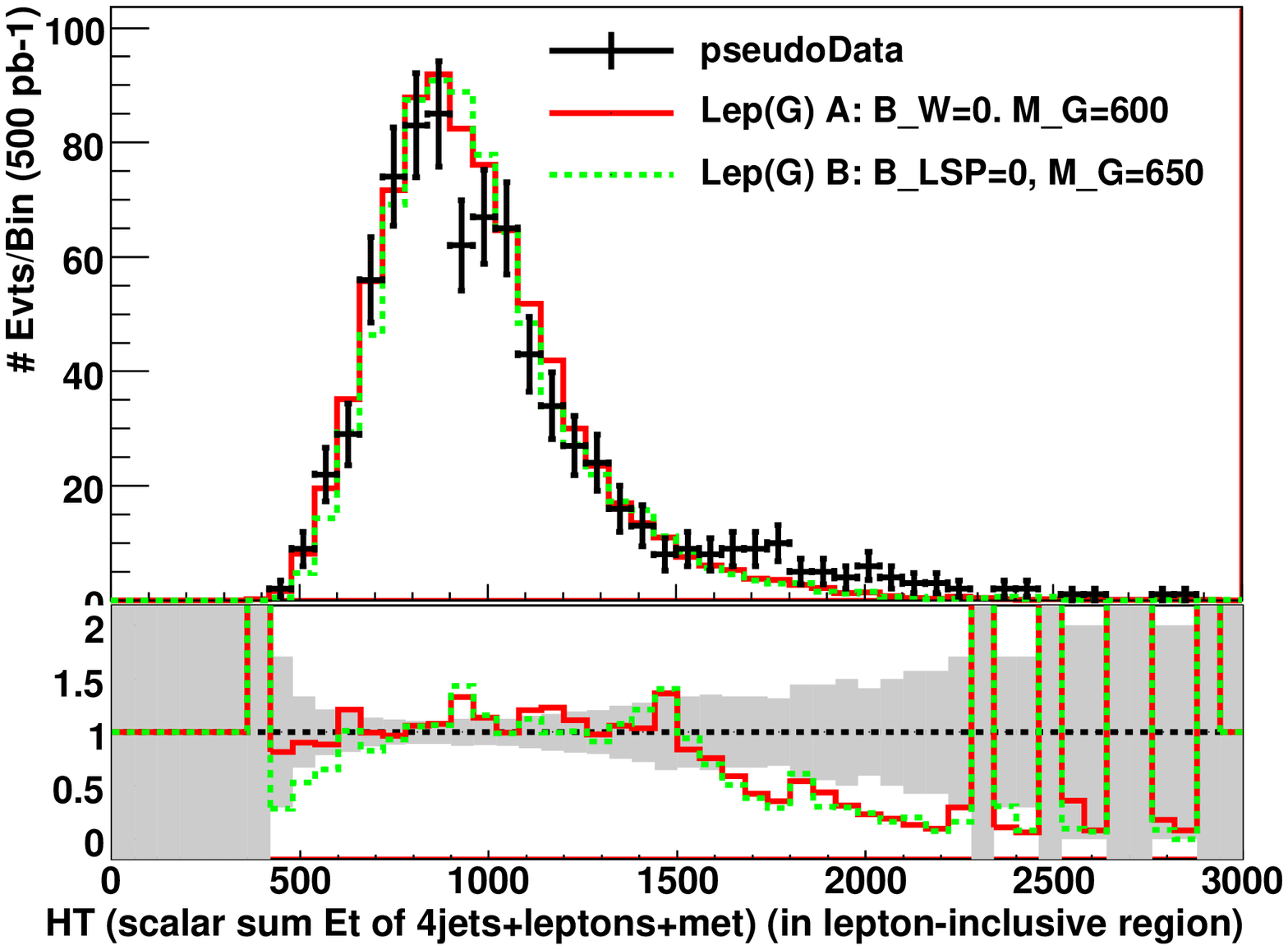}
\caption{ Left: Effect on $H_T$ distribution from varying only
  branching fractions in the Lep(G) model, while maintaining
  consistency with lepton counts.  The red solid and green dashed lines on the left correspond
  respectively to models A and I in Table
  \ref{tab:ex1_G-LCM_bestFitTable}, which have identical spectra.
  Right: The effect can be compensated for by changing the spectrum
  with the branching fractions (lines A in red and B in green from the table).
 \label{fig:HT_varyBranching}
}
\eef

\subsubsection{Comparison of Lep(Q) Models}\label{sec:ex1_Q-LCM}
Constraints on the Lep(Q) models are quite similar to those discussed
above.  Again, we can reproduce lepton counts and event kinematics for
suitable mass choices, with or without $W$ decays, as shown in Figure
\ref{fig:ex1_Q-LCM_goodFits} (the model parameters used are summarized
in Table \ref{tab:ex1_Q-LCM_bestFitTable}).  The jet structure,
however, is inconsistent with production dominated by quark partners,
for both extreme cases, as illustrated
by Figure \ref{fig:ex1_Q-LCM_jets}. The discrepancy is most clearly
seen in the 2-lepton region, which even in the $W$-rich scenario has
few hadronic $W$'s.
\bef
\includegraphics[height=2.0in]{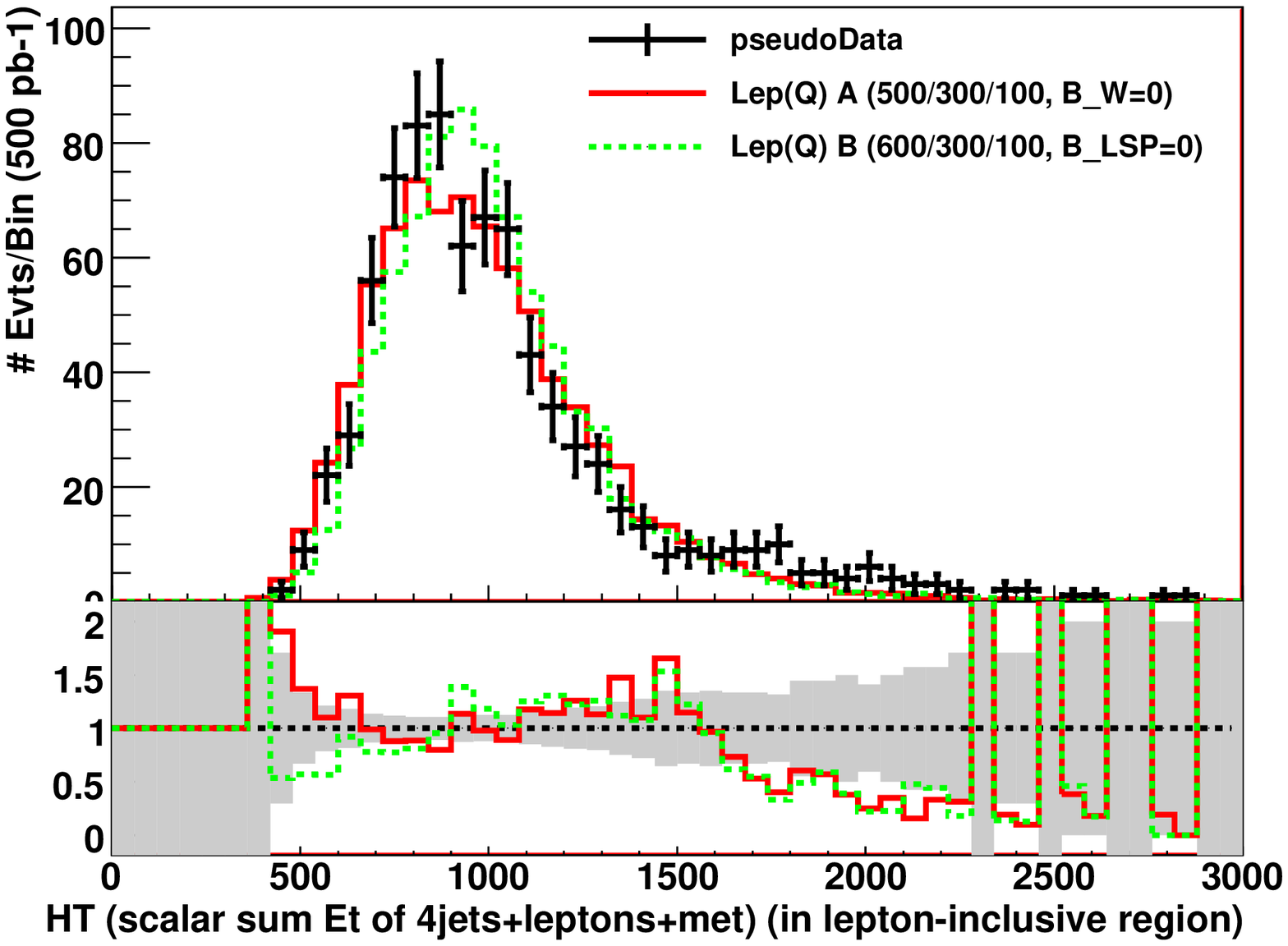}
\includegraphics[height=2.0in]{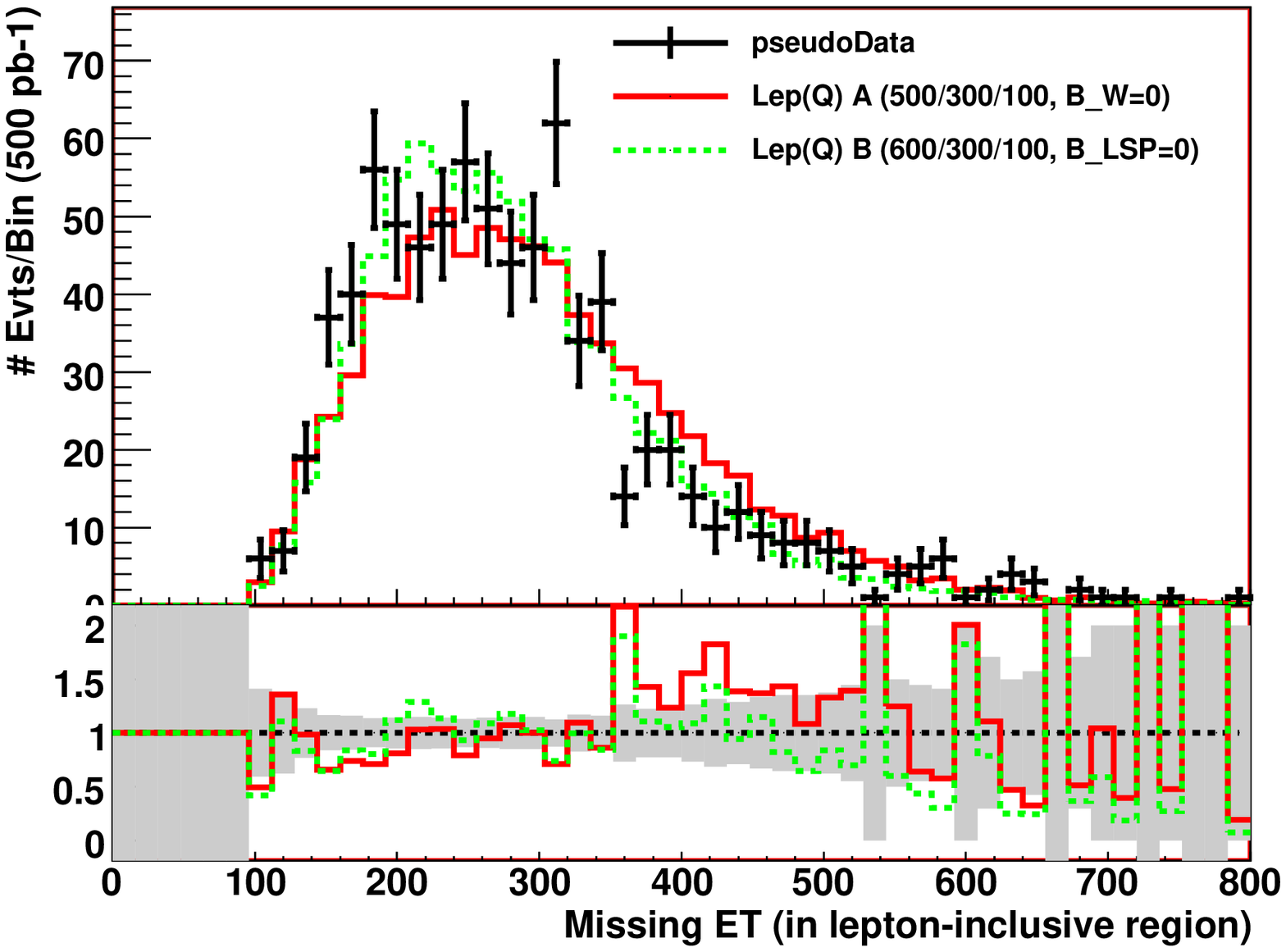}
\includegraphics[height=2.0in]{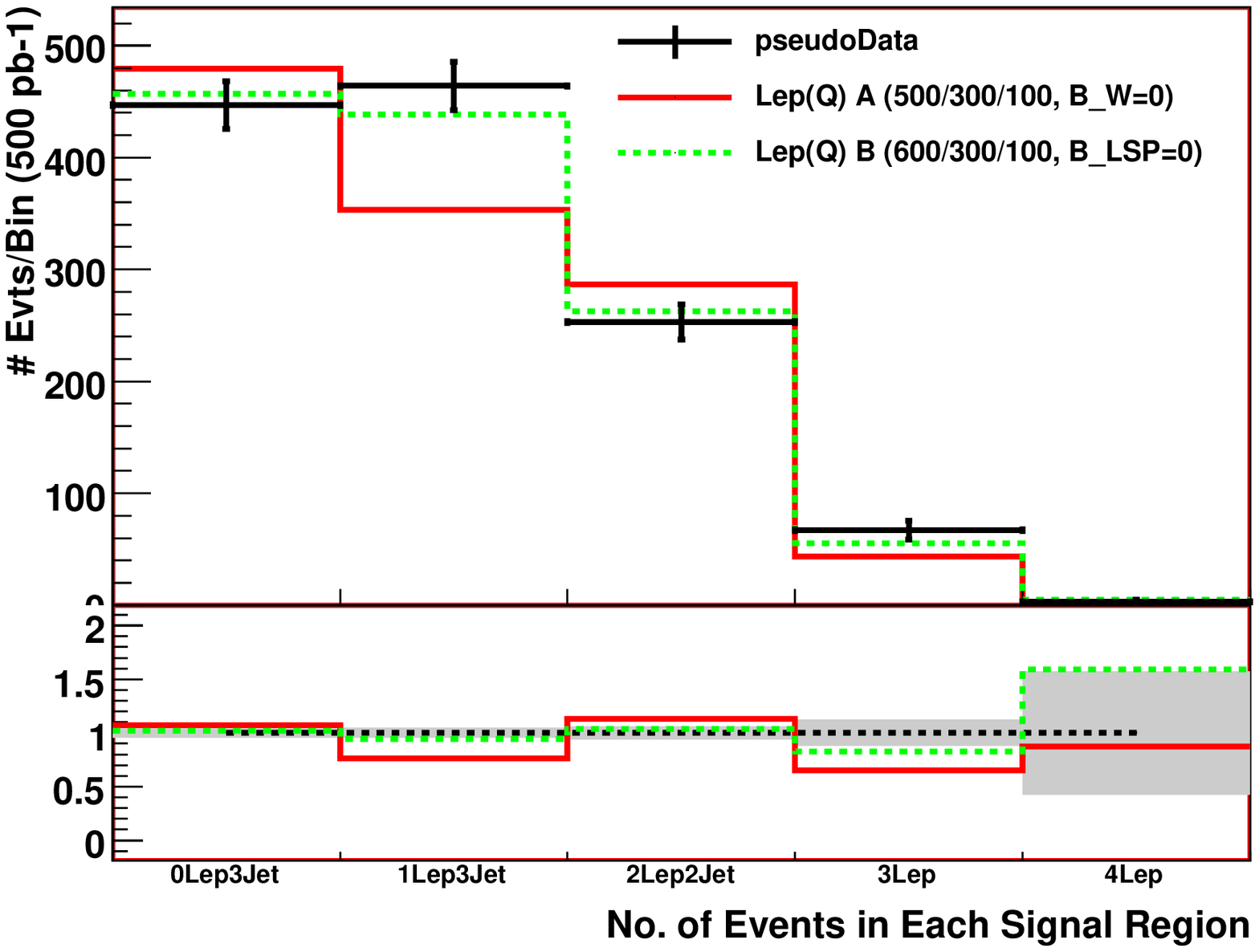}
\includegraphics[height=2.0in]{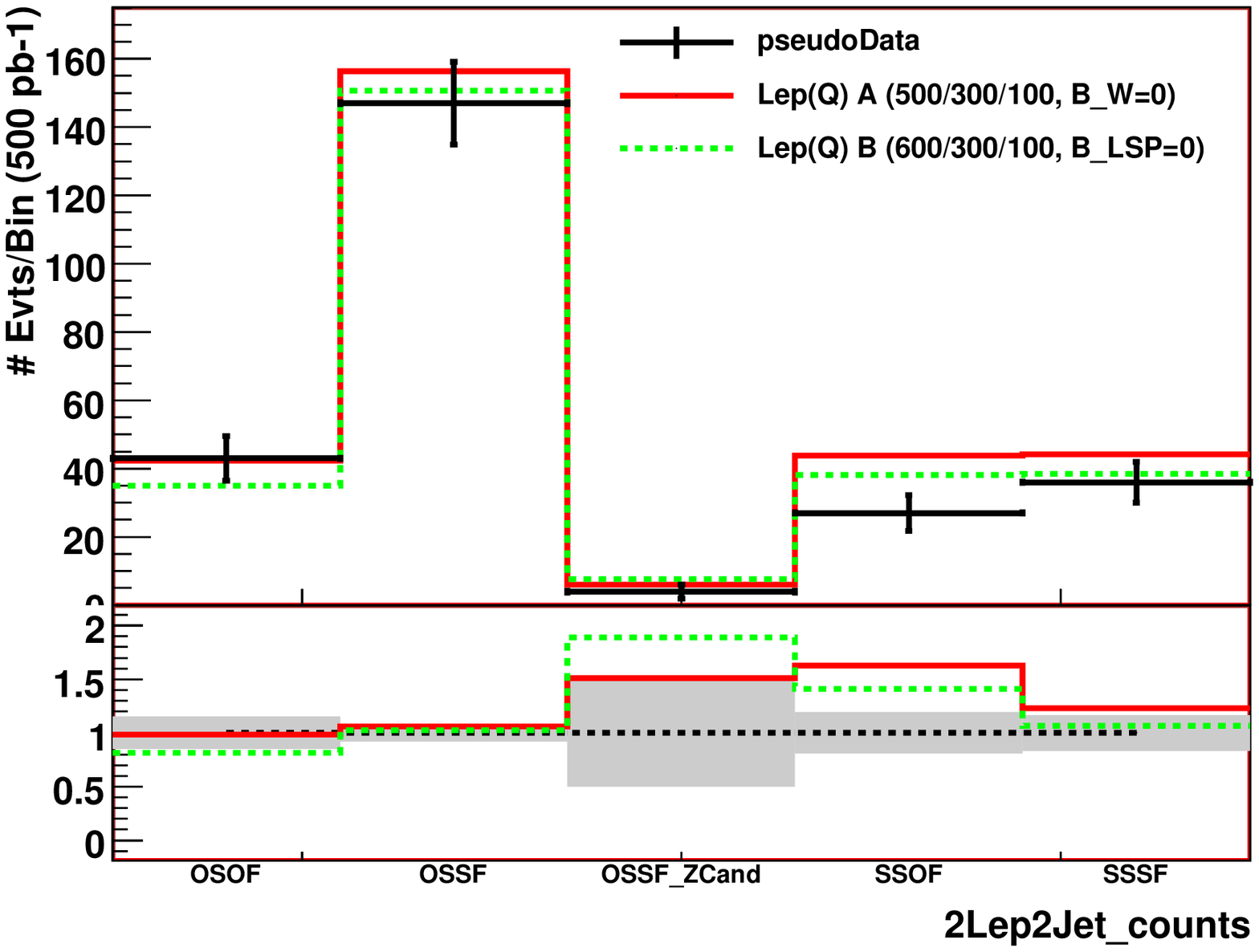}
\caption{Comparisons of basic kinematics and counts between the ``data''
  (error bars) and
  Lep(Q) simplified models at the two parameter choices from Table
  \ref{tab:ex1_Q-LCM_bestFitTable} (first, second, third lines in red,
  green, and blue respectively). Top: $H_T$ and missing energy in the
  lepton-inclusive region.  Bottom: lepton multiplicity in lepton-inclusive region, and di-lepton counts.
  \label{fig:ex1_Q-LCM_goodFits}}
\eef

\bet
\begin{tabular}{|c|l|c|c|c|c|c|c|c||c|c|c|}
\hline
\multicolumn{12}{|c|}{\textbf{Leptonic Decay Models for Quark Partners (Lep(Q))}}\\
\hline
Label& Description & $M_Q$/$M_I$/$M_{LSP}$ & 
          $\sigma_Q$ (pb) & $B_{\ell\ell}$ & $B_{\ell\nu}$ & $B_{LSP}$  &
$B_W$   & $B_Z$  & Leptons & Jets & Kin.  \\
\hline
\multicolumn{12}{|l|}{Rate variations with uniform kinematics}\\
\hline
A &  $B_W=0$ & 500/300/100 & 10.7 & 0.064 & 0.40 & 0.54  &
---   & --- &  + &  - & + \\
B &  $B_{LSP}=0$ & 600/300/100 & 6.1 & 0.11 & 0.19 & ---  &
0.69  & --- &  + &  - & + \\
\hline
& Approx. error & N/A  & 0.1   & 0.005 & 0.01 & 0.04 & 0.05
& 0.02 &  \multicolumn{3}{|c|}{}\\
\hline
\end{tabular}
\caption{Best-fit parameters for the leptonic decay model for quark
  partners, as determined by fitting to the count 
  information in Appendix \ref{app:fitData} (see also Figure
  \ref{fig:ex1_Q-LCM_goodFits}). \label{tab:ex1_Q-LCM_bestFitTable}}
\eet

\bef
\includegraphics[width=3in]{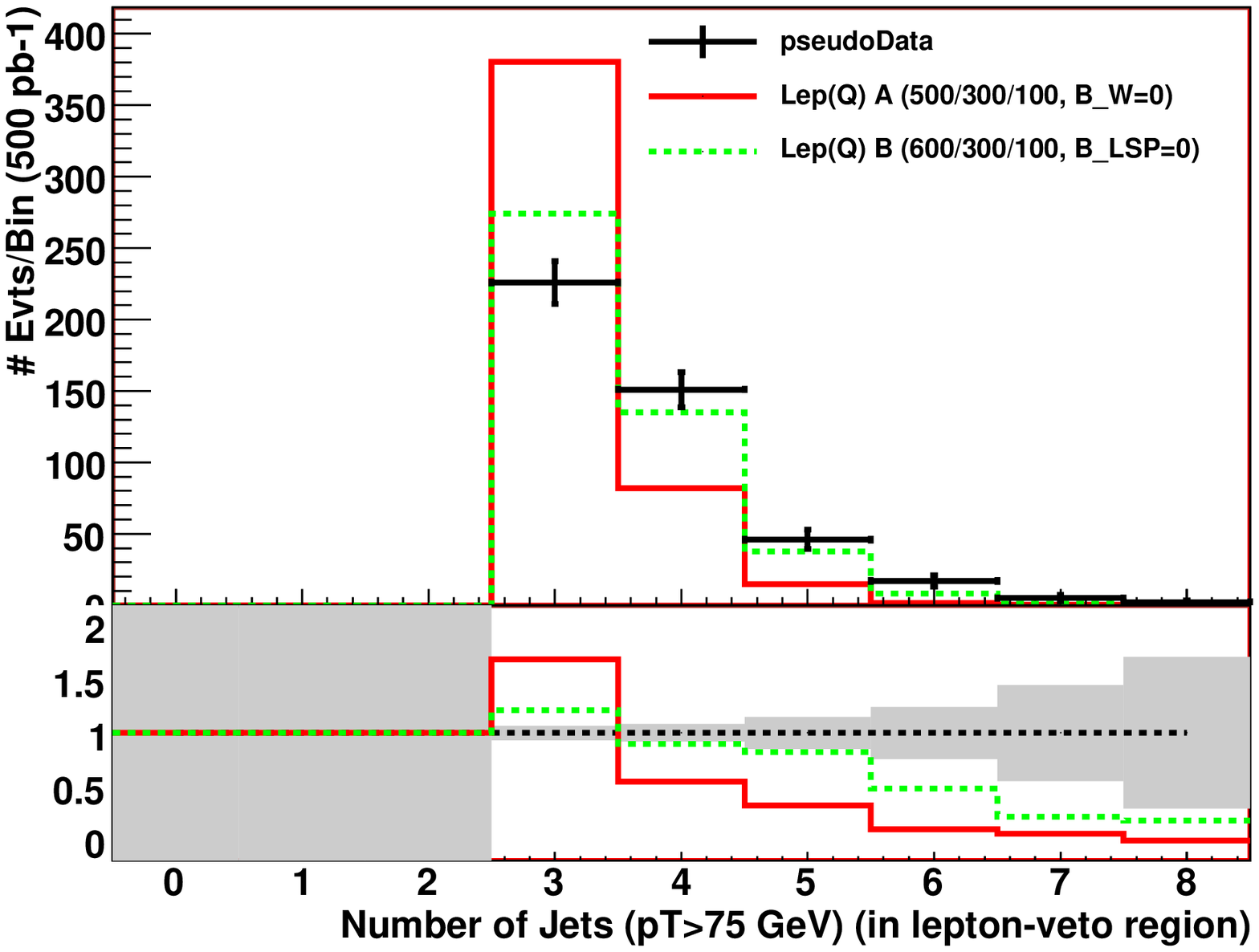}
\includegraphics[width=3in]{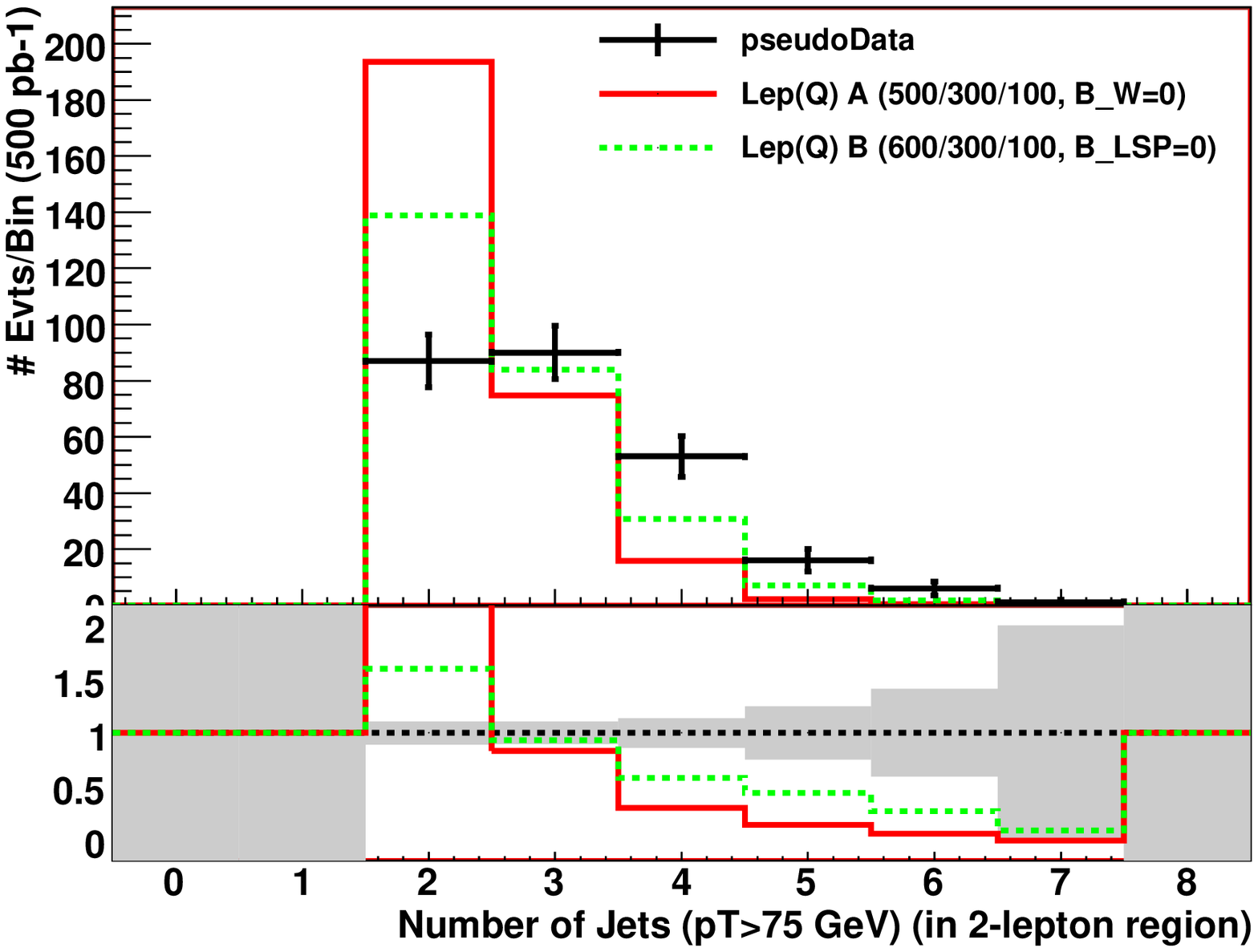}
\caption{Number of Jets with $p_T > 75$ GeV in the 0-lepton and
  2-lepton regions, for ``data'' (error bars) and  the simplified model
  Lep(Q) A, with $B_W=0$ (full) and Lep(Q) B, with $B_{LSP}=0$
  (dashed) with best-fit parameters from  Table 
  \ref{tab:ex1_Q-LCM_bestFitTable}.
  \label{fig:ex1_Q-LCM_jets}}
\eef

\subsubsection{Comparison of Btag(G) Models}\label{sec:ex1_HFM2}
In studying heavy flavors, we will focus on Btag(G), which reproduces
the jet multiplicities in the sample far more accurately than Btag(Q).
Two lines of questioning are assisted by a comparison of the data to
Btag(G) models: what is the heavy flavor fraction in decays, and can all
the leptons be accounted for by $W$'s from top quarks (i.e., is the data
consistent with a model with \emph{no} electroweak cascades, even
though there are leptons)?  The second question is less relevant here
because we already have strong evidence for electroweak cascades (the
$\ell^+\ell^-$ edge) and a strong argument that not all leptons come
from $W$'s (from the Lep(G) analysis in Section
\ref{sec:ex1_G-LCM_WvsLNU}).

Therefore, we include only the $b\bar b$ and $q \bar q$ modes in
Btag(G), omitting the $t \bar t$ mode, and we tune the parameters by
fitting to b-tag multiplicities in the lepton-inclusive multi-jet
region (see Appendix \ref{app:signalRegions}).  Figure \ref{fig:ex1_HFM2_bjets} shows the best-fit to
$b$-tag multiplicity and some tagged jet kinematics, for the best-fit
choice in Table~\ref{tab:ex1_Btag_bestFitTable}.
\bet
\begin{tabular}{|l|c|c|c|c|c|}
\hline
\multicolumn{6}{|c|}{\textbf{B-tag Model for Gluino Partners (Btag(G))}}\\
\hline
Parameter      & $M_{G}$, GeV & $M_{LSP}$, GeV &  $\sigma_G$ (pb)  & $B_{qq}$  &
$B_{bb}$   \\
\hline
Value &  600 & 100 & 4.0 & 0.84  & 0.16 \\
\hline
Approx.\ error & \multicolumn{2}{|c|}{---} & 0.2 & 0.03 & 0.03 \\
\hline
\end{tabular}
\caption{Best-fit parameters for the b-tag model for gluino
  partners, with $B_{tt}=0$. \label{tab:ex1_Btag_bestFitTable}}
\eet

\bef
\includegraphics[width=3in]{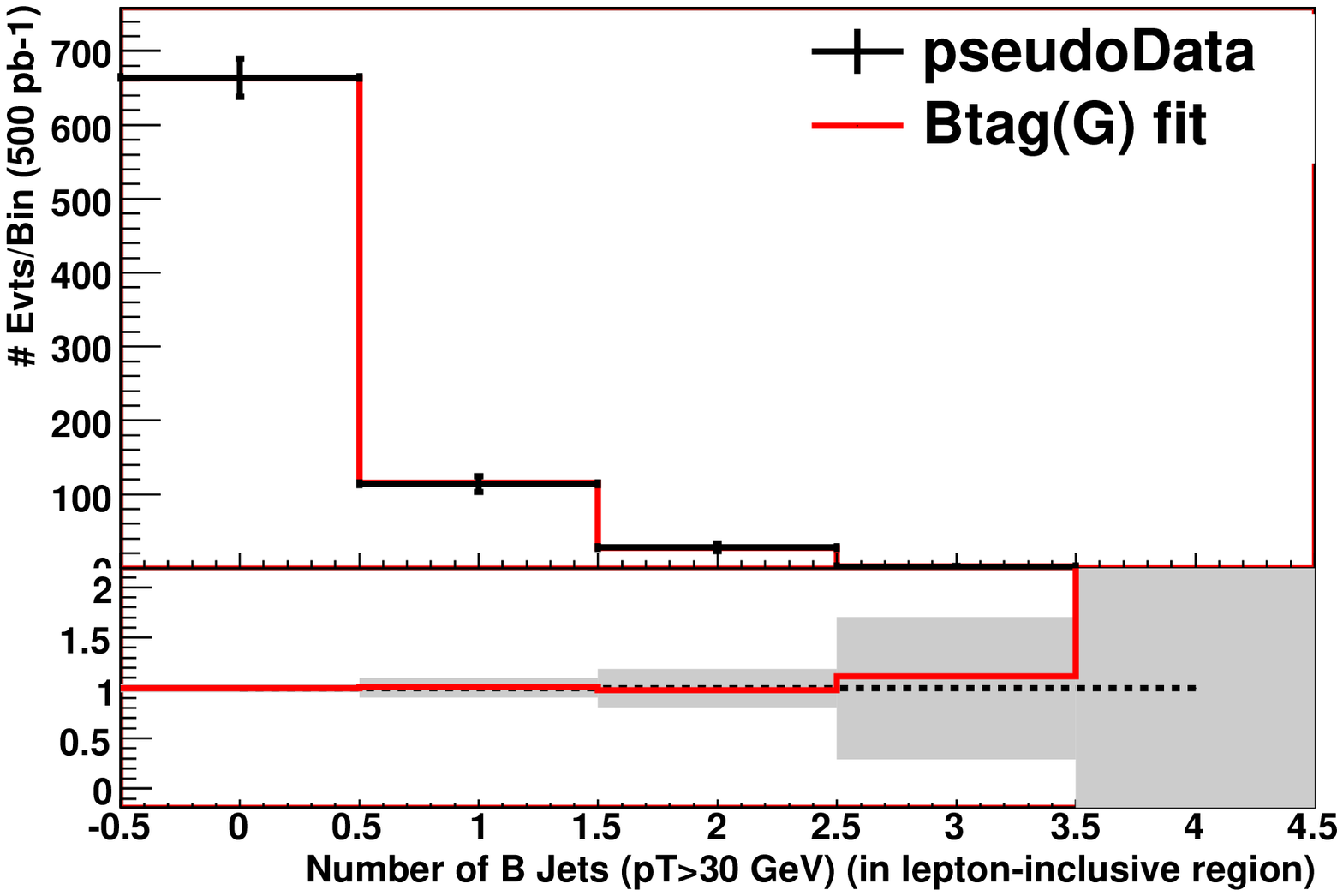}
\includegraphics[width=3in]{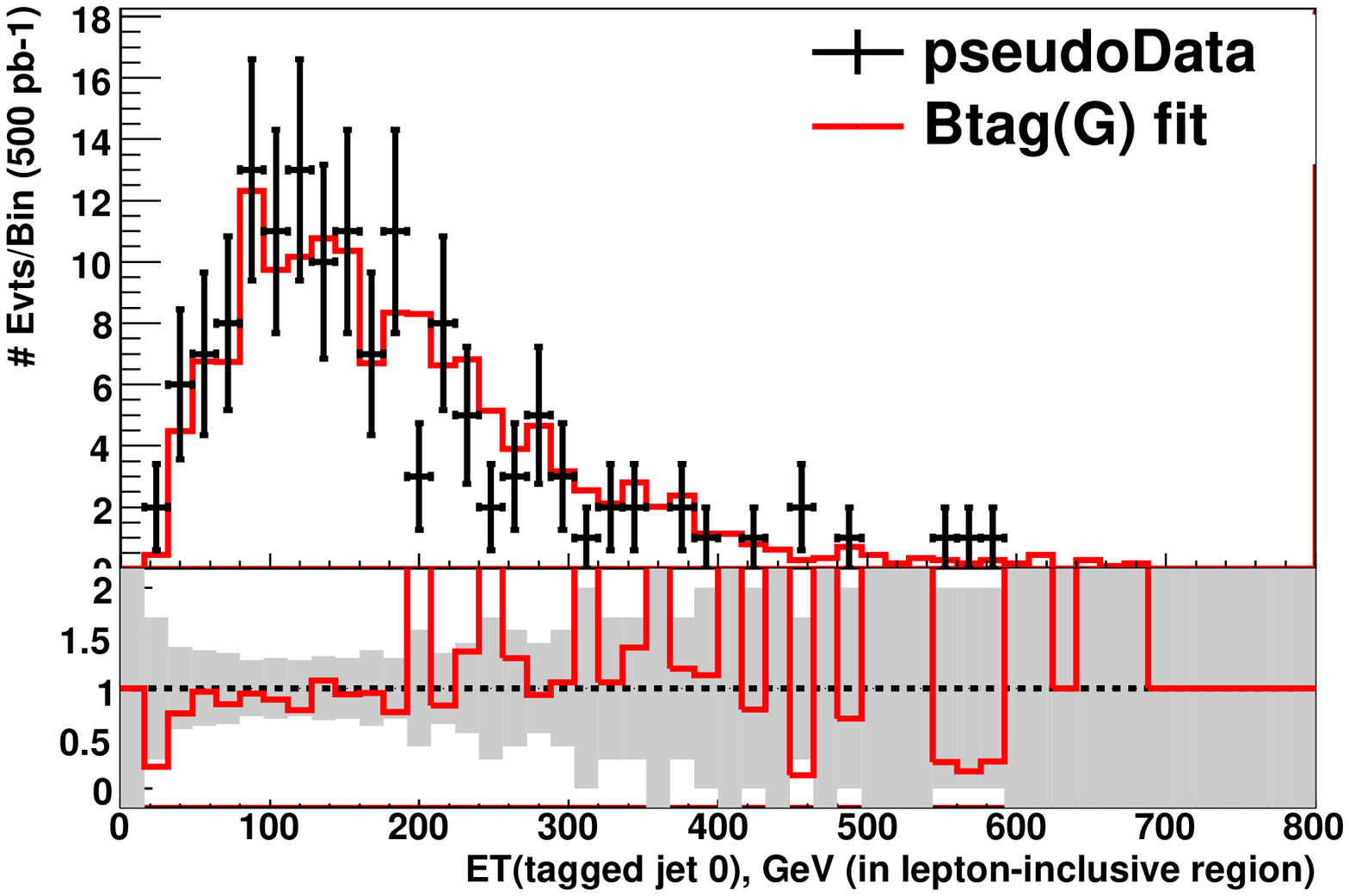}
\caption{Left: Number of $b$-tagged Jets with $p_T > 30$ GeV,
  Right: $E_T$ of leading  $b$-tagged jet (when present) in ``data''
  (error bars) and  the simplified model 
  Btag(G) (red) with parameters given in text.  All plots are taken in
  the lepton-inclusive multi-jets+$\MET$ signal region
  defined in Appendix \ref{app:signalRegions}.  Errors have been estimated only for rate parameters.
  \label{fig:ex1_HFM2_bjets}}
\eef

The agreement of our model with all multiplicities of $b$-tags
supports the two simplifying
assumptions in the Btag(G) model: that there is only one
initial state (or all initial states decay to $b$ jets in the same
fraction), and that heavy-flavored jets are produced in pairs.
The kinematics of these $b$-jets is consistent with the expectation
for $G \rightarrow b \bar b\, LSP$, and favors this over production
through top or Higgs decays (which were not included in the model, but
are expected to produce softer $b$ jets).  

The fit fraction consistent with $\approx 20\%$ heavy flavor is
suggestive of nearly universal decays to 5 generations, with top quarks
suppressed (perhaps by phase space).  This leads us to ask: are the
leptonic cascades in gluino decay independent of quark flavor?  This
question is beyond the reach of the simplified models, but is simple
to study qualitatively, by comparing the $b$-tag multiplicities in
different \emph{exclusive} regions --- as shown in Figure
\ref{fig:ex1_bcount_lep_correl}, there is no obvious correlation.

\bef
\includegraphics[width=3.5in]{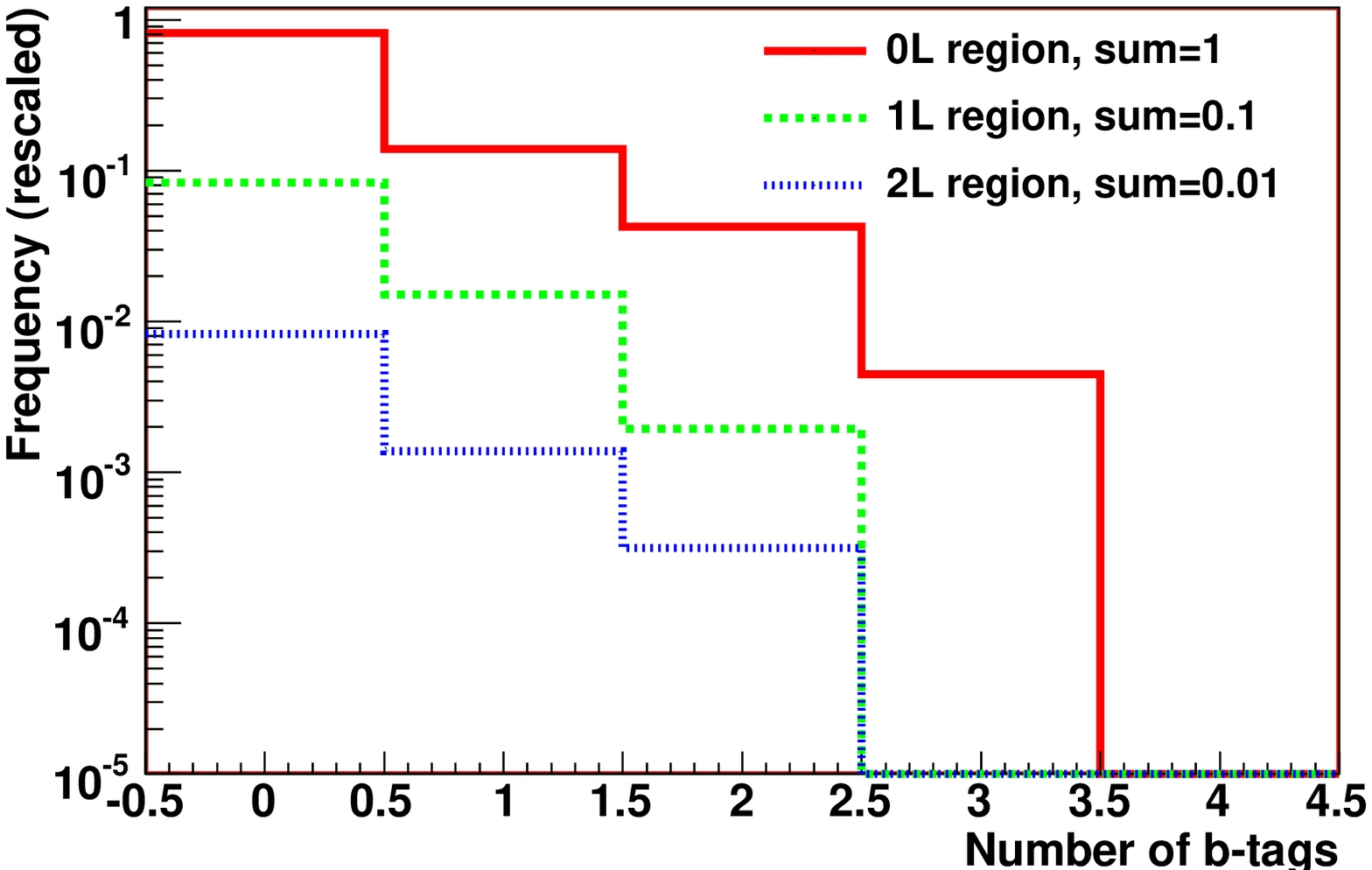}
\caption{
B-tag multiplicity in three signal regions: lepton-veto (top, red
solid line), 1-lepton (middle, green dashed line), and 2-lepton
(bottom, blue dotted line).  The three multiplicity distributions have
been re-scaled to fit on the same graph, to a total normalization of
1, 0.1, and 0.01 respectively.
\label{fig:ex1_bcount_lep_correl}}
\eef

\subsection{Interpreting the Simplified Model Comparisons}\label{sec:ex1_mssm}
\subsubsection{Summary of Conclusions from Plots and Simplified Models}
We begin by summarizing what is known (and suspected) about the
model.  Some of what we know can be inferred from plots alone:
\begin{itemize}
\item There is an on- or off-shell dilepton cascade with appreciable
  rate, with known edge/endpoint location
\item There is evidence for events with no leptons, and for a
  prominent decay  mode involving one lepton.  
\item There are $b$-tagged events; by a ball-park estimate that 1/6 as
  many events have 1 tag as 0 tags; assuming a $b$-tag efficiency of
  30-50\%, the average number of $b$-jets per event is $\approx 1/3 -
  1/2$. 
\item There is no significant correlation between b-tag and lepton
  multiplicities.
\end{itemize}

The simplified model comparisons have given us more quantitative
results, as well as information from jet modeling that could \emph{not} have
been derived by simply looking at plots:
\begin{itemize}
\item Jet structure is quite consistent for the leptonic decay model
  for gluon-partners, Lep(G), provided the hadronic $W$/$Z$ fractions are small.
  The leptonic model for quark-partner decay, Lep(Q), is inconsistent
  with observed jet multiplicities.
\item The dilepton branching fraction is $\approx 12\%$, and the
  combined single-lepton branching fraction (accounting for both
  slepton-mediated $\ell\nu$ decays and leptonic $W$'s) is $\approx
  40-50\%$.  Most of the remaining decays do not emit more
  jets with $p_T > 30$ GeV (to a good approximation they are
  invisible).
\item The heavy flavor fractions are consistent with pair production
  of a single particle that decays to light quarks $q \bar q$ 80-85\%
  of the time and to heavy quarks $b \bar t$, $b \bar b$, or $t \bar
  t$ 15-20\% of the time.  The $b$-tagged jet kinematics is consistent
  with direct $b$ emission, and probably \emph{not} consistent with a
  dominant $t \bar t$ mode, or with emission from a Higgs in a
  cascade.
\end{itemize}

We now see what we can deduce about the underlying physics; in this
section we will focus on analytical estimates within the MSSM; for
example 2, in
Section \ref{sec:6Interpretation}, we will use
a more quantitative method, namely numerically simulating
models and comparing them to simulations of the simplified models.
For definiteness, in both cases, we will take the position of a
theorist trying to explain the excesses seen at the LHC in the
context of a SUSY model within the MSSM.  

\subsubsection{Discussion: Consistent MSSM Parameter Space} \label{sec:ex1_mssm_ana}

As jet multiplicities are 
quite consistent with a four-parton topology, we will assume that
the signal is dominated by gluino pair-production.  There is a
variation to keep in mind: squarks that are slightly heavier than the
gluinos (and decay to them) may also be produced, but the additional
jets may be fairly soft.  

What is the origin for the $\ell\nu$ decays?  The study of rates in
the Lep(G) model showed that they cannot \emph{all} consistently
result from $W$'s, if (as in the leptonic models Lep(Q/G)) at most one
$W$ is produced in each cascade.  Jet counts disfavor a significant
$W$ mode, and would disfavor multi-$W$ cascades even more.  So to
explain the high rate of leptons in the signal, we assume that a large
fraction come from a slepton-mediated mode.  We have not obtained a
lower bound on the fraction of $\ell\nu$ coming from sleptons, but the
good agreement with jet multiplicities in the limit $B_W=0$ leads us
to ask what new-physics scenarios could be consistent with $B(\ell\nu)
\approx 45\%$ as in the Lep(G) model point A.

This suggests a roughly four-to-one ratio between $B_{\ell\nu}$ and
$B_{\ell\ell}$, which is rather striking.  The ratio can be achieved
in several ways for off-shell sleptons, but with a mass splitting
$M_I-M_{LSP}\approx 200$ GeV it is difficult to engineer couplings such that
three-body decays dominate over $W$ and $Z$ emissions.  With on-shell
sleptons, it is difficult to account for such a ratio.  If the LSP is
an $SU(2)$ singlet with no charged partner, and a $\nu\nu$ mode is
also open, then charged intermediate particles always decay to $\ell\nu$,
while neutral intermediate particles may be evenly split between $\ell\ell$
and $\nu\nu$ modes.  If charged and neutral parents are produced in
equal rates, this gives one factor of two.  For another factor of two,
we must produce more charged than neutral states.  This is
achieved \emph{in gluino decays through off-shell squarks}, where two
modes for gluino decays to $\widetilde W^+$ interfere constructively.
This is also consistent with our results based on
jet multiplicities in Lep(G) and Lep(Q), and it has significant
implications for the spectrum.  Firstly, in the MSSM, the LSP must be
a bino and the NLSP either wino or higgsino.  Secondly, the
right-handed sleptons must be too heavy to play a large role in
decays.

From Btag(G) comparisons, we learn that the $b$ fraction is
consistent with a single production mode, that produces $b\bar b$
pairs $\approx 20\%$ of the time.  A higgsino NLSP is implausible for
two reasons: it would probably enhance the heavy flavor fraction well
above flavor-universal rates, \emph{and} would introduce an
anti-correlation between leptonic decays and $b$-tag multiplicity
(decays involving light flavors would go dominantly to the bino LSP,
without a cascade) which we did not observe.  With a wino NLSP, this
rate is likely consistent with universal squark masses, with $\tilde g
\rightarrow t\bar b\widetilde W$ decays suppressed by phase
space.  

We are fortunate, in this case, to be led to a \emph{unique} ordering
of relevant weakly interacting species, fully determined by the
cascade and fractions we have measured (it is only unique if we take
all the hints from the data seriously, such as the 45\% branching
fraction to $\ell\nu$, though it was only loosely constrained).  
These conclusions are indeed correct.  The spectrum of the model is shown
schematically in Figure \ref{fig:ex1_spectrum} (the full Pythia
parameter set is in Appendix \ref{app:susyPythiaCardExample1}).  

\bef
\includegraphics[height=3.5in]{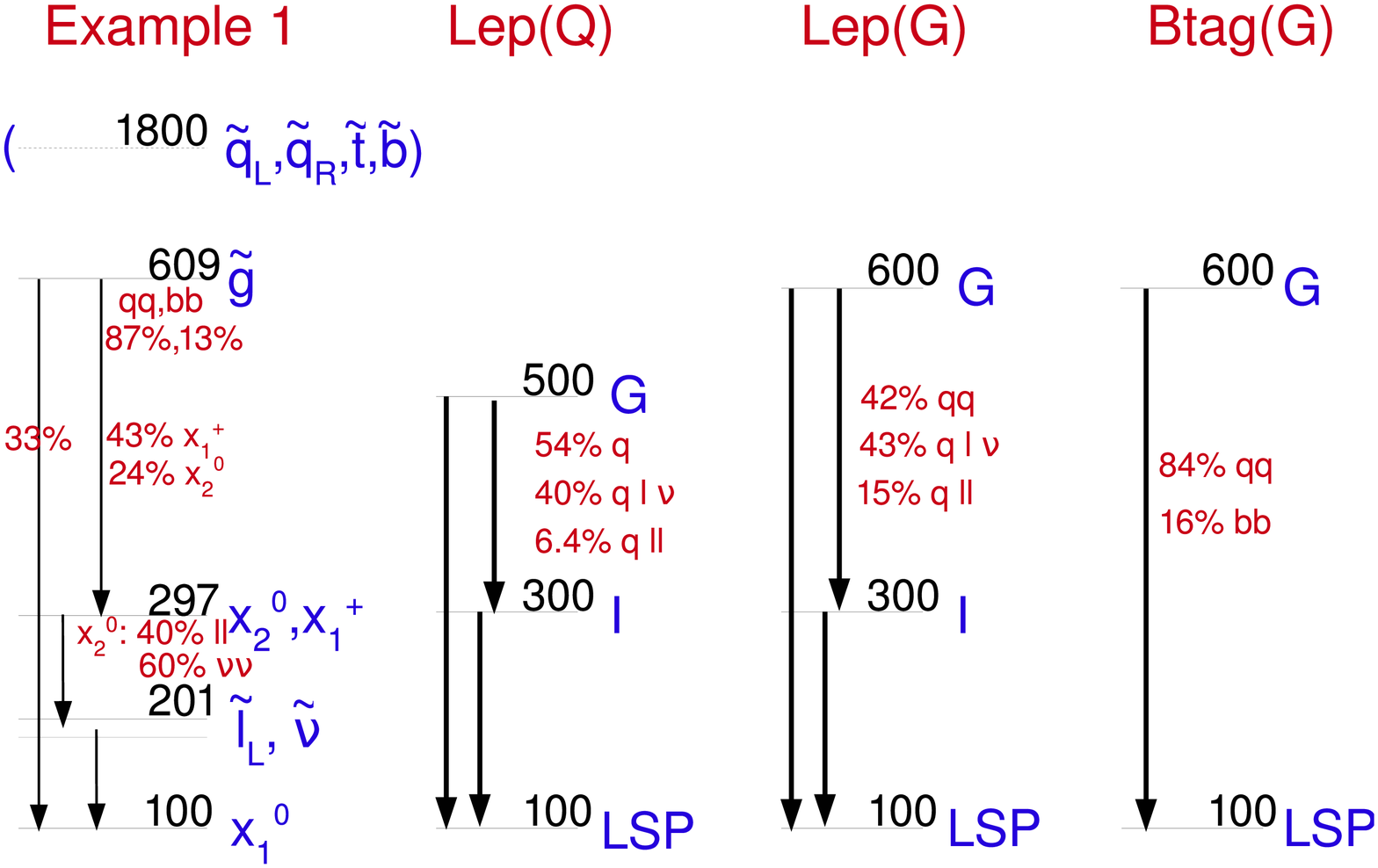}
\caption{Left: Spectrum cartoon for the model used in Example 1
  (parameters in Appendix \ref{app:susyPythiaCardExample1}).  Right:
  best-fit Lep(Q), Lep(G), and Btag(G) spectra (with $M_{LSP}$ fixed
  at 100 GeV).
\label{fig:ex1_spectrum}}
\eef

This ``back-of-the-envelope model-building'' is sufficient here only
because the underlying new physics is almost as simple as the
simplified models we have compared it to, but illustrates that
constraints on the simplified models translate into quite strong
constraints on new physics, and the model-independent statement of
this constraint makes it easily usable in the context of any SUSY-like
model.  In Section \ref{sec:example2}, we will see specifically how
the simplified models can be used even when the true structure of the
new physics is significantly more complex.

 \section{Probing Physics Beyond the Simplified
   Models}\label{sec:beyondSimple}

The simplified models have a rigid minimal structure, with only one
pair-produced species and a limited set of one-stage electroweak decay
chains.  The kinematics of decay products in all modes are determined
by uniform initial and intermediate particle masses; the
pair-production assumption also leads to a specific ``quadratic''
correlation between the rates of different processes.  These
simplifications of both kinematic shapes and rates make the simplified
models quite easy to constrain, but restricted.  If new physics has a
more complex structure, either kinematic shapes or rates may differ
from simplified model predictions; these deviations suggest what
additional structure is necessary to explain data.

In this section and the case-study of Section \ref{sec:example2}, we
study several limits of new physics with structure beyond the
simplified models.  In these examples,
deviations of observables from the simplified model predictions are
quite statistically significant, but still smaller than one might
expect.  Indeed, the degree of success of simplified models in these
cases suggests that more complex models needed to capture their
structure would have very poorly constrained parameters.  This
justifies studying and presenting best-fit simplified models carefully
even when they do not fully reproduce data, as a well-constrained
coarse-grained description of tne new physics, in addition to seeking
extensions consistent with all data.

We focus here on two of the most generic deviations in a SUSY context:
multiple production modes, and particles that decay through a series
of cascades.  In the first case, the rapid fall-off of production
cross-sections with particle mass (see Figure
\ref{fig:productionXSecByParticle}) simplifies the situation
dramatically: because particles of much higher mass than the lightest
produced particle (those that affect shape most significantly,
leading, for instance, to a visible bump in $H_T$) are strongly
suppressed. These rare production modes do not change rates enough to
make the non-quadratic structure apparent.  In the opposite situation,
when multiple particles of comparable mass are produced, the effects
on rates can be significant, as discussed in
Sec.~\ref{sec:LCM_LR} below. Fortunately,
multiple production modes near the same mass scale are benign where it
concerns shapes. When different jet production modes originate from a
similar mass scale, the simplified models can match the broad
kinematic structure of the trigger jets with just an overall and
intermediate mass scale. So long as the gross structure of the jet
counts are matched, which they usually are within the range of
topologies in the simplified models, the qualitative shapes of jet and
lepton structures in the data can be matched.  Therefore, multiple
production modes can significantly affect either rates or kinematic
shapes, but not both.

The second effect, the presence of double lepton cascades, can
have an impact on both rates and shapes, if the two chained decays
have very different kinematics.  Here, shape corrections are
particularly easy to diagnose, as they can be seen in leptons (see the
example in Section \ref{sec:example2}), though the inability to model
these shape discrepancies in jets may be a concern.  We have found in
a large number of examples that the best fit to rates within the
simplified models is remarkably good. The example of Section
\ref{sec:chainedCascades} is representative.  

A first characterization of data need not account for all correlations
in multiplicities or kinematics of final state objects.  However, when
the correlation is large, it is desirable to describe it
quantitatively.  This can be done either by finding a consistent point
in e.g. MSSM parameter space or by extending the simplified models to
a larger on-shell effective theory (OSET)\cite{ArkaniHamed:2007fw}.
It is ideal to do both, with the MSSM point providing proof of concept
and the larger OSET describing the consistent range of phenomenology
in a model-independent way.  In the latter case, the appropriate
generalization of the simplified models depends on observations, and
is beyond the scope of this paper, but we discuss one case of
particular interest --- correlations between lepton multiplicity and
$b$-counts --- in Section \ref{sec:leptonBCorrel}.

It should be emphasized that the basic count, object $p_T$, and $\eta$
distributions are \emph{not} the most sensitive means of finding
deviations from the simplified models.  But they are the distributions
that govern modeling of detector response to objects in an
event. Therefore, in order for any model to provide a meaningful
approximation to the underlying physics, $p_T$ and $\eta$ signatures
on the objects that are triggered on must be described well. Having
done this, the approximate description can be used as a target for
vetting models without having to worry significantly about systematic
errors introduced by trigger rate mis-modeling. So the figure of merit
for determining if the simplified models are good enough for
approximating complex physics is how well $p_T$, $\eta$ and very basic
count observables are modeled. A very simple model that passes this
test can be used for meaningful initial comparisons to other models.

\subsection{Left/Right or Isospin Differences and Lep(Q)}
\label{sec:LCM_LR} 

The first deviation we consider is very generic when light quark
partners dominate new-physics production: whereas the Lep(Q)
simplified model contains one triplet, in SUSY we expect two new
triplet scalars for each of the six flavors of quarks!  Disregarding
the third generation, we can expect approximate flavor universality
across the first two generations, but decays of $\tilde q_L$, $\tilde
u_R$ and $\tilde d_R$ can be quite different from one another (a related complication occurs when both quark and gluon partners are
produced, and favor very different decay modes).  An
extreme example is the case when the bino is the LSP and the winos
have masses between the bino and the squarks, while the higgsinos are
heavier than the squarks. The left-handed squarks couple dominantly to
the winos, and therefore cascade decay emitting $W/Z$ or lepton
pairs/lepton + neutrino, while the right-handed squarks only decay
directly to the LSP. Depending on the squark and gluino masses, the
left- and right-handed squarks can be produced together or
predominantly in pairs of particle-antiparticle. The assumption in
Lep(Q), that there is only one particle species produced with a set of
branching ratios to leptons and weak bosons, can give a better or
worse description of the data depending on the mix of production
processes.

The particular way in which the description fails gives important
hints as to how the model can be amended. If in particular the
associated production is absent, we would see an excess of
different-flavor and same-sign two-lepton events as compared to
single-lepton events, and vice versa if associated production
dominates. Reliably modeling the actual mix of production modes, as
well as the branching ratios and decay modes of the different species
produced requires constraining at least 11 parameters (three
cross-sections for pair production of two types of quark-partner and
their associated production, and four free branching fractions for
each species of quark partner).  Without large statistics, the risk of
having fits pulled by statistical fluctuations is also large once an
increased number of parameters is introduced.  Moreover, the
interpretation of these effects is ambiguous, as similar features
could instead point towards multiple-stage decay chains as discussed
in the next section.  We therefore recommend using such features in
the fits of the model to early data as hints, and publishing the
relevant comparison plots and pulls, rather than try to publish less
stable fits with an enlarged parameter space.

\bef
\includegraphics[width=3in]{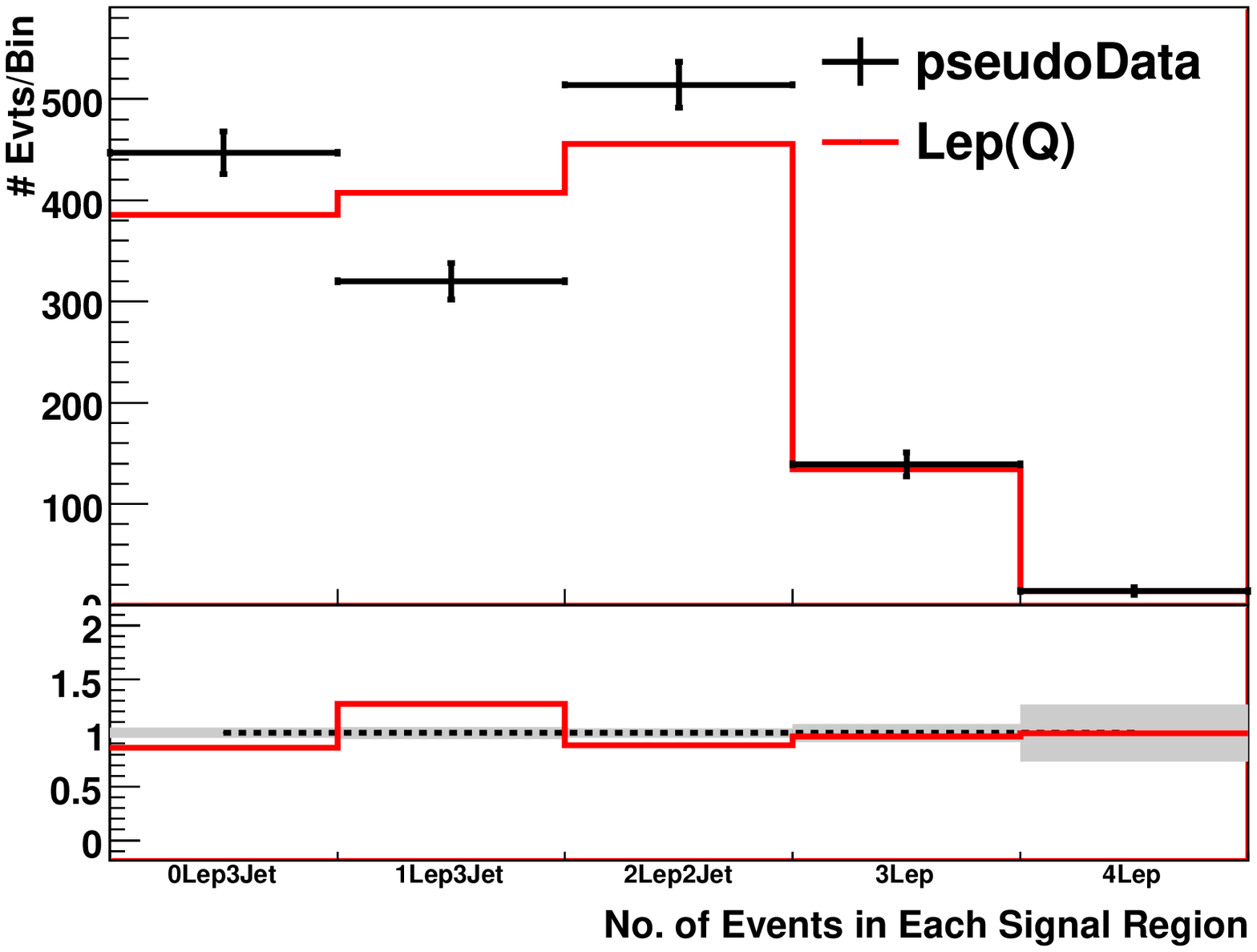}
\includegraphics[width=3in]{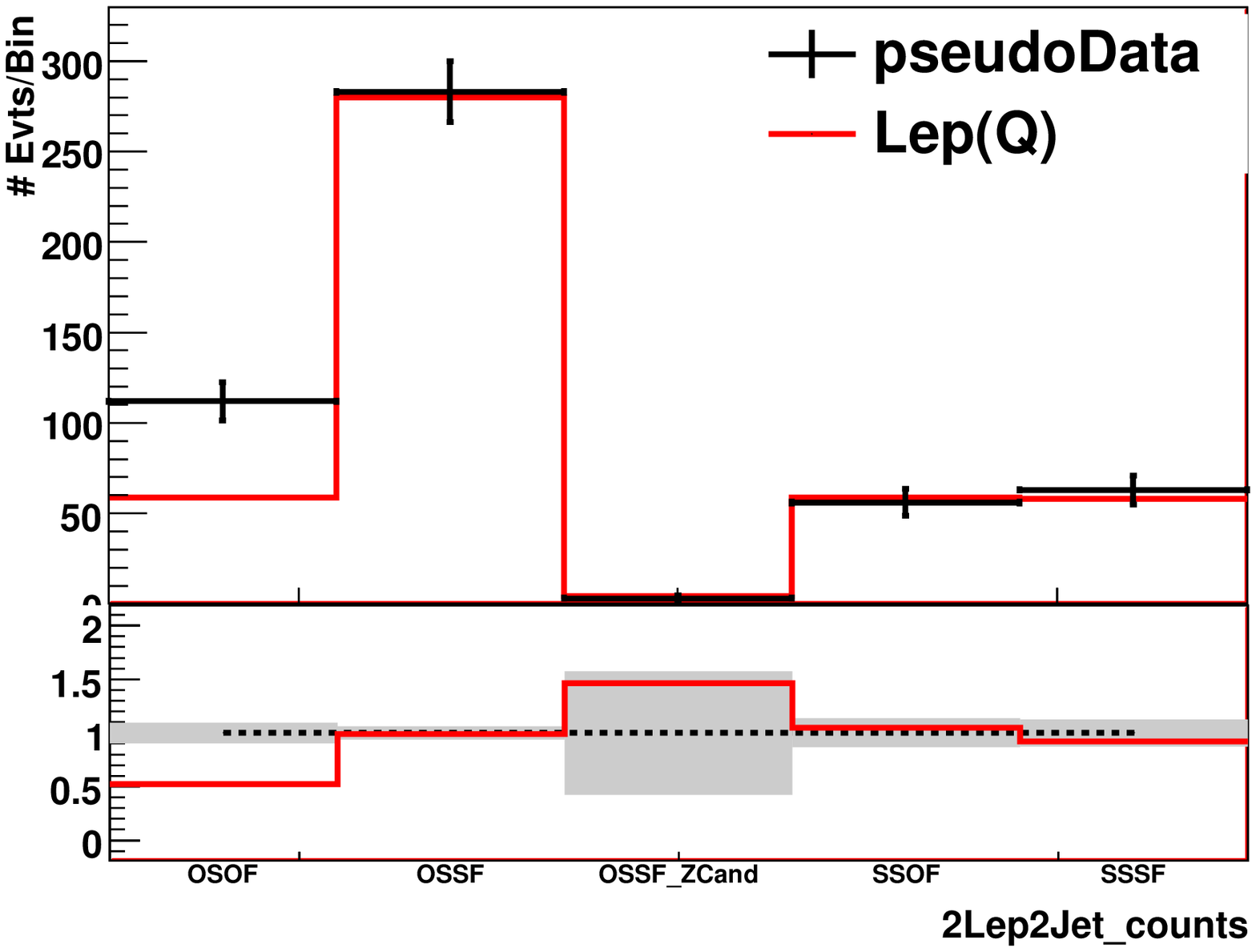}
\caption{Comparisons of basic kinematics between the ``data''
  (error bars) and simplified model Lep(Q) (red), in a case where the
  data has two pair-produced species, one decaying only to quark + LSP
  and the other to lepton pairs or lepton + neutrino. From left to
  right, the number of events in the different lepton signal regions,
  showing the fitted excess of one-lepton events, and the lepton
  counts in the 2-lepton signal region, showing the deficit of
  different-flavor lepton production (OSOF). See
  sec.~\ref{sec:LCM_LR}.\label{fig:Example51}}
\eef

An example of diagnostics plots for the case outlined above, with
$\tilde q_L$ decaying to lepton + neutrino or lepton pairs through
intermediate winos, and $\tilde q_R$ only decaying directly to the
LSP, is shown in fig.~\ref{fig:Example51}, together with the best-fit
Lep(Q). The best fit balances the lack of different-flavor
events in the two-lepton region with an overpopulated one-lepton
region. The deficit of different-flavor events can be seen in
the OSOF bin in the 2-lepton signal region lepton counts. Pseudo-data and Lep(Q)
fit parameters can be found in Appendix \ref{app:Example51}.

It should be noted that, in the early running of the LHC, before tau
tagging is fully functional, a similar effect might be due to an
over-representation of tau lepton decays. It is fairly generic to have
tau lepton partners lighter than the electron and muon lepton
partners. In this case, gauge boson partner decay into tau leptons
will be enhanced with respect to light-flavor leptons, especially if
the tau lepton partners are the only ones kinematically accessible
below the gauge boson partners. Before the hadronic taus from these
decays can be reliably tagged, only the leptonic tau decay will be
noticed, leading to an enhancement in different-flavor leptons (and
corresponding depletion of same-flavor pairs) with respect to the
assumption of (most of) the light-flavor leptons coming from decay
through light-flavor lepton partners. This situation will immediately
be resolved once the rate of hadronic taus can be reliably estimated.

\subsection{Multiple Intermediate-State Masses and Chained Cascades}
\label{sec:chainedCascades}

\bef
\includegraphics[width=3in]{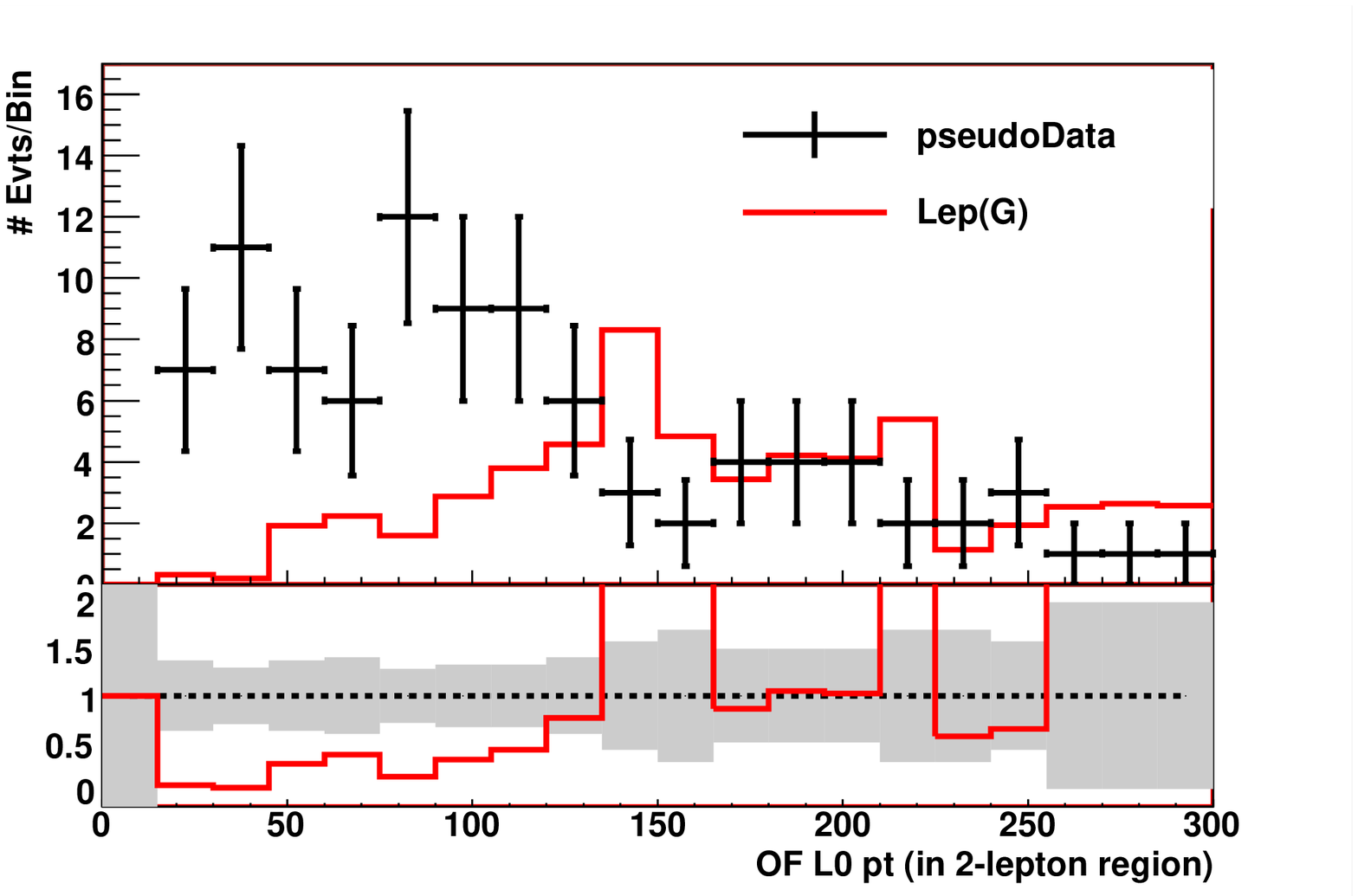}
\includegraphics[width=3in]{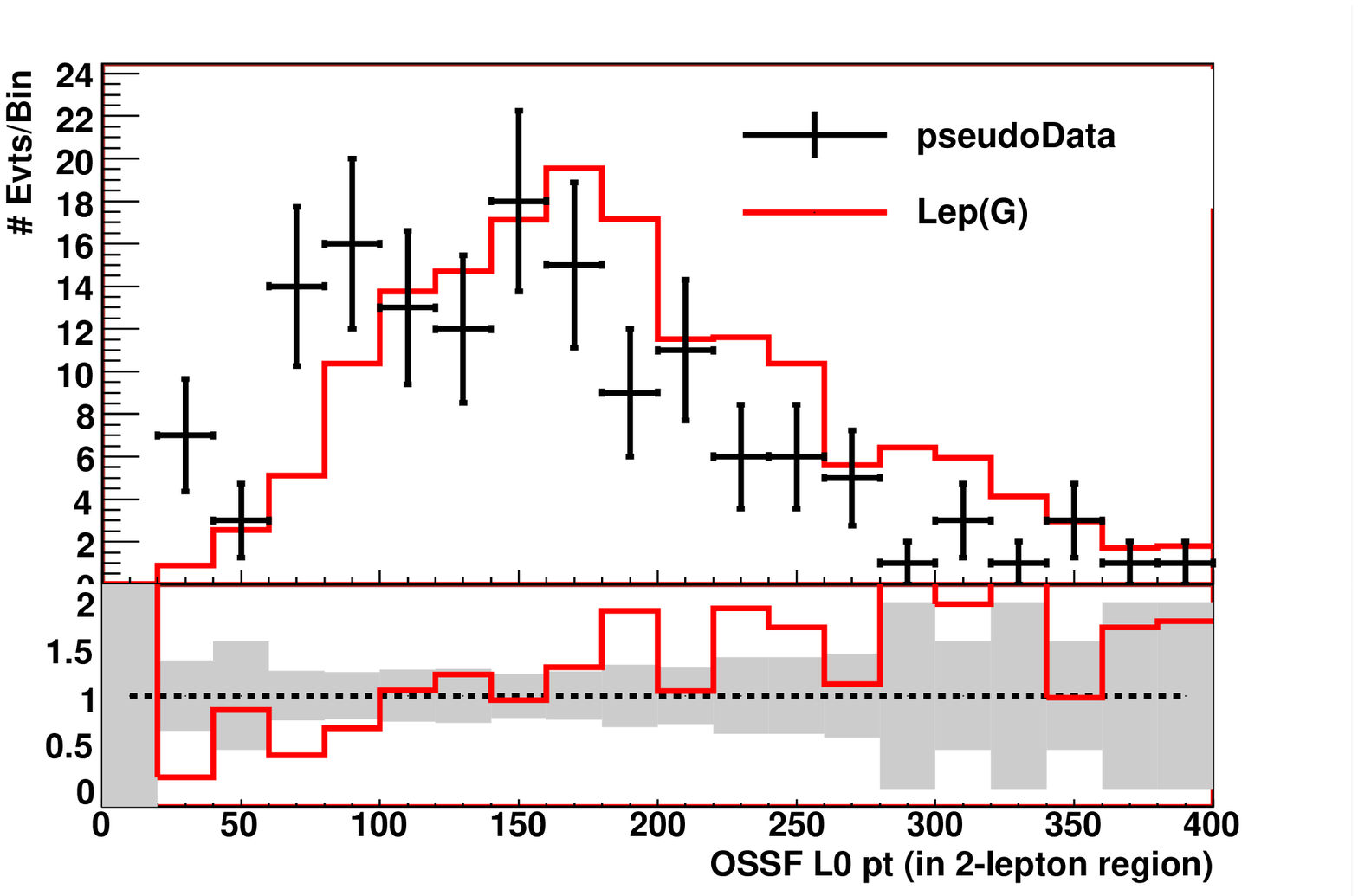}
\includegraphics[width=3in]{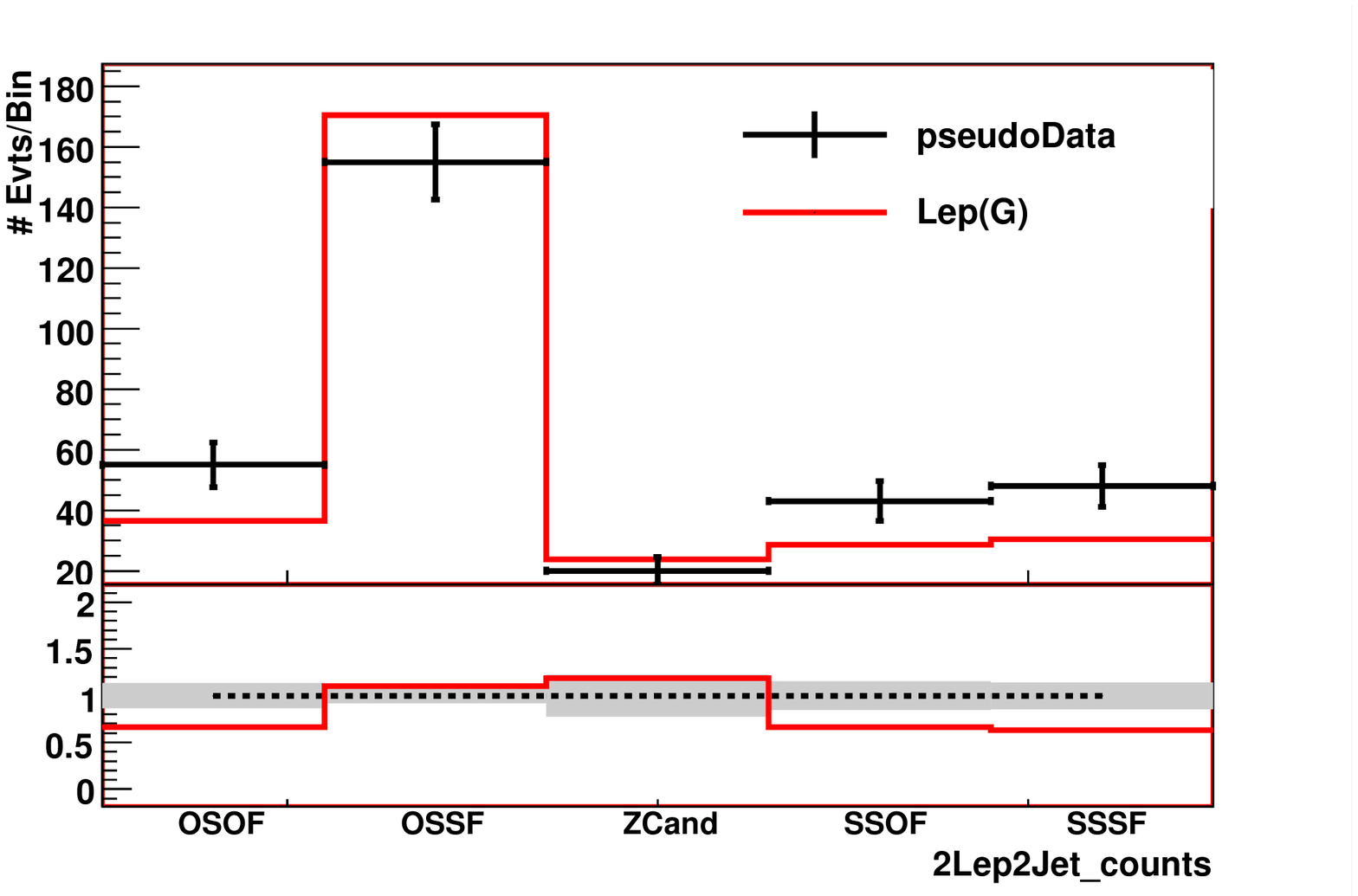}
\caption{Comparisons of di-lepton kinematics and counts between the ``data''
  (error bars) and simplified model Lep(G) (red), in a case where the
  data has a high fraction of $WW$, $WZ$, $ll+W$, and $\ell\nu+W$
  decay modes. A good diagnostic for this in the Lep(Q/G) fits is an
  excess of opposite flavor di-leptons relative to single lepton or
  same flavor events. Typically in such cases, there is also a
  difference in the lepton kinematics between opposite flavor and same
  flavor. See sec.~\ref{sec:chainedCascades}\label{fig:MultipleCascades}}
\eef

Another effect that we have so far omitted from the discussion is the
possibility of cascades chained one after the other (up to two within
the MSSM, or more if higgsinos or winos have large splittings).  The
effects of a double cascade can be \emph{partially} modeled by simply
increasing the rates for cascade modes in a single-cascade model like
Lep(Q/G).  This will suffice so long as branching fractions $> 1$ are
not required to obtain the observed frequencies of leptonic events.

The success of describing long cascades by these ``flattened'' models
relies on low leptonic branching fractions of $W$ and $Z$ bosons, such
that --- in early data --- the statistical uncertainties in
the rate of many-lepton events are likely to be quite large.  So any
optimized fit will be pulled most by the bins that are populated by
only one leptonic decay.  However, chained cascades that produce more
weak gauge bosons have enhanced rates for multiple bosons to decay
leptonically (because combinatoric factors are higher than for bosons produced
singly).  So chained cascades should first appear as
\emph{excesses in multi-lepton events} over what is expected from the
rate of events with fewer leptons.

Note however that these effects can arise also due to the production
of multiple species, one decaying to weak bosons and one which
doesn't, as described in Sec.~\ref{sec:LCM_LR} above. A
distinguishing feature might here be lepton kinematics. An example of
this is shown in Fig.~\ref{fig:MultipleCascades}.

\subsection{Lepton and Heavy Flavor Correlations and Extensions of the Simplified Models}\label{sec:leptonBCorrel}

In our simplified models, we chose to model lepton and heavy flavor
observables separately, in order to keep the number of parameters
down. The reason for this is that the number of extensions of the
simplified models necessary to account for possible combinations is, even
in just the MSSM, too large to be tractable. Furthermore such models
would in general have too many parameters to be uniquely constrained
by early data, which reintroduces flat directions and arbitrary
fits. There are however certain situations where conclusions about
lepton and heavy flavor correlations can be drawn from very simple
extensions or combinations of the simplified models.

One such case, which can be seen directly from the simplified models,
is when all leptons come from top decays. In this case, the Btag(Q/G) models
including top decays will by themselves properly model all the lepton
counts and kinematics. Such a case will be indicated in the Lep(Q/G) by an
absence of $\ell^+\ell^-$ modes (or $Z$s), and a jet structure
compatible with $W$ decays rather than $\ell\nu$. 

A second case is exemplified in Sec.~\ref{sec:example1}. Here, we find
that the Btag(G) fit is consistent with flavor-independent decay of
gluon partners (with top decays suppressed by kinematics). In such a
case, it is very natural to include $G\to b\bar b+\text{color singlet}$
decay modes, with branching ratios for the color singlets constrained
to be identical to those in the decay to light flavor quarks.

If, alternatively, the Btag(Q/G) fits show that lepton and $b$ jet kinematics
(such as invariant masses) are well described by the top
hypothesis, but top quarks by themselves fail to explain all leptons
in the data, another simple extension would be to add a direct top
quark decay ($G\to t\bar t+\text{LSP}$ or $T\bar T$ production with
$T\to t+\text{LSP}$) to the Lep(G/Q) models.

Each of these extensions only introduce one extra parameter to the Lep(Q/G)
models, describing the $b$- or $t$-rate, respectively. They can be
used in a similar way as the four basic simplified models, to
investigate to which extent data can be described, except that the
relevant data now is correlations between leptons and $b$-tags, such as
the lepton counts for different number of $b$-tags. Deviations from
the expected fits can then be a basis for further conclusions about
the spectrum and couplings. It is important, however, not to add
progressively more complexity to account for every feature of observed
deviations, since the uniqueness of descriptions of features then
soon will be lost. It is also very important not to create extensions to
model deviations that are not statistically significant. We therefore
recommend, once again, to publish fits to the unextended simplified models
alongside any extensions, the ones suggested here or others.

\section{Example 2: Complex New Physics}\label{sec:example2}
Having considered a relatively simple example in section 4 to
illustrate how simplified models can characterize and then represent
the data, we now move on to a more intricate example. As pointed out in
section 5, the allowed new-particle spectra --- and hence the allowed
decays --- in ``SUSY-like'' physics can be much more complex than
those of the simplified models. In section 5, we discussed common ways
in which our simplifications can have an impact on the fits of the
simplified models to data. We also commented on signatures that can be
helpful for detecting what simplifications are violated by the
underlying model, though we don't expect that process to be very
straightforward at low luminosity.

Here, we will consider an example where the underlying model is
significantly more complicated than the simplified models. We will see
that most basic signatures are well modeled by many limits of the
simplified models. There are some sources of tension, mainly
kinematical. While we won't be able to clearly diagnose what's
different between the underlying model and each of the simplified
models, we will be able to draw qualitative and quantitative
conclusions about the structure of production and decay that will
offer an excellent starting point for model building. We will
highlight those aspects that cannot be simply read directly off plots
of data alone, and illustrate the procedure of vetting more detailed
model hypotheses against fits to the four simplified models, Lep(Q/G) and
Btag(Q/G).

We will consider the SUSY model generated using the Pythia
parameters of Appendix \ref{app:susyPythiaCardExample2} in Pythia
6.404 \cite{Sjostrand:2006za}. As before, the parameters are provided for
reference, but we will treat this as an unknown signal for the
remainder of this section. We also use the same set of ``signal
regions'', as described in Appendix \ref{app:signalRegions}, that we
used in Section \ref{sec:example1}. 

In the following subsection, we provide a summary of the simplified
model fits and main areas of agreement and 
tension with the data. We will discuss the Lep(Q/G) fits in detail in
subsection \ref{sec:6Leptons} and the Btag(Q/G) fits in
\ref{sec:6Btag}. In subsection \ref{sec:6Interpretation}, we will
investigate how to use the simplified models for interpreting data.  

\subsection{Summary of Model-Independent Results}\label{sec:6summary}
In summarizing the main results of the simplified model fits, we will choose
particular masses. We will not discuss the question of mass estimation
in any detail for this example, as the emphasis is on how to use the
fits. We have set the LSP mass to $100$ GeV, and then
estimated the remaining mass parameters using $H_T$, jet $p_T$, and
lepton $p_T$.  In table \ref{tab:ex2-LCMfits}, we
summarize fits to on- and off-shell leptonic models
Lep(Q/G). Likewise, table \ref{tab:ex2-HFMfits} presents fits to the
$b$-tag-study models Btag(Q/G). 

We now highlight features of the data evident from studying plots, and
the refinements that are made possible by quantitative comparison to
the simplified models.  As they are closely related, we will list them
together, with the conclusions from distributions alone in italics:
\begin{itemize}
\item Gluon or quark partner models alone do not give a good
  description of the jet structure, suggesting that a combination of
  production modes is required. Fits to the total event rates suggest
  a cross section in the range of $10-14$ pb. A lower bound estimate of the mass scales is, $M_{Q,G}\sim
  600-700$ GeV, with $M_{LSP}=100$ GeV. Referring to figure
\ref{fig:productionXSecByParticle}, this strongly supports the
hypothesis of production of particles charged under SU(3).  
\item \emph{There is an OSSF dilepton decay mode.  We can conclude
  this from the excess of OSSF events over OSOF (and other dilepton
  events). There is also a di-lepton invariant mass structure that suggests
  either on- or off- shell lepton partners.}
\begin{itemize}  
  \item From the leptonic simplified model comparisons, we conclude that an $ll$ decay mode occurs in
  $\approx 4-6\%$ of decay chains. 
  \item There is also strong evidence for a sizable $\ell\nu$
  channel, with branching fraction $B_{\nu l+l\nu}\approx
  30\%$. 
  \item The observed $Z$ fraction appears small, in the range
  of $B_Z\approx 2-3\%$.
  \item It is difficult to obtain enough opposite flavor di-lepton
  events without overpopulating single lepton events. In addition, the
  shape of the lepton $p_T$ signatures, as compared to the fits to
  Lep(Q/G), suggests that there is a missing source of relatively soft
  leptons. The data must include some source of leptons not included
  in the simplified models.
\end{itemize}  
\item \emph{There is a preponderance of $b$-jets, with extremely high
tag rates.}  We learn significantly more detail from the
  fits to Btag(Q/G) simplified models: 
\begin{itemize}
  \item The distribution of $b$-jet counts is pretty well accounted
    for by pair production of a gluon partner $G$ that  decays
    to a pair of 3rd generation quarks $\approx 60\%$ of the time, and
    a pair of light-flavor quarks the remaining $\approx 40\%$ of the
    time.
  \item When the heavy-flavor decays are all to $t\bar t$, we
    correctly reproduce both $b$-jet $p_T$ distributions and the
    lepton/b-count correlations in events with more than one $b$-tag.
  \item There is slight disagreement between the $b$-tag multiplicity
    predicted by Btag(G) and the data --- in particular, we cannot
    account for all the 1- and 2-tag events without over-estimating
    the number of 3- and 4-tag events.  This is only an $\approx
    2\sigma$ effect at the statistics shown.  If we take it at face
    value, the simplest interpretation is that there is a distinct
    production process that produces \emph{up to two} heavy flavor
    quarks (for example, either stop or sbottom production or
    associated production of the gluon partner with a light-flavor
    quark partner.
\end{itemize}
\item \emph{There is a qualitative trend in the $b$-count
  distributions as we move across lepton regions: the lepton-rich
  events have fewer $b$-jets.}  From the quantitative comparison to
  Btag(G), we saw also that the difference is approximately compatible
  with adding sources of leptons to zero-$b$ events.  This gives
  evidence that the $4-6\%$ $\ell^+\ell^-$ mode appears dominantly in
  light-flavor decays (either of the gluon partner or of some other
  state).  Using the $\sim 30\%$ light-flavor fraction from the Btag(G)
  fit, we are led to hypothesize an $\ell\ell$ decay mode in $\sim
  15\%$ of these light-flavor decays.  This number was inferred
  indirectly and should not be trusted too much.
\item \emph{The jet multiplicity in 2-lepton events seems
  significantly lower than in 0- or 1-lepton events --- but there are
  many interpretations: A decrease in lepton ID efficiency in events
  with many jets? $W$'s that produce more jets when they don't decay
  leptonically? Or evidence that 2-lepton events are dominated by a
  mode with fewer partons from the $SU(3)$ decay?}  The approximate
  consistency of Lep(G) (with $W$'s) with the jet counts shown in
  Figure \ref{fig:ex2-BestPlots} suggest that ID efficiencies and $W$
  decays are sufficient to explain this trend.  Other interpretations
  are also possible (these jets could be radiation, or products of
  heavier states decaying to gluon-partners).
\end{itemize}

Some other features are beyond the resolution of the simplified
models --- for example, we can't repeat the fit to leptonic branching
fractions in the presence of a top-quark decay mode.  We have,
however, built evidence for the basic structural components of
the new physics, and found a characterization of the new physics to
which we can compare any model.  

In the following sections, we discuss in more detail the structure of the Lep(Q/G) and Btag(Q/G) characterization of the data. However, the above summary is sufficient for discussing how to use the fits presented in tables \ref{tab:ex2-LCMfits} and \ref{tab:ex2-HFMfits}. We therefore recommend that the reader interested in this topic skip to subsection \ref{sec:6Interpretation}.

\subsection{Comparisons to Leptonic Decay Models}\label{sec:6Leptons}

\bet
\begin{tabular}{|l|c|c|c|c|c|c|c|r}
\hline
Model / Limit & $M_{Q/G}$-$M_I$-$M_L^*$-$M_{LSP}$ & $\sigma (pb)$   & $B_{ll}$   & $B_{\nu l+l\nu}$  ($\frac{B_{\nu l}}{B_{\nu l+l\nu}}$) & $B_{LSP}$  & $B_W$   & $B_Z$ \\
\hline
Lep(Q) / $B_W=0$            & 500-440- -- -100   & 46.1 & 0.0151 & 0.4155/-- & 0.5274  & --   & 0.0420      \\
Lep(Q) / $B_{\ell\nu}=0$     & 650-440- -- -100   & 12.8 & 0.0485 & --   & 0.0  & 0.9244 & 0.0270     \\
Lep(G) / $B_W=0$            & 650-440- -- -100   & 13.6 & 0.0507 & 0.2928/-- & 0.5840  & --   & 0.0725      \\
Lep(G) / $B_{\ell\nu}=0$     & 700-440- -- -100   & 11.5 & 0.0636 & --   & 0.0  & 0.8710 & 0.0654      \\
\hline
Lep(Q)$_{on}$ / $B_{\ell\nu}=0$       & 650-440-240-100   & 12.8 & 0.0464 & --   & 0.0  & 0.9224 & 0.0312       \\
Lep(G)$_{on}$ / $B_W=0$ (a)          & 625-440-240-100   & 14.2 & 0.0474 & 0.3012 (0.0) & 0.5702  & --   & 0.0812       \\
Lep(G)$_{on}$ / $B_W=0$ (b)          & 625-440-240-100   & 14.4 & 0.0465 & 0.3129 (0.5) & 0.5561  & --   & 0.0845       \\
Lep(G)$_{on}$ / $B_W=0$ (c)          & 625-440-240-100   & 14.6 & 0.0473 & 0.3221 (1.0) & 0.5465  & --   & 0.0841       \\
Lep(G)$_{on}$ / $B_{\ell\nu}=0$       & 700-440-240-100   & 11.6 & 0.0637 & --   & 0.0  & 0.8682 & 0.0680      \\
\hline
Approx. error & N/A & $\pm$ 2 & $\pm$ 0.005 & $\pm$ 0.05 & $\pm$ 0.05 & $\pm$ 0.05 & $\pm$ 0.005 \\
\hline
\end{tabular}
\caption{Example 2: A summary of the fits to simplified models of
  leptonic structure. ``No $W$ mode'' fits have $B_W$ set to
  zero. Likewise, ``No $\ell\nu$ mode'' has $B_{\nu l+l\nu}$ set to
  zero. Those denoted as Lep(Q/G)$_{on}$ have on-shell sleptons in the
  $\ell\nu$ and $\ell\ell$ modes.  In this case, there are two types
  of $\ell\nu$ kinematics, corresponding to $I \rightarrow \ell L
  \rightarrow \ell\nu$ and $I \rightarrow \nu L \rightarrow \nu\ell$.
  The lines labelled (a), (b), and (c) correspond to different fixed fractions of $\nu
 \ell$ versus $\ell \nu$ decays, set to $0.0$, $0.5$, and $1.0$
  respectively.
\label{tab:ex2-LCMfits}}
\eet

\bef
\includegraphics[width=3in]{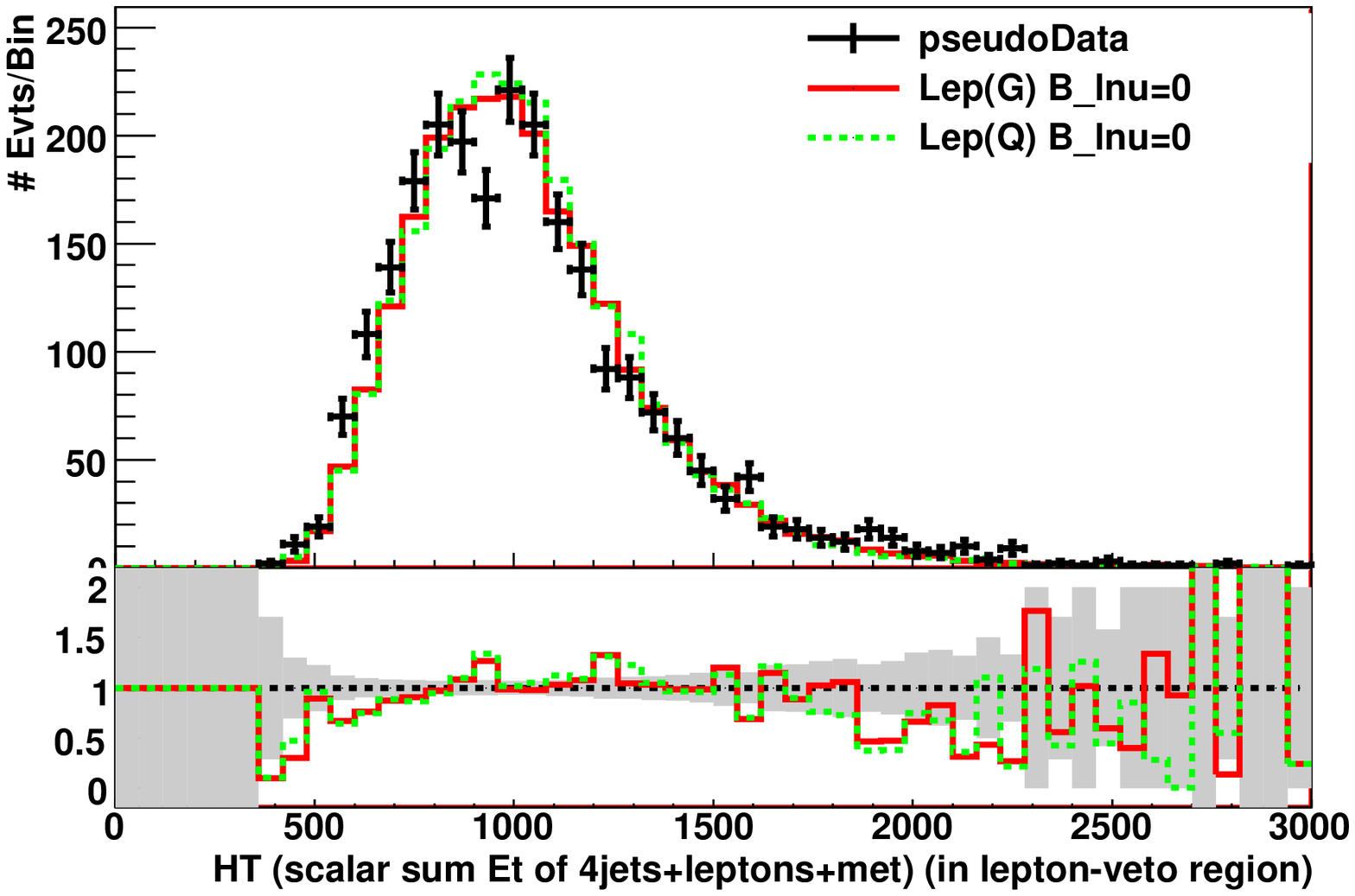}
\includegraphics[width=3in]{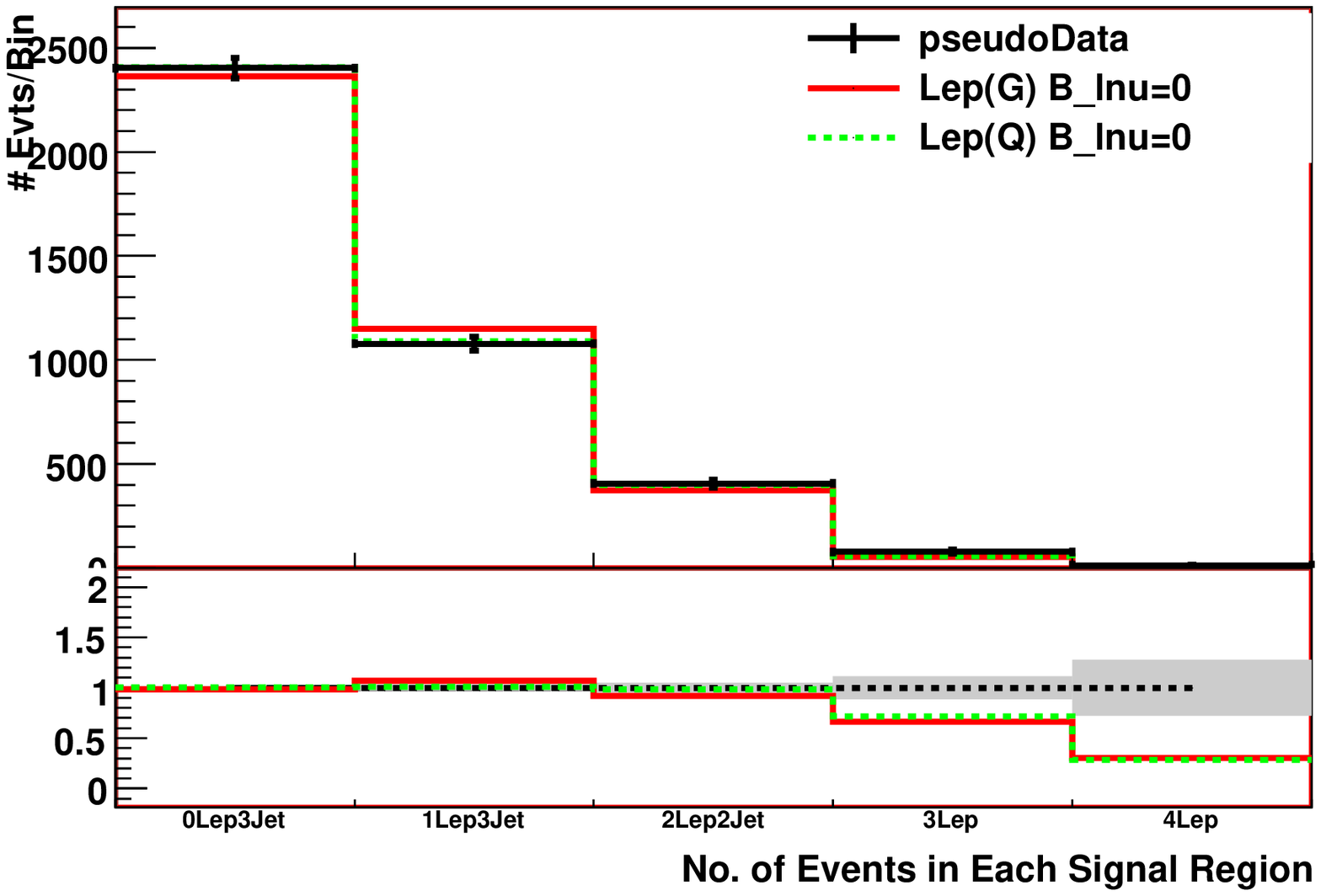}
\includegraphics[width=3in]{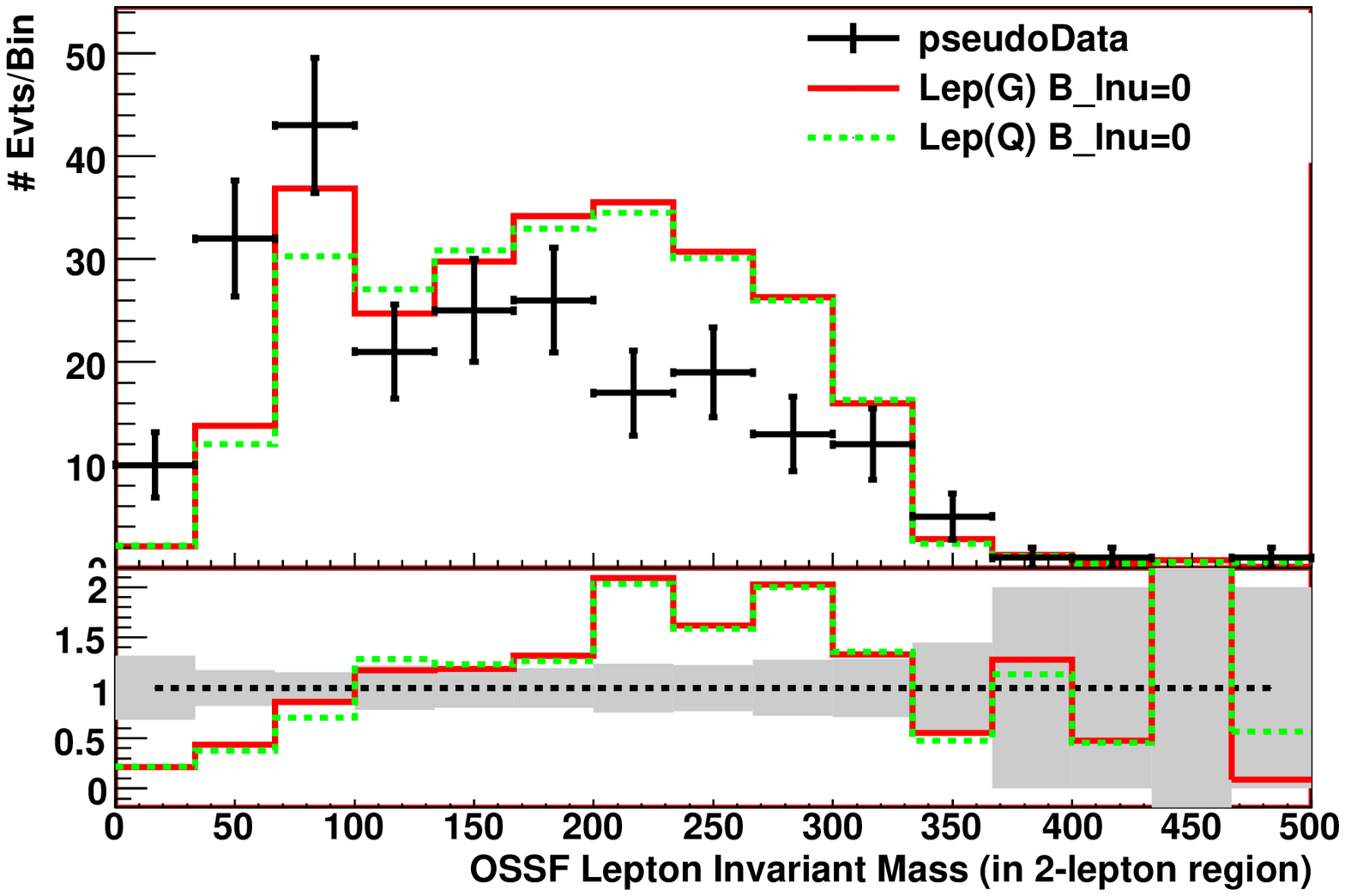}
\includegraphics[width=3in]{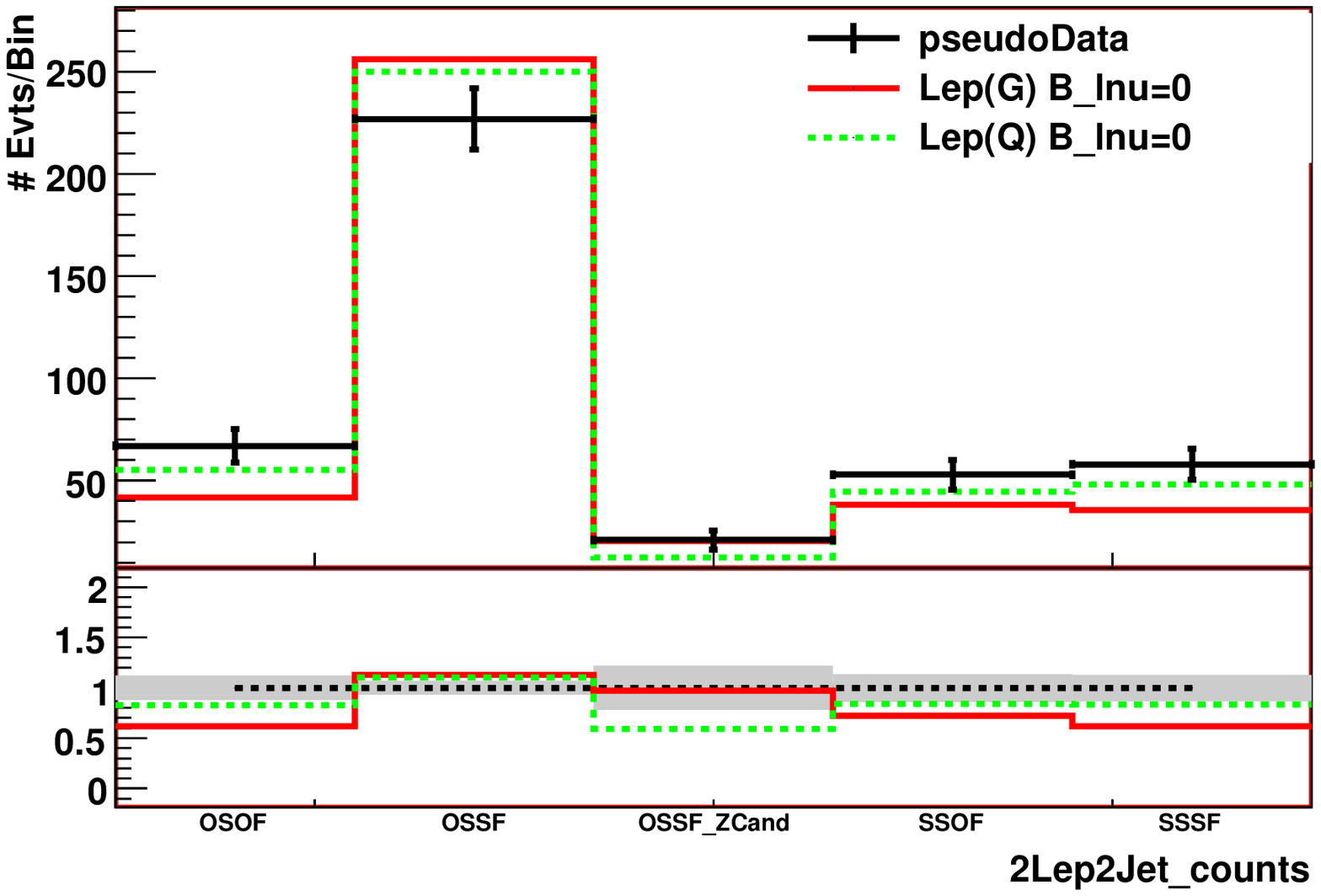}
\includegraphics[width=3in]{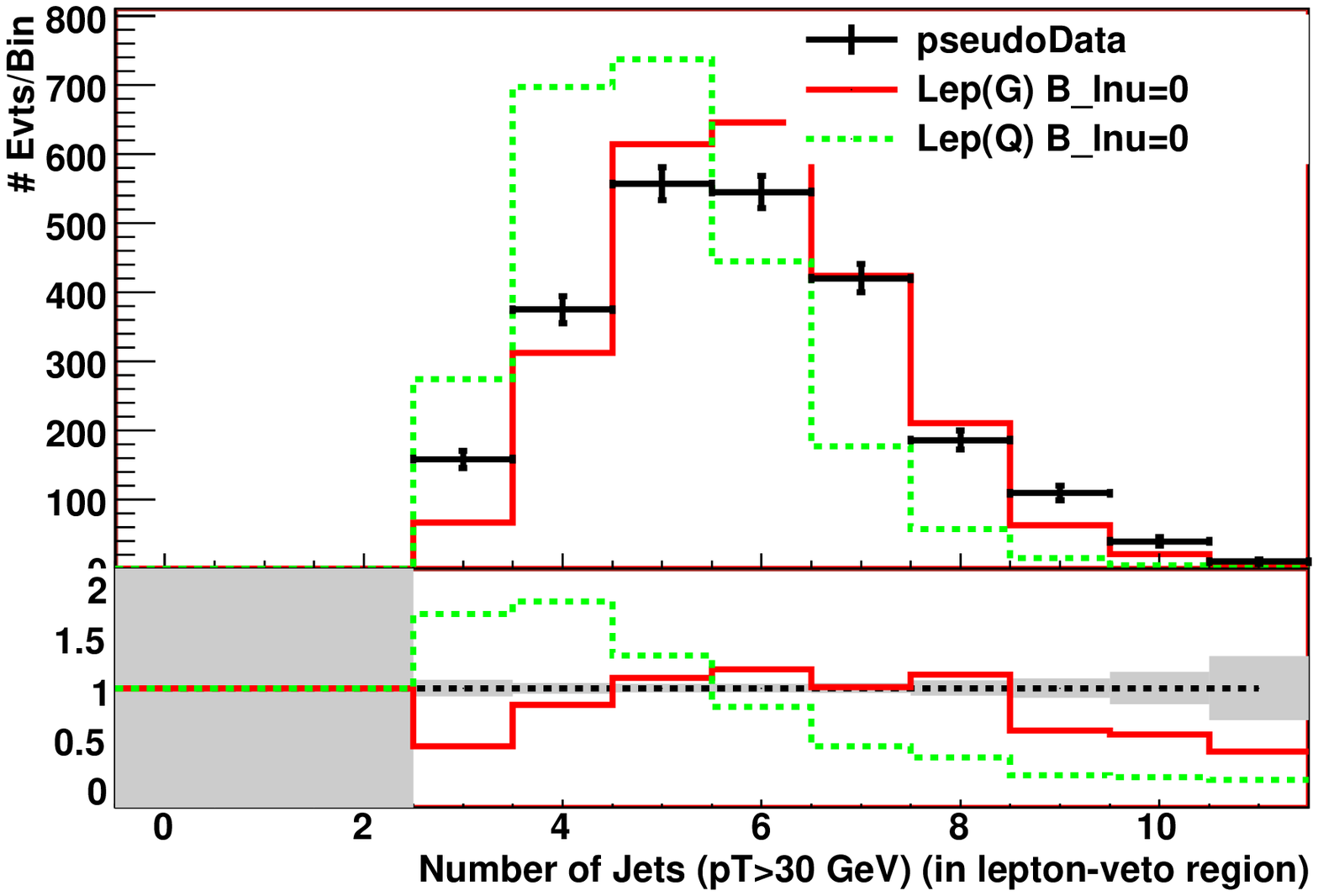}
\includegraphics[width=3in]{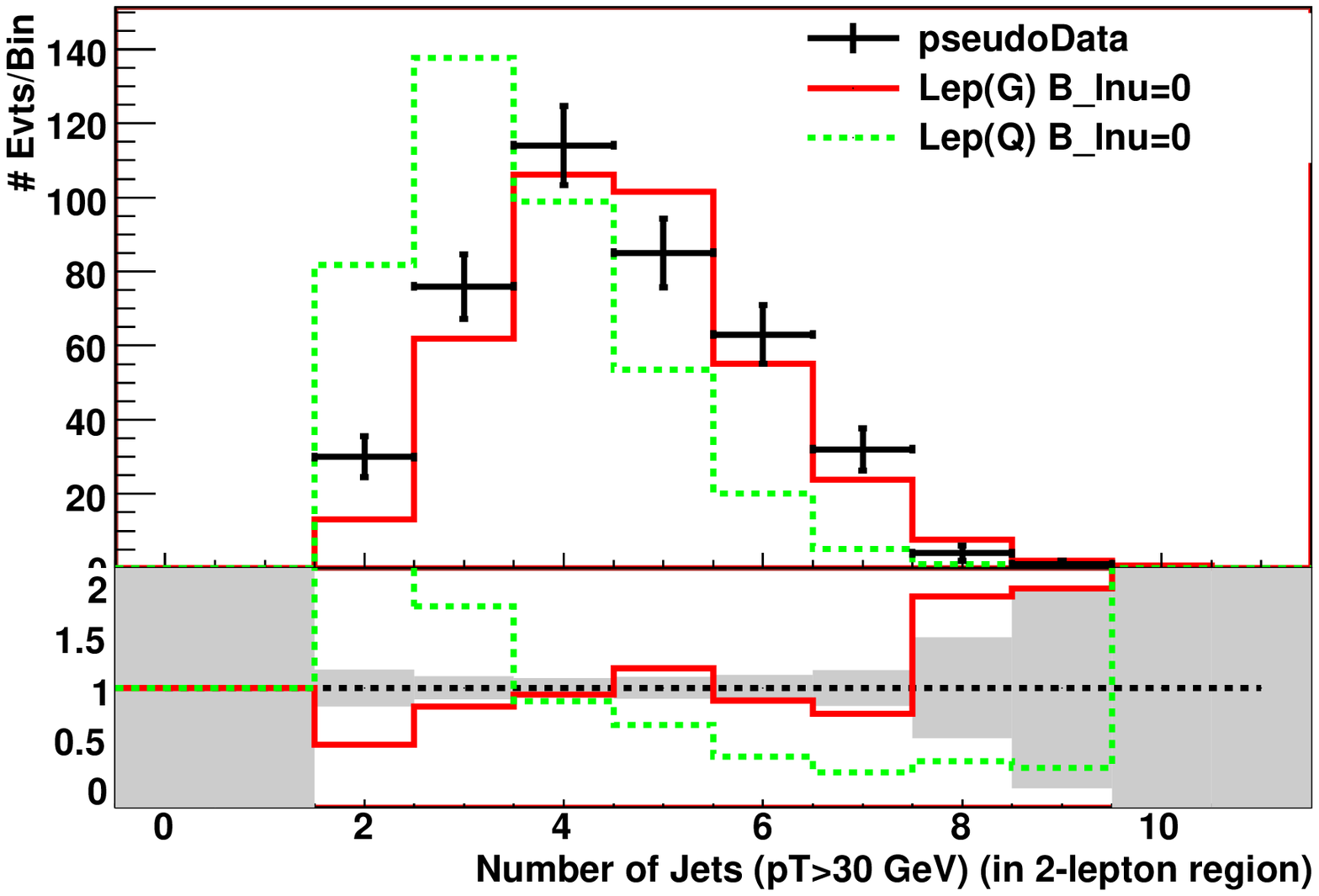}
\caption{Example 2: A subset of signatures as described by the Lep(G)
($B_{l\nu}=0$) and Lep(Q) ($B_{l\nu}=0$) fits. Jet counts and
kinematics are well-approximated by the Lep(G) fits with $W$. All fits
have difficulties modeling the di-lepton correlations, such as the
opposite sign same flavor di-lepton invariant mass shown here. We will
comment on other sources of tension in subsection
\ref{sec:6tension}.}
\label{fig:ex2-BestPlots}
\eef

As explained above, masses were not fit for any of the four
simplified models, but were estimated by setting $M_{LSP}=100$ GeV,
and then using jet and lepton kinematics and $H_T$ to estimate the
other mass scales. The mass estimates do depend on the type of fit
-- fits with the W fraction set to zero require different masses from
those with the primary $l\nu$ decays set to zero. Consequently, table
\ref{tab:ex2-LCMfits} (and table \ref{tab:ex2-HFMfits} for the b-tag
fits) shows results for different mass choices. As the fit results in
table \ref{tab:ex2-LCMfits} indicate, the di-lepton, single-lepton
(with or without W's), and Z rates for the Lep(Q) and Lep(G) models
are consistent with one another.

Two types of fits give a fairly good description of the data across
most channels. One good fit is the Lep(G) assuming $B_{l\nu}=0$ ($W$ boson rich), with a
lower bound mass estimate of $M_G\approx 700$ GeV, $M_I\approx 440$
GeV, and $M_{LSP}=100$ GeV. Another decent fit is the Lep(Q) assuming no
primary $l\nu$ decay mode with masses of $M_Q\approx 650$ GeV,
$M_I\approx 440$ GeV, and $M_{LSP}=100$ GeV. Also shown in table
\ref{tab:ex2-LCMfits} are on-shell variants of these fits. The fit
cross sections are in the range of $\approx 11-14$ pb for Lep(G) fits, and $\approx 45$ pb for Lep(Q) fits without $W$'s. 

A subset of important signatures for a subset of fits (the best of the fits) are shown in figure
\ref{fig:ex2-BestPlots}. In this figure, the $H_T$ distribution demonstrates the overall consistency of the masses for the choice of
decay parameters. The di-lepton invariant mass distribution, though not modeled very well, exhibits
an edge- or endpoint- like structure, which gives rise to a $ll$ decay
mode fraction in the range of $5\%$. The single lepton decay
fractions in these fits are high, in the neighborhood of $\approx
30\%$ (or a $W$ decay fraction close to $\approx 95\%$). Additionally,
a $Z$ decay fraction of $2-8\%$ is required. Combined, these
decay structures account for the overall lepton counts and di-lepton
flavor texture. Finally, the jet ($p_T\geq 30$ GeV) count
distributions in the $0$ and $2$ lepton regions show that Lep(G) with $W$ rich decays is preferred. 
All other fits are qualitatively worse in modeling the $30$ GeV jets. From this fact, it is worth emphasizing that the other fits (other than Lep(G) with $B_{l\nu}=0$) must be interpreted with care because there is reason to suspect that the trigger rates based on jet kinematics may be biased for those fits. 

For example, Lep(Q) (with no $W$) fit has a trigger efficiency that is
very sensitive the the mass choices. This is because $M_Q-M_I$ is
small in these fits and Lep(Q) is under-producing jets relative to the
data.  As a result, the trigger efficiency is also lower than in the
other models, and so the fit cross section is higher. Before
discussing fits of heavy-flavor production in the Btag(Q/G) models,
let's analyze the lepton and jet structure in a bit more detail, and
comment on sources of tension.

\subsubsection{$W$ Versus Primary $l\nu$ Decay and Jet Counts}

\bef
\includegraphics[width=3in]{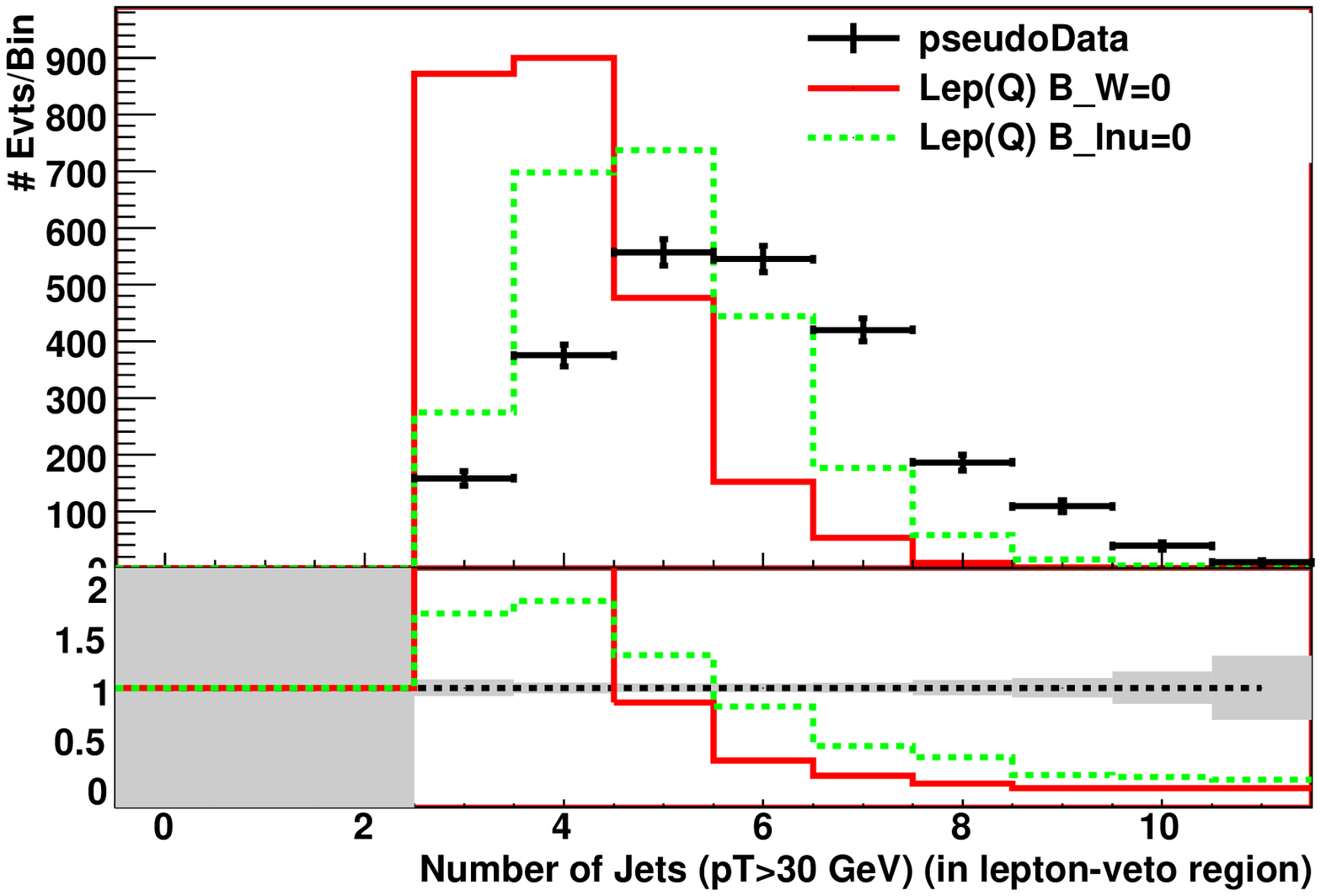}
\includegraphics[width=3in]{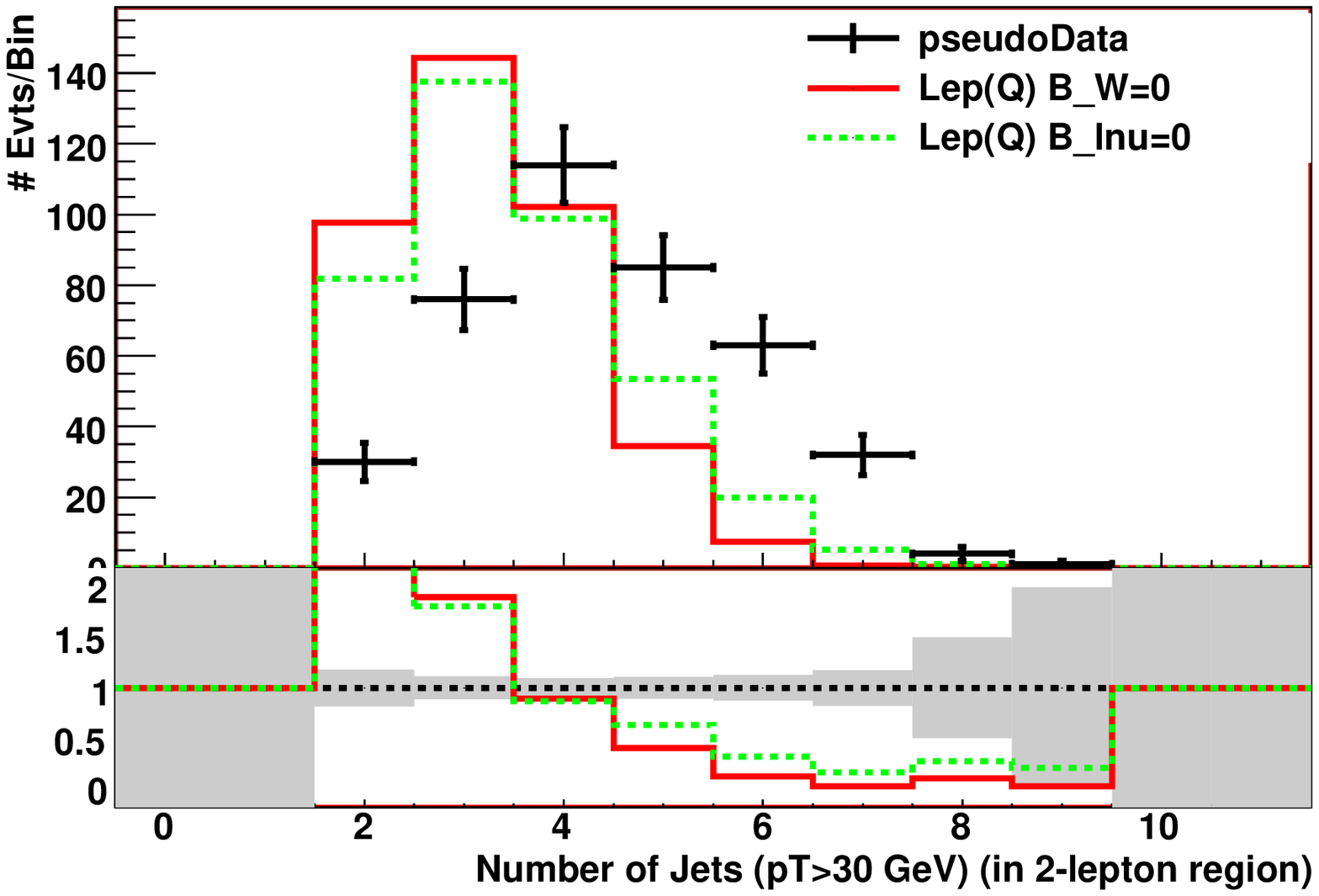}
\includegraphics[width=3in]{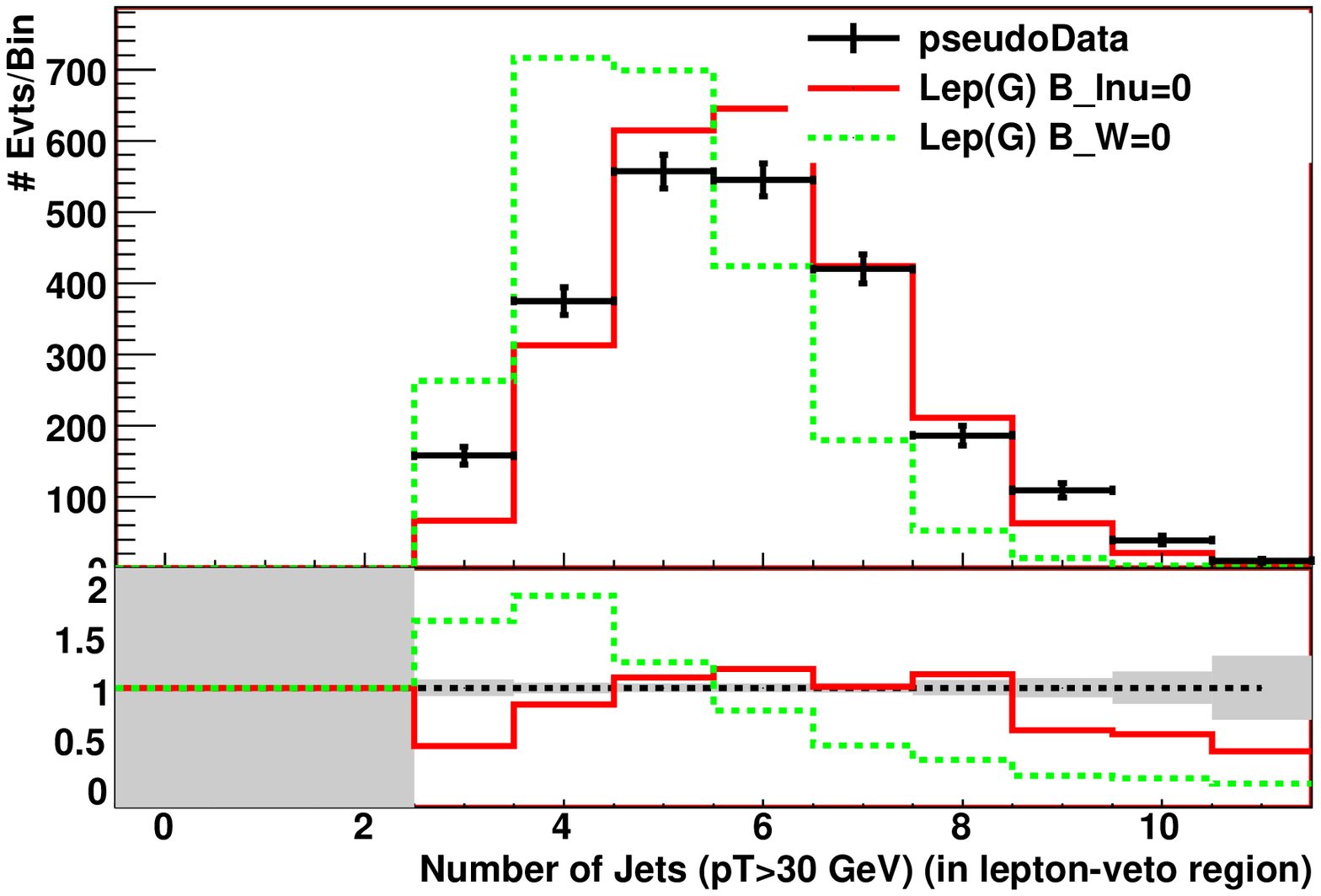}
\includegraphics[width=3in]{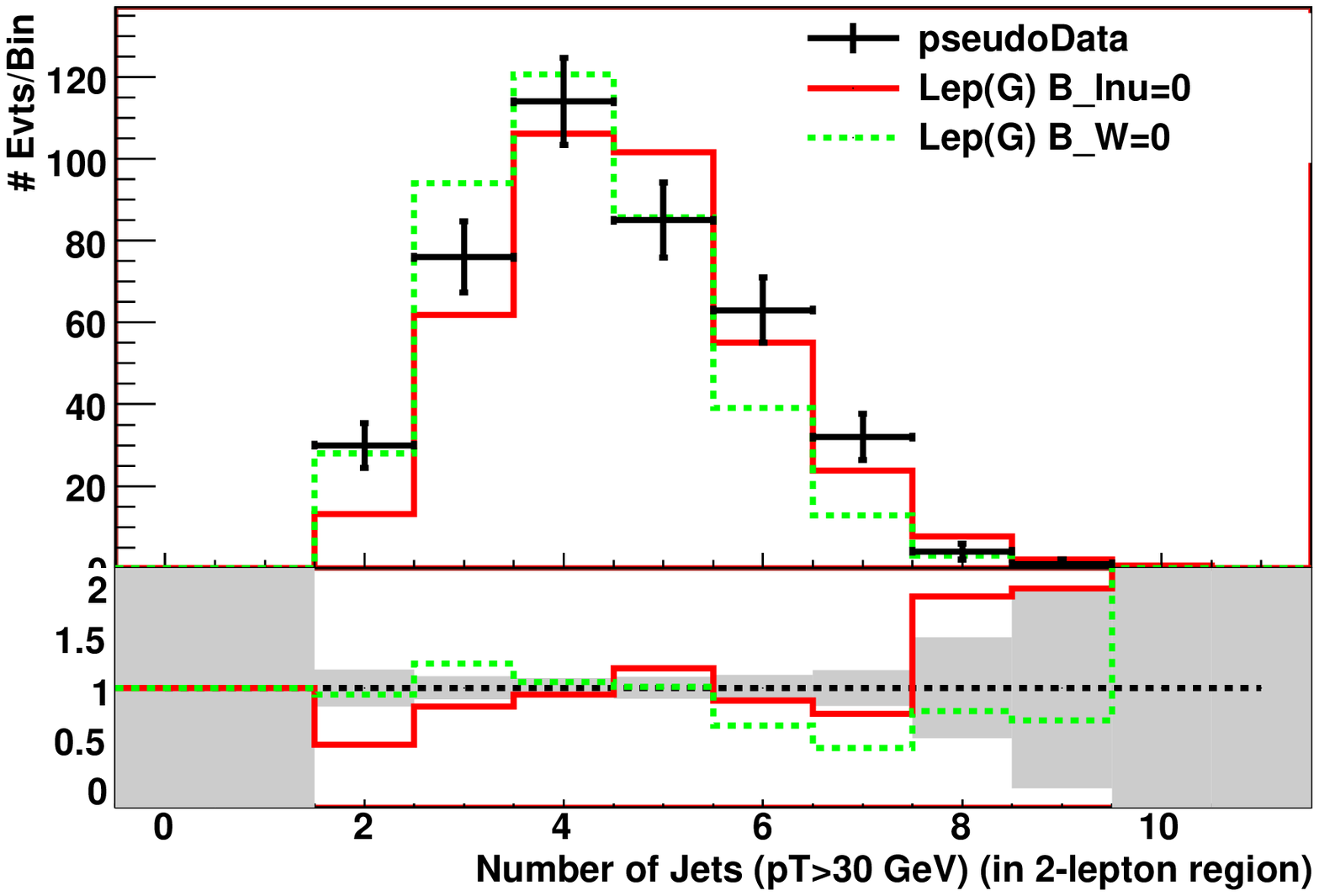}
\caption{Example 2: Jet count distributions of jets with $p_T\geq 30$
  GeV. The $0$ lepton region is shown on the right, while the $2$
  lepton region is shown on the left. This comparison is meant to
  highlight any jet-lepton correlations that exist in the data or the
  fits to leptonic models. The top row shows Lep(Q) fits, while the bottom row
  shows Lep(G) fits.}
\label{fig:ex2-WvsLNuA}
\eef

Given the high fraction of single lepton events required by the fits,
it's important to look in more detail at the impact of jet-lepton
correlations. In particular, a high $W$ fraction will necessarily have
an impact on jet counting, and our fits can give us some idea for what
combinations of quark/gluon partner production and W fractions are
consistent. This in turn will provide important clues later about the
underlying model.

Consider counts of jets with $p_T\geq 30$ GeV, as shown in
figure \ref{fig:ex2-WvsLNuA}. Neither of the Lep(Q) fits, with or without $W$ rich decays (which also have more jets) has enough jets. Though not shown, jet multiplicities of harder jets, with $p_T\geq 75$ GeV for example, look somewhat consistent with the Lep(Q) with $W$ fits. For the Lep(G) fits, the fit with
\emph{only} $W$ boson decays is clearly the most consistent, while the
$B_W=0$ fit give slightly too few jets. This general trend remains true, even as the jet $p_T$ threshold is increased, though mild tension accounting for the highest multiplicity (5, 6, or 7 jet) bins is apparent as the threshold is increased. This is mostly above the trigger threshold, so we do not expect significant trigger bias systematics in this case.  

The correlation of jet counts and lepton counts, shown here by comparing the jet counts in $0$ and $2$
lepton regions, again appears most consistent with a $W$ hypothesis for Lep(G) with the statistics available. 

\subsubsection{On- Versus Off- Shell Lepton Partners}

\bef
\includegraphics[width=3in]{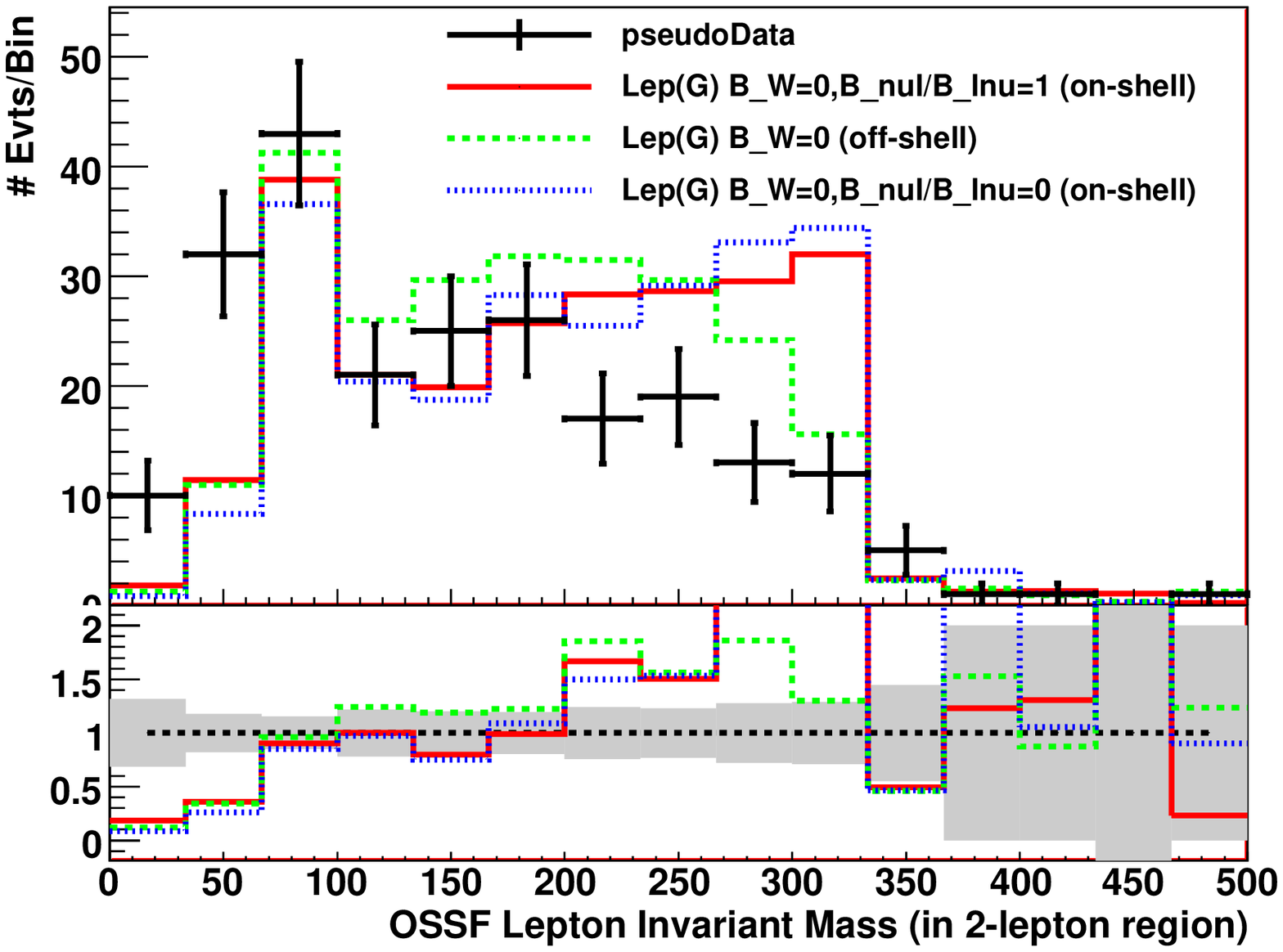}
\caption{Example 2: On- and off- shell Lep(G) fits comparing the structure of the OSSF di-lepton invariant mass.}
\label{fig:ex2-OnOffShellPlots}
\eef

Another question concerns trying to discern if the di-lepton invariant
mass structure is an edge or an endpoint. For this, we compare the on-
and off-shell variants of the leptonic models (the two variants of
Lep(G) are overlayed in Figure \ref{fig:ex2-OnOffShellPlots}; Lep(Q)
is similar). With the statistics available, neither on- nor off-shell
slepton models are fully consistent with the signal distribution.
This suggests multiple sources of di-lepton pairs, such as from
chained cascades, as is confirmed in Sec.\ \ref{sec:6tension}.  The
milder inconsistency of the off-shell variant should not be taken as
evidence that the underlying physics has off-shell lepton pairs.  For
instance, in a model with a second source of di-lepton pairs with
$m(\ell^+\ell^-)< 200$ GeV, in which only $\approx 50\%$ of observed
di-lepton pairs came from the $\ell^+\ell^-$ source modeled in the
simplified model, then the expectations for an on-shell slepton decay
in the range $200 < m_{\ell\ell} < 350$ would be reduced by half, and
statistically consistent with the data.  However, as we cannot
discriminated between the two options, and the off-shell scenario does
better model the lepton kinematics, we will consider only the
off-shell fits in the rest of this section (the on-shell fits are
included in Table \ref{tab:ex2-LCMfits} to illustrate the weak
kinematics-dependence of best-fit rates).

\subsubsection{Sources of Tension and Kinematics}\label{sec:6tension}

\bef
\includegraphics[width=3in]{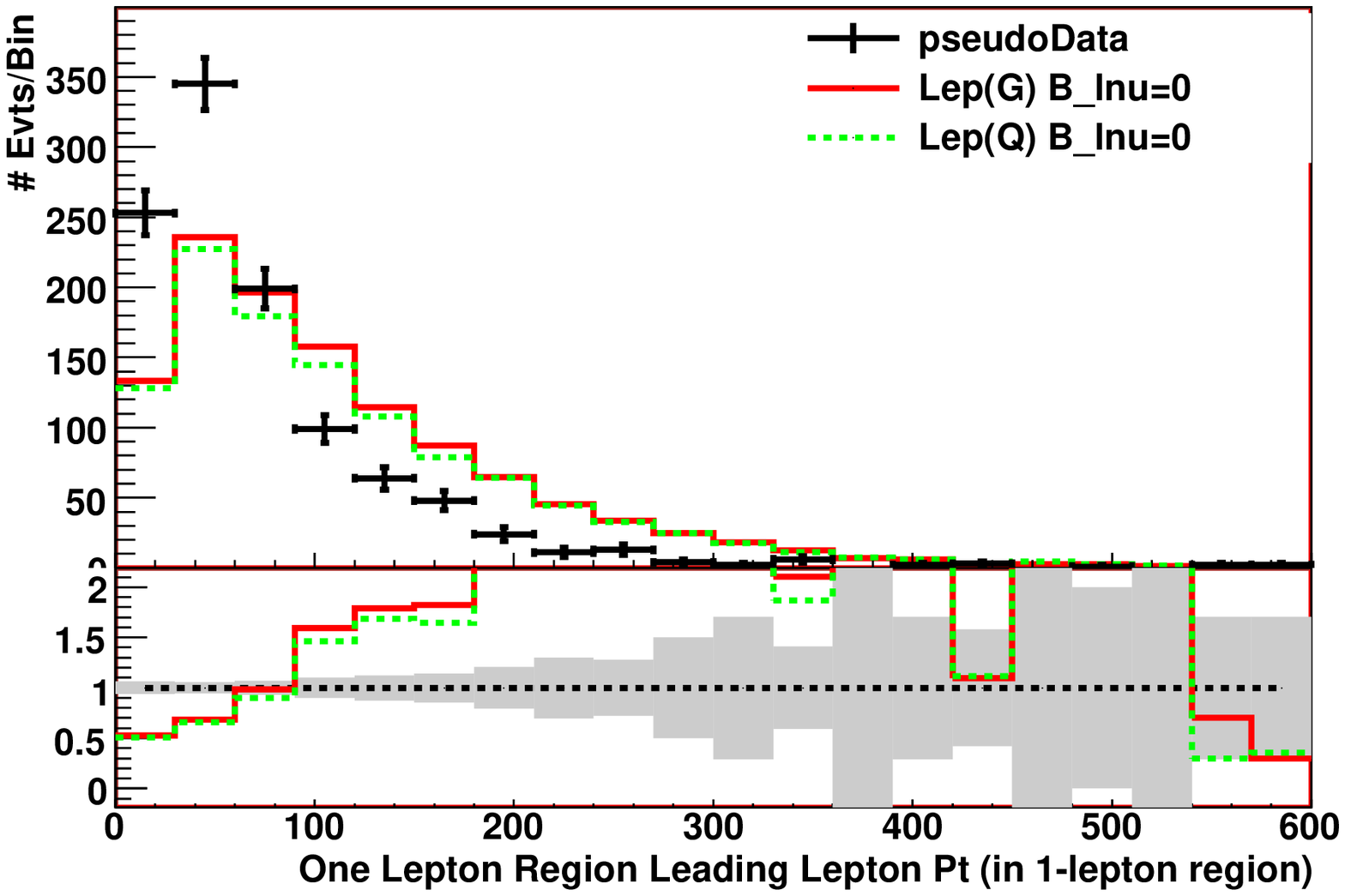}
\includegraphics[width=3in]{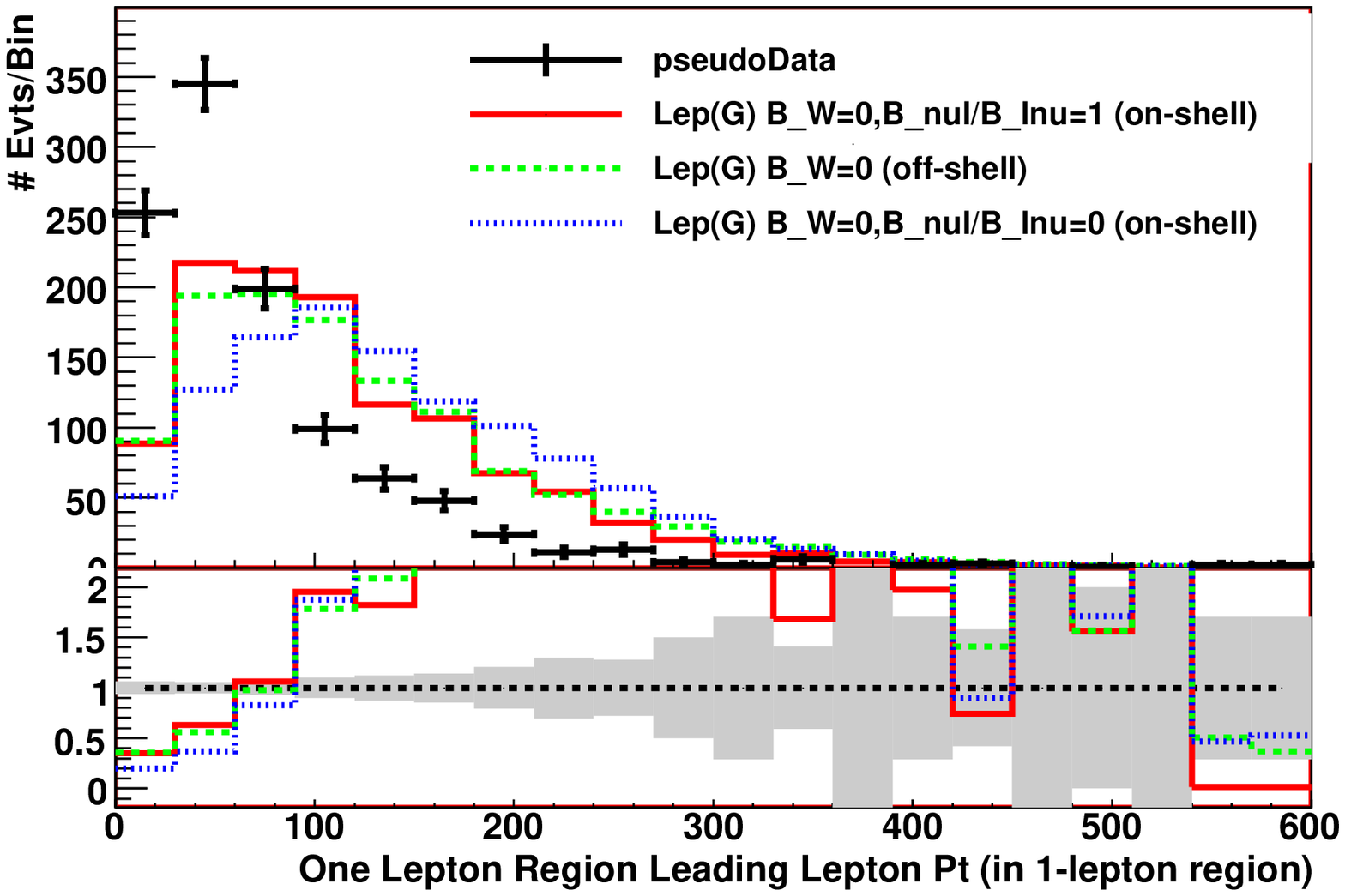}
\includegraphics[width=3in]{Example2Plots/LCM-BestFits-Overlay__2Lep2Jet_OSSF_2Lmass.eps}
\includegraphics[width=3in]{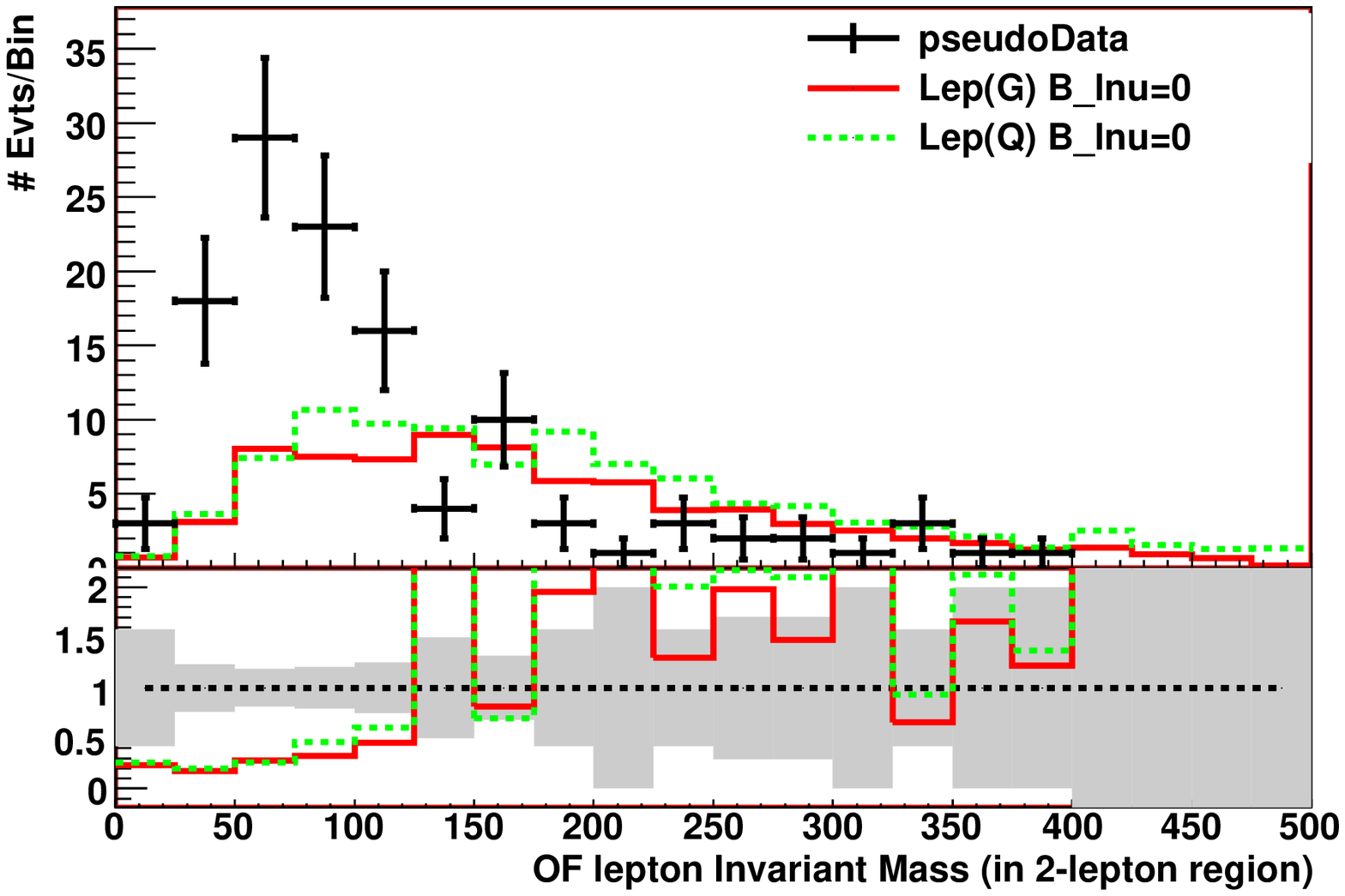}
\caption{Example 2: Representative lepton signatures where the
Lep(Q/G) fits exhibit tension accounting for the data.}	
\label{fig:ex2-TensionPlots}
\eef

Before studying heavy flavor sources in the Btag(Q/G) models, we
comment on a few persistent sources of tension with the Lep(Q/G)
fits. The most dramatic source of tension is with the lepton
kinematics. In figure \ref{fig:ex2-TensionPlots}, we show both the
lepton $p_T$ distribution in the $1$ lepton region, and the opposite and same flavor di-lepton mass distributions in the $2$ lepton region. We see that there is a deficit of leptons below $p_T\approx 75$ GeV, and that in general the
lepton $p_T$ distribution is too hard. This problem persists for both
on- and off- shell kinematics in the leptonic models. While not
justified in detail here, varying the masses in the Lep(Q/G) models
does not appreciably help this structural problem. For opposite flavor
events, the Lep(Q/G) fits give rise to harder than observed
leptons. This is reflected in the bulge of events at an invariant mass
of $\approx 30-100$ GeV relative to either simplified model. Again, these
structural problems cannot be completely resolved within the
simplified models. We should note that the signatures shown in figure
\ref{fig:ex2-TensionPlots} are representative. We do not explicitly
show here other signatures with similar problems correlated with the
lepton kinematic problems.

\subsection{B-tag Comparisons}\label{sec:6Btag}

\bet
\begin{tabular}{|l|c|c|c|c|c|r}
\hline
Btag(G)/ Parameter & $M_{G}$-$M_{LSP}$ & $\sigma (pb)$   & $B_{uu}$   & $B_{bb}$ & $B_{tt}$  \\
\hline
Btag(G) Inclusive Lepton      & 700-100   & 11.2 & 0.3836 & 0.6164 & --  \\
Btag(G) Exclusive Lepton     & 700-100   & 11.8 & 0.3541 & 0.0275 & 0.6184  \\
\hline
Approx. error & N/A & $\pm$ 2 & $\pm$ 0.05 & $\pm$ 0.05 & $\pm$ 0.05 \\
\hline
Btag(Q)/ Parameter & $M_{Q}$-$M_{LSP}$ & $\sigma_{uu} (pb)$   & $\sigma_{bb} (pb)$ & $\sigma_{tt} (pb)$ & -- \\
\hline
Btag(Q) Inclusive Lepton      & 600-100   & 1.2 &  21.4 & -- & --  \\
Btag(Q) Exclusive Lepton     & 600-100   & 0 & 0 & 16.8 & --  \\
\hline
Approx. error & N/A & $\pm$ 2 & $\pm$ 2 & $\pm$ 2 & N/A \\
\hline
\end{tabular}
\caption{Example 2: Summary of fit parameters for the Btag(Q) and Btag(G) models.}
\label{tab:ex2-HFMfits}
\eet

\bef
\includegraphics[width=3in]{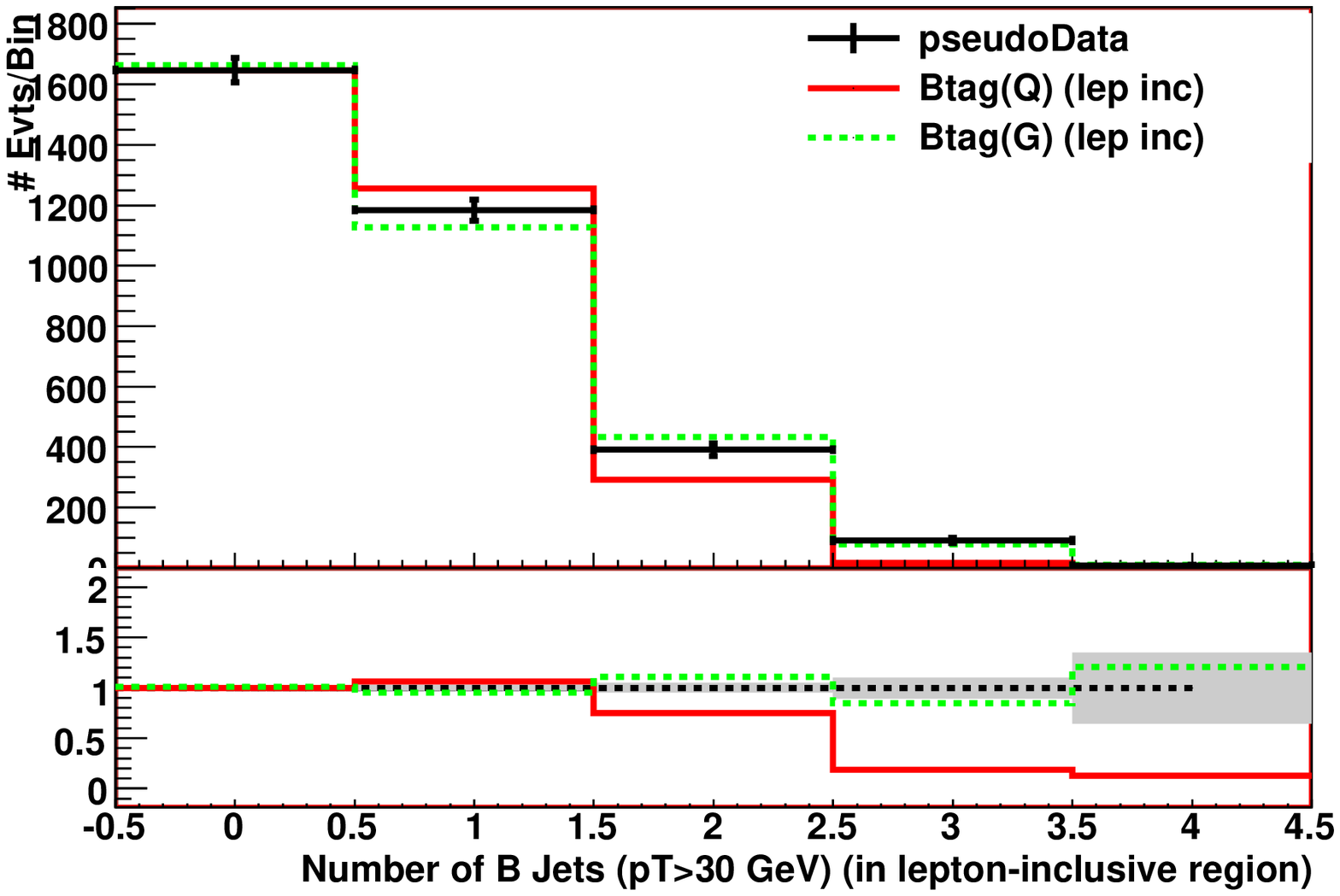}
\includegraphics[width=3in]{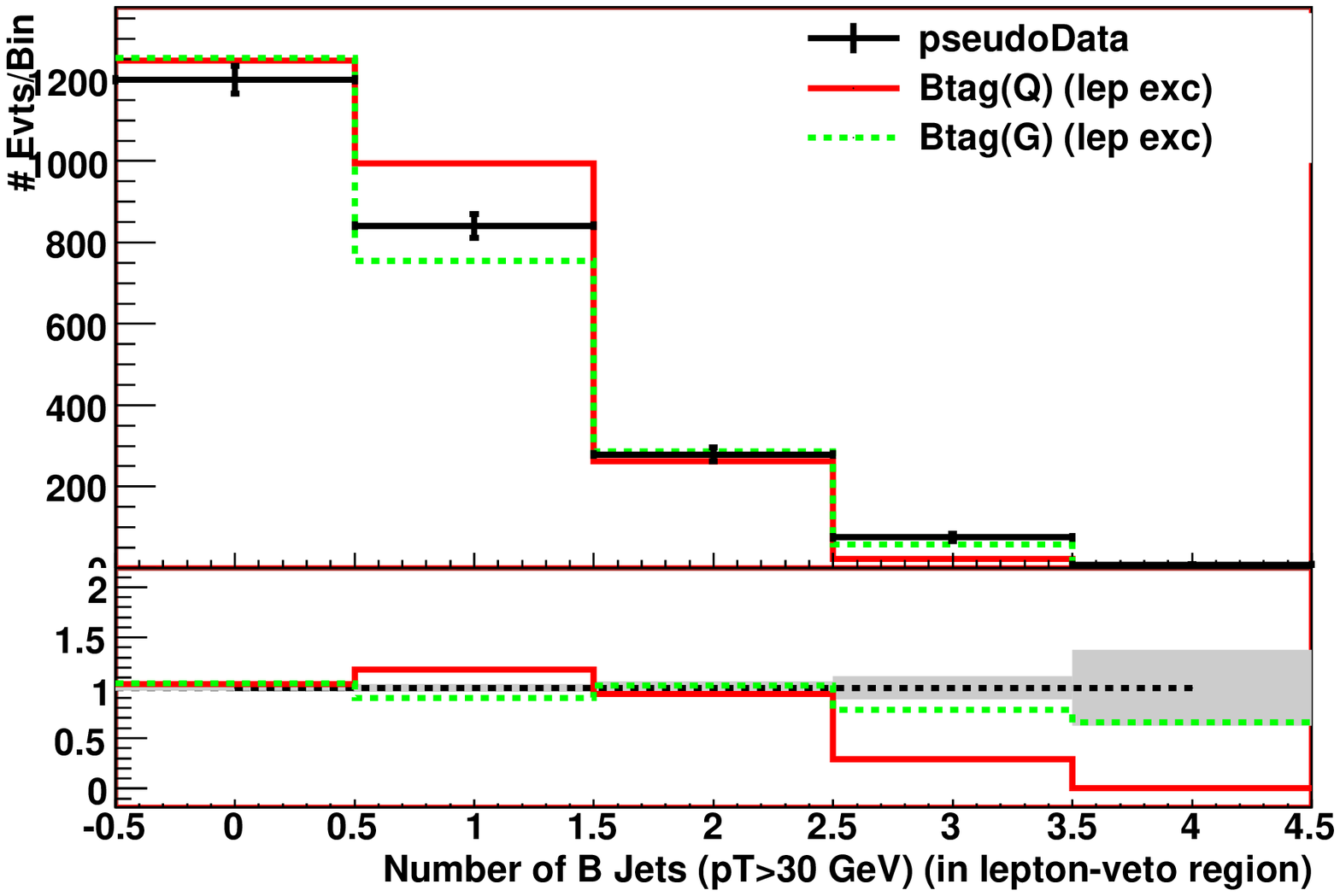}
\includegraphics[width=3in]{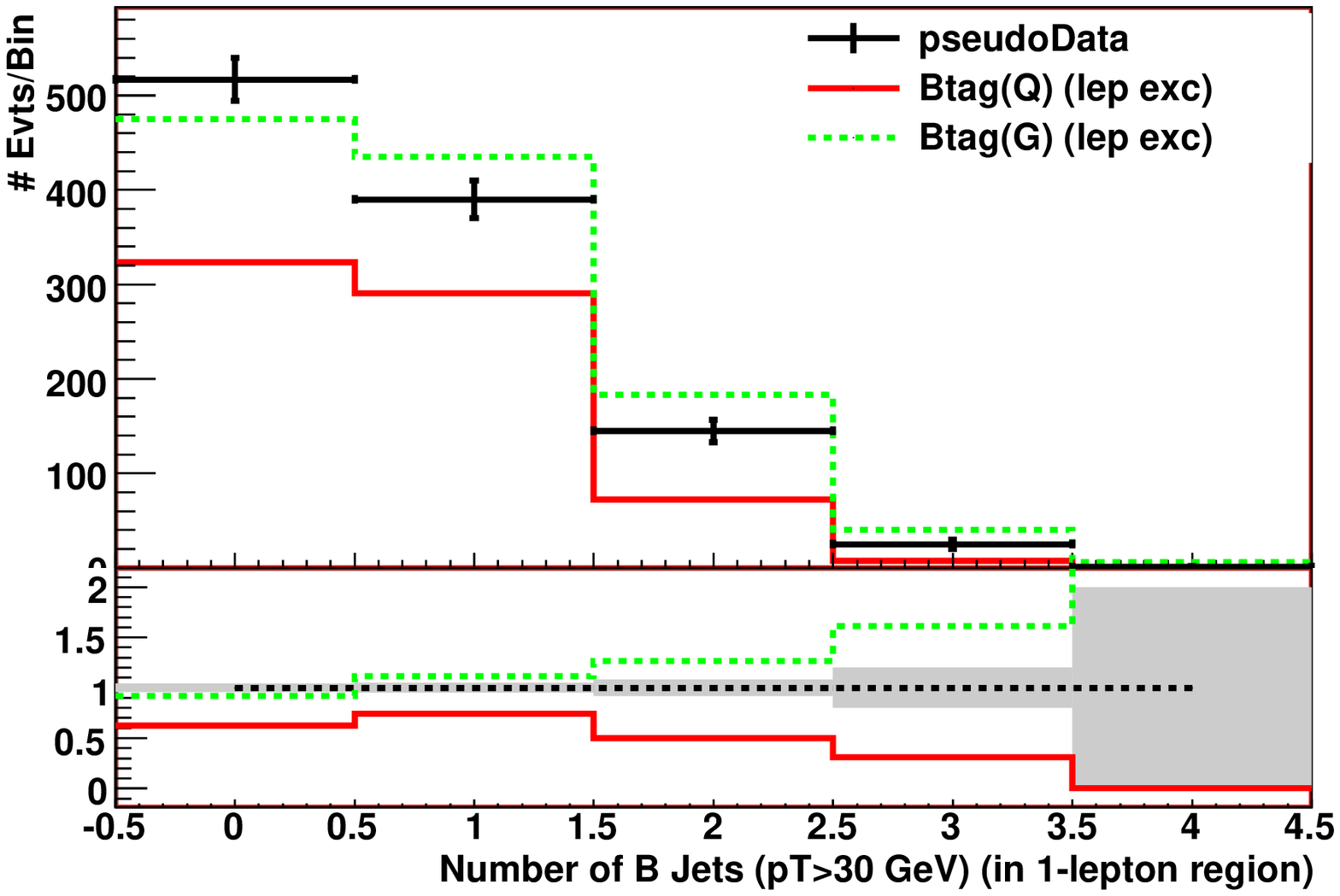}
\includegraphics[width=3in]{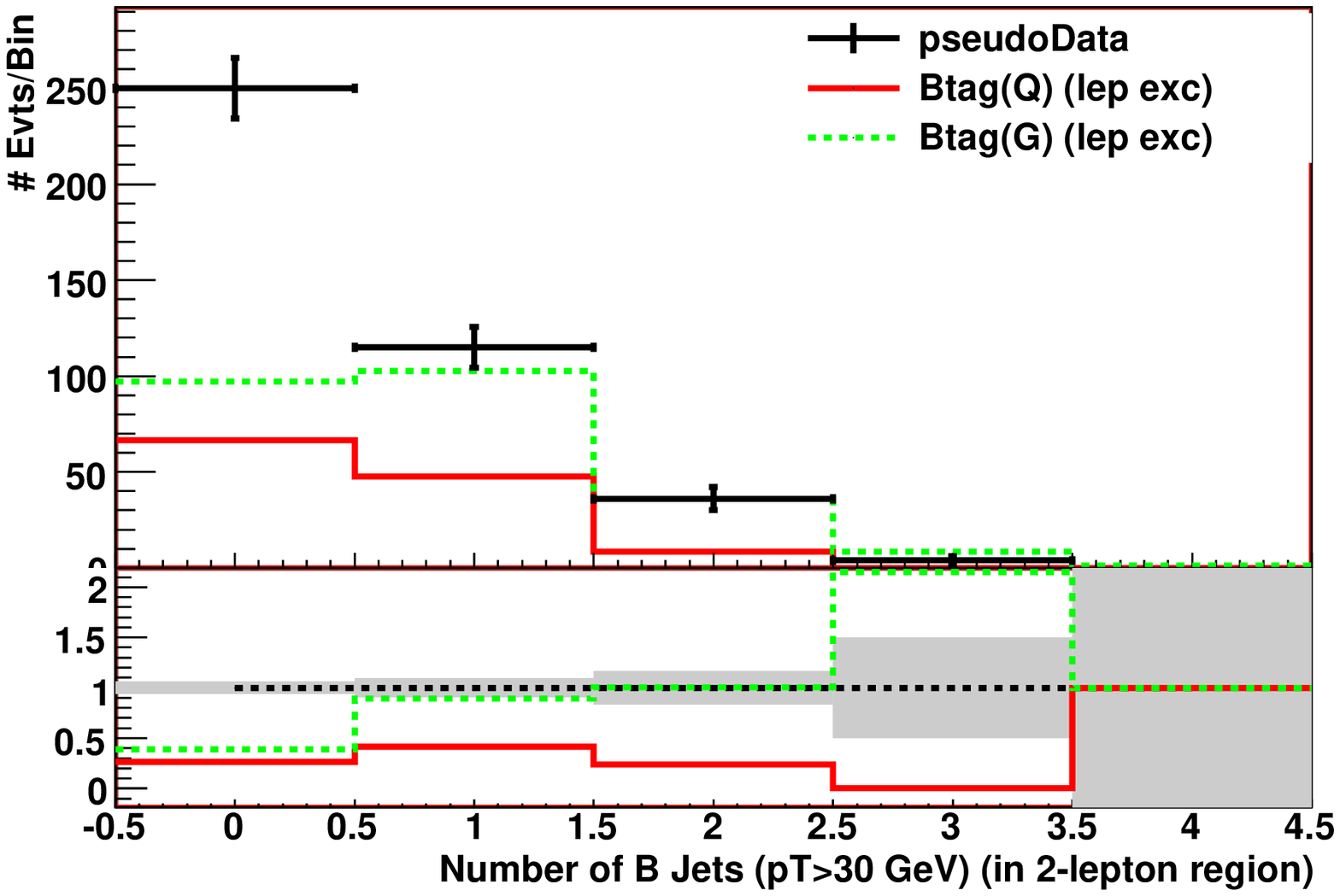}
\caption{Example 2: $b$-jet count distributions for the lepton
  inclusive, $0$, $1$, and $2$ lepton regions. Note that Btag(G)
  provides a better overall description of these signatures. Also note
  that the lepton exclusive fits, in which $W$'s from top-channels are
  used to account for leptons, fail to account for all the lepton in
  the $0$ $b$-jet regions. This is strong evidence for lepton channels
  beyond those that may accompany any third generation channels.}
\label{fig:ex2-BCounts}
\eef

As with our discussion of leptonic simplified models in this example,
the masses shown in our comparisons are lower-bound estimates, based
on $H_T$, jet and lepton $p_T$ signatures, with $M_{LSP}=100$ GeV. The
resulting lower bound estimate is $M_Q\approx 600$ GeV and $M_G\approx
700$ GeV. Results of fitting the Btag(Q/G) models to lepton inclusive
$b$-jet counts, and lepton exclusive $b$-jet counts are shown in table
\ref{tab:ex2-HFMfits}. The distribution of counts, inclusive in
leptons, is shown in figure \ref{fig:ex2-BCounts}. The deficit of the
$\geq 3$ $b$-jet count for the Btag(Q) fits is persistent across a
variety of more exclusive channels, and so we'll focus our discussion
on the Btag(G) fits. The fit cross sections for Btag(G) are consistent
with the Lep(G) cross sections. However, we now see that a rather high
$b$-jet decay fraction of $50-60\%$ is needed. So jet-flavor
universality appears to be violated. Moreover, if $b$-counts across
the lepton channels are simultaneously fit, top decay modes
dominate. What this fit tests more precisely is the consistency of
assuming all lepton come from top. This hypotheses does fail to
account for all the leptons in the $0$ $b$-jet regions. This can be
seen in the Btag(G) lepton exclusive fits of $b$-jet counts shown in
figure \ref{fig:ex2-BCounts}. So we now have robust evidence for
leptons, primarily $ll$ decays, correlated with light
jet-flavor channels.

\bef
\includegraphics[width=3in]{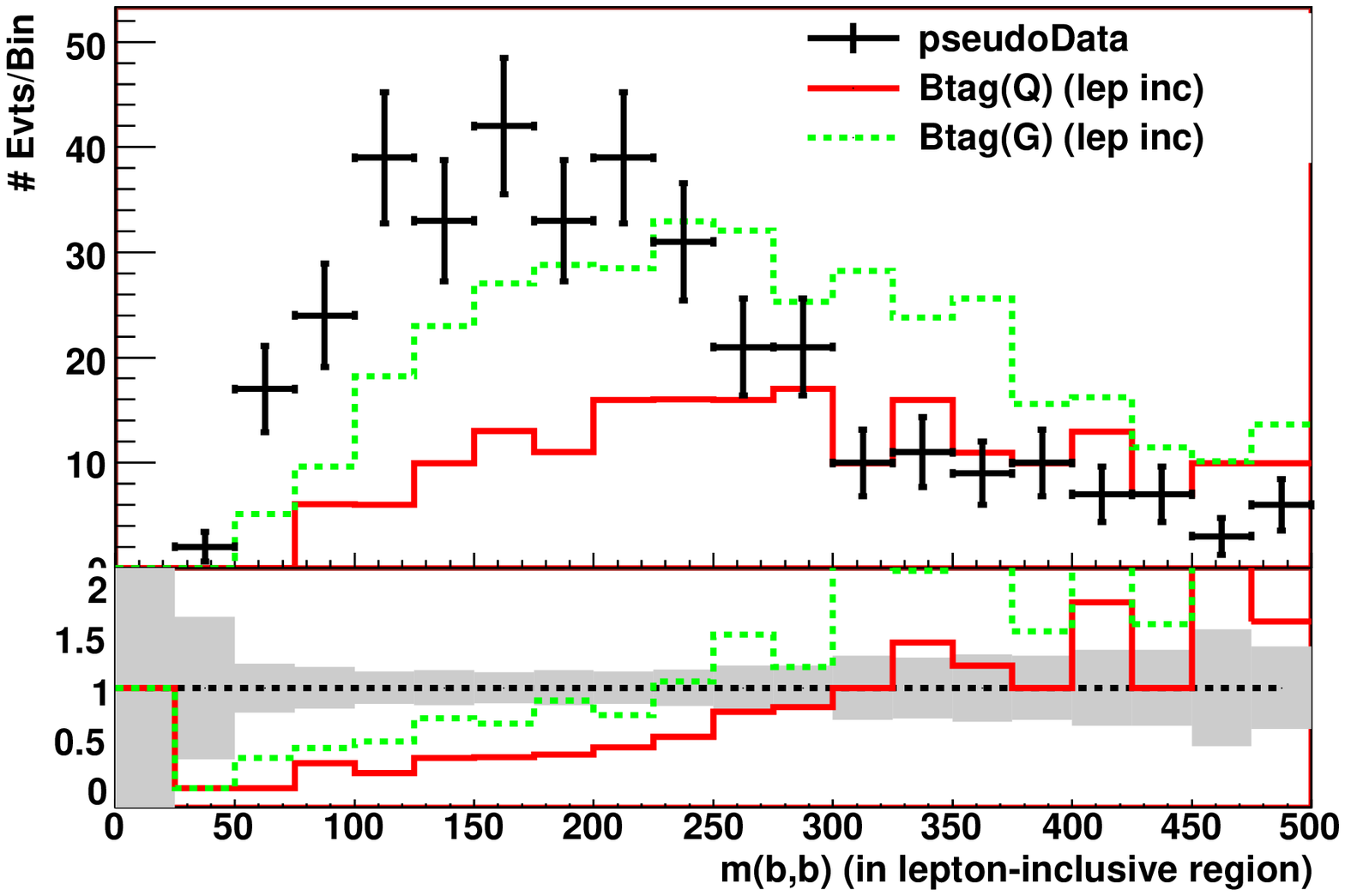}
\includegraphics[width=3in]{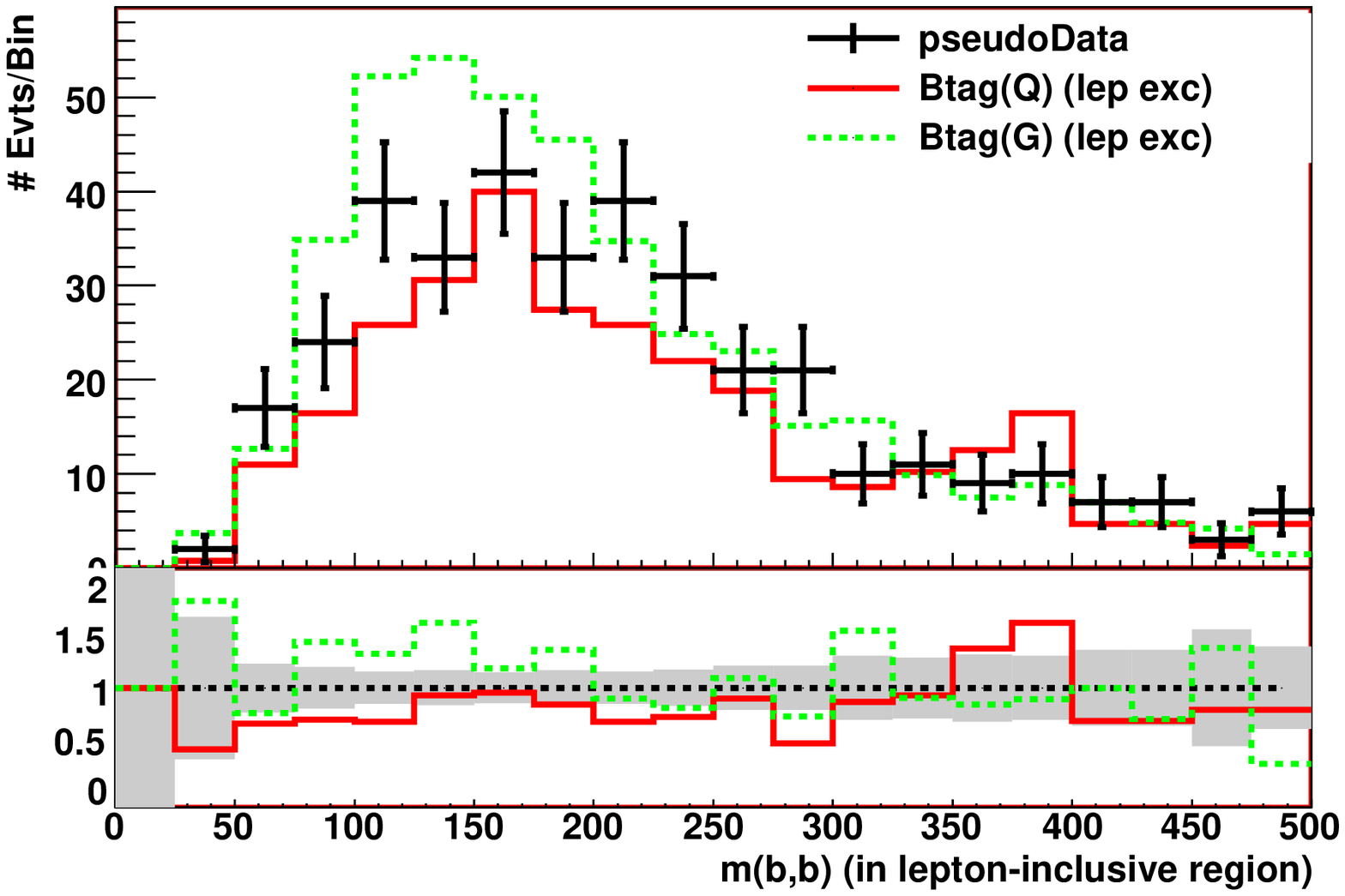}
\includegraphics[width=3in]{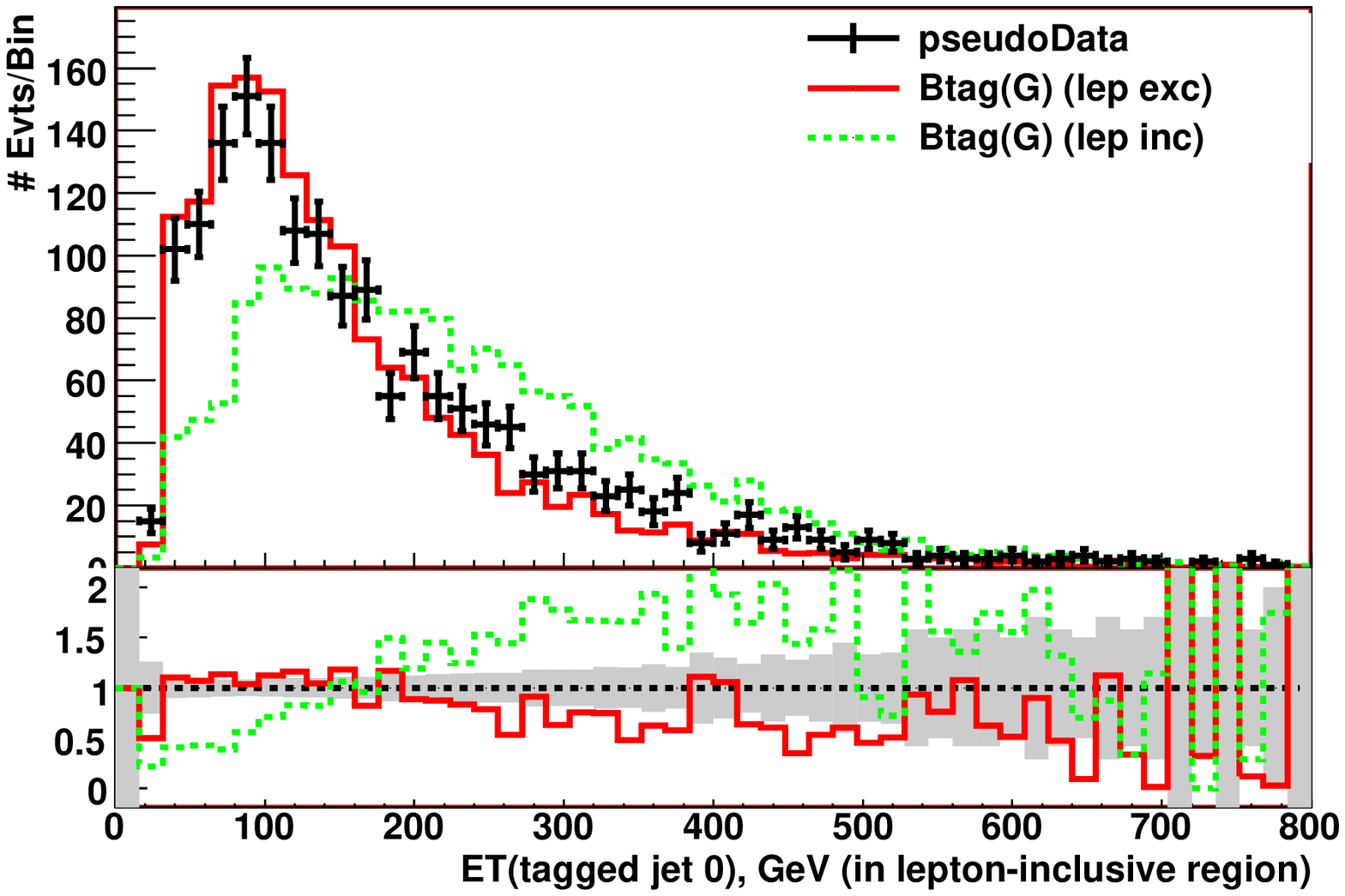}
\includegraphics[width=3in]{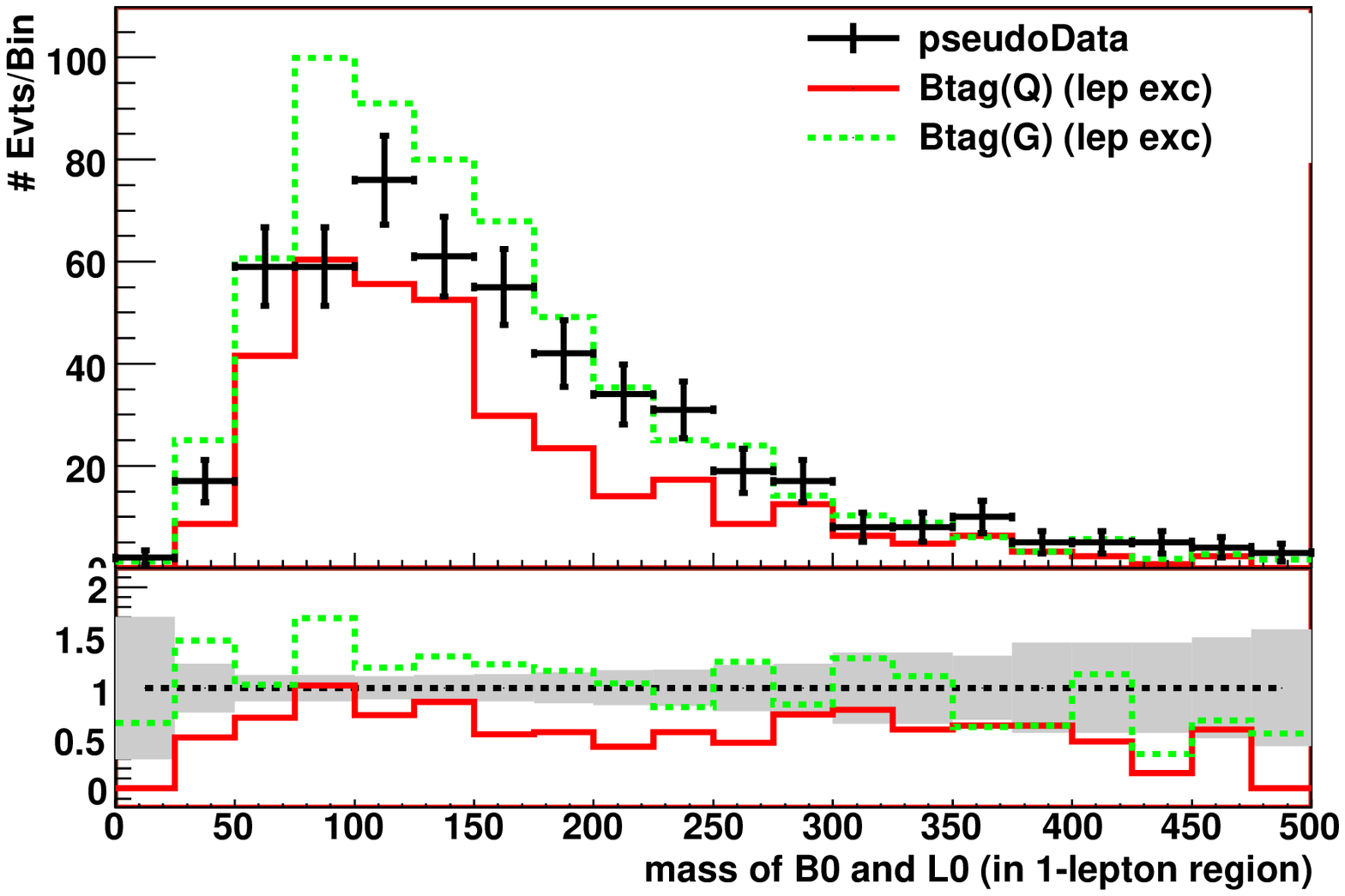}
\caption{Example 2: The $bb$ invariant mass is shown in the top
  row. Note that the lepton exclusive fits, in which top dominates,
  describe the data the best. The leading $b$-jet $p_T$ and
  lepton-$b$-jet invariant mass signatures (in the bottom row) are
  also more consistent with the top rich lepton exclusive fits.}
\label{fig:ex2-HFMPlots}
\eef

Figure \ref{fig:ex2-HFMPlots} shows several other useful comparisons
of the b-tag fits. For the range of masses considered, both $b$-tag
models describe the $bb$ invariant mass signature quite well when it
is dominated by top decay modes, as is the case with the lepton
exclusive fits. Moreover, the kinematics of the $b$-jets themselves
are better modeled by the lepton exclusive fits, in which the primary
source of $b$-jets is from top. These comparisons do not directly
imply a preponderance of top decays (an attempt at direct top
reconstruction might be a better source of evidence for this), but
they certainly support a top-rich hypothesis.

The data is globally well described by a subset of limits of the
four simplified models, as presented above, despite the simplicity
of these models. As we'll see, the qualitative and quantitative
information from the above fits is good enough to directly motivate
model-building. However, due to tensions in the fits, more precise
information about the underlying description can only be obtained by
comparing models directly to these fits. We now turn to this topic.

\subsection{Interpreting Simplified Model Fits}\label{sec:6Interpretation}

In this subsection, we will demonstrate how one can compare any model (in
this case, a set of parameter points in the MSSM) to the simplified
model results presented earlier. We will illustrate how the comparisons can be done, by exhibiting three partially
consistent MSSM parameter points. We emphasize that this is possible only after the simplified models have been fit experimentally. This allows complicated detector corrections to be folded in properly by experimentalists while carrying out the analysis and fit. The reader is referred back to subsection \ref{sec:6summary} for a summary of salient features of the fits, as we've performed them in this paper. 
We'll start by outlining possible mechanisms for reproducing the characteristics of the fit simplified models in the MSSM.  

\subsubsection{Plausible SUSY Models}
Two of the properties of the data identified above seem especially
telling about how it could be modeled.  The first is the
enhancement of heavy-flavor decays (but not to 100\%).  There are three
ways of achieving this, starting from a gluon partner initial state:
\begin{description}
\item[Off-shell decays/enhancement from spectrum:] All electroweak
  states have flavor-universal couplings ($\widetilde W/\tilde B$), and
  all squarks are heavier than the gluino; stop decays dominate
  because $m({\tilde t}) \ll m({\tilde q})$.  
\item[Off-shell decays/enhancement from couplings:] All squarks are
  heavier than the gluino and have comparable mass; stop decays dominate
  because $y_t \gg g_2, g'$ is the largest coupling among the
  electroweak-inos. $b$ decays can also be significantly enhanced at
  large $\tan\beta$.
\item[On-shell decays/enhancement from phase space:] all squarks are
  lighter than the gluino, and gluino decays are on-shell, but
  $m({\tilde g}) - m_({\tilde t}) \gg m({\tilde g}) - m({\tilde q})$,
  and reduced phase space shuts off decays to the first two
  generations.  Direct and associated production of the light-generation squarks also contribute to the effective $\tilde g \rightarrow q \bar q \MET$ (i.e. non-$b$) mode.
\end{description}
We will focus here on the latter two.

The second interesting feature is the presence of significant leptonic
modes: the overall dilepton branching fraction (which we should
attribute to an on- or off-shell slepton) and the single-lepton
fraction (which could come from a combination of slepton cascades and
$W$'s, including $W$'s from top quarks.  The energy of the leptons
suggests a mass splitting between color-singlet intermediate states
that is big enough to allow $W$, $Z$, or Higgs emission; it is
unlikely that off-shell slepton cascades would be competitive with
these modes, so we are led --- not by kinematics, but by the large
branching fractions --- to consider regions of MSSM parameter space
with \emph{on-shell} intermediate sleptons.

\subsubsection{MSSM Comparison Points}
\label{sssec:susymodels}
In this section, we explore the qualitative possibilities mentioned
above in more detail by comparing each model to the best-guess
simplified models.  Model parameters are tuned to reproduce the
features of the simplified models.  We will present comparisons of
three qualitatively different SUSY models with the simplified models.
Our goal is not to study MSSM parameter space exhaustively, but to
demonstrate the process of model/simplified-model comparison.

Most theorists will have at their disposal at best a simple detector
simulator with roughly the same behavior as the real detector (for
example, $b$-tagging efficiency correct to within 20\%, roughly
comparable jet energy resolution).  This is certainly not adequate
for generating distributions to compare to observed data!  If the
simplified models are truly a good representation of the data, in that
both the distributions of interest and distributions that affect their
efficiencies are well modeled, then a theorist can simulate the
best-fit simplified model with the limited tools at his or her
disposal, as well as the models they are trying to compare to data.
One can reasonably assume that, when a model reproduces features of
the fitted simplified models in the crude detector simulator, it will
\emph{also} reproduce the same features in the actual detector, and so is a
reasonable candidate explanation for the observations.

It is important to check this intuition by comparing best-fit
simplified models to different new-physics models in a full detector
simulator for CMS or ATLAS, and again checking their consistency in an
untuned simulator such as PGS, with care taken to make PGS objects
``analogous'' to those used in the full detector simulator (e.g. using
the same cone size and isolation criteria).  This is a subject for
future work.  

We summarize the three MSSM parameter points below
(Pythia parameters are in Appendix \ref{app:susyPythiaCardExample2}) and
compare them to the simplified models in Figures
\ref{fig:ex2_and_models_LCM} and \ref{fig:ex2_and_models_HFM}.
\begin{description}
\item[SUSY A (the correct model)] has split left- and right-handed
  squarks, with the right-handed squarks 160 GeV heavier than gluinos,
  and the left-handed squarks just lighter than the gluinos.  Both
  $\tilde g \rightarrow \tilde q_L q$ (a $\approx 15\%$ decay mode)
  and associated $\tilde q_L \tilde g$ production contribute to the
  non-b fraction.  As anticipated from $b$ kinematics in the Btag(Q/G)
  fits, decays involving top quarks dominate the third-generation
  gluino decays, of which about 1/3 are $t\bar t$ , and 2/3 $b \bar t$
  or $t \bar b$.
\item[Off-Shell B] has 700 GeV gluinos decaying through off-shell
  squarks of all generations.  The squarks of the first two
  generations are very near in mass to the gluino (720-750 GeV);
  third-generation squarks are nearly degenerate (the right-handed
  stop at 575 GeV is lighter than the gluino, but $m(\tilde t_R)+m(t)
  > m(\tilde g)$ so the decay is still off-shell). With these masses
  and a light higgsino (the LSP near 100 GeV), the Gluino decays to
  $b$ and $t$ modes $\approx 75\%$.  The remaining 25\% of gluinos
  decay to a wino, which decays through an intermediate left-handed
  slepton.  The value of $\tan\beta$ and the precise 3rd-generation
  squark masses determine the relative rates of $b\bar b$, $t \bar b$,
  and $t \bar t$ decays of the gluino; a particular combination is
  tightly constrained by lepton multiplicities.  The bino is light,
  and approximately degenerate with the neutral higgsinos.  We have no
  kinematic evidence for this --- rather, it was necessary to
  reproduce the observed frequencies of different kinds of di-lepton
  events (specifically, models considered without a light bino had 
  over-produced OSSF dilepton events by a factor of two).  This fact
  is noteworthy for another reason: a light pure-higgsino LSP is
  inconsistent with standard cosmology; a light higgsino/bino mixture
  can be fully consistent.
\item[On-Shell C] has a very similar gaugino and higgsino spectrum to
  B.  However, in this case the 700 GeV gluino decays almost
  exclusively to lighter sbottoms ($m(\tilde b_R, \tilde t_R, \tilde
  q_3) = 620$ GeV); the light-flavor modes result from associated
  production of the light-flavor squarks ($m(\tilde q) =720-740$ GeV),
  which decay frequently to the wino and bino because phase space
  suppresses their decays to the gluino.
\end{description}

A comparison between the spectrum of the model used to 
generate pseudo-data and the two incorrect comparison models (B and C) is given in
Fig.~\ref{fig:ex2_spectrum}.  The Pythia parameters are also given in
Appendix \ref{app:susyPythiaCardExample2} for the correct model and 
\ref{app:susyPythiaConjecturesExample2} for the two guesses.

\bef
\vspace{-0.7in}
\includegraphics[width=3in]{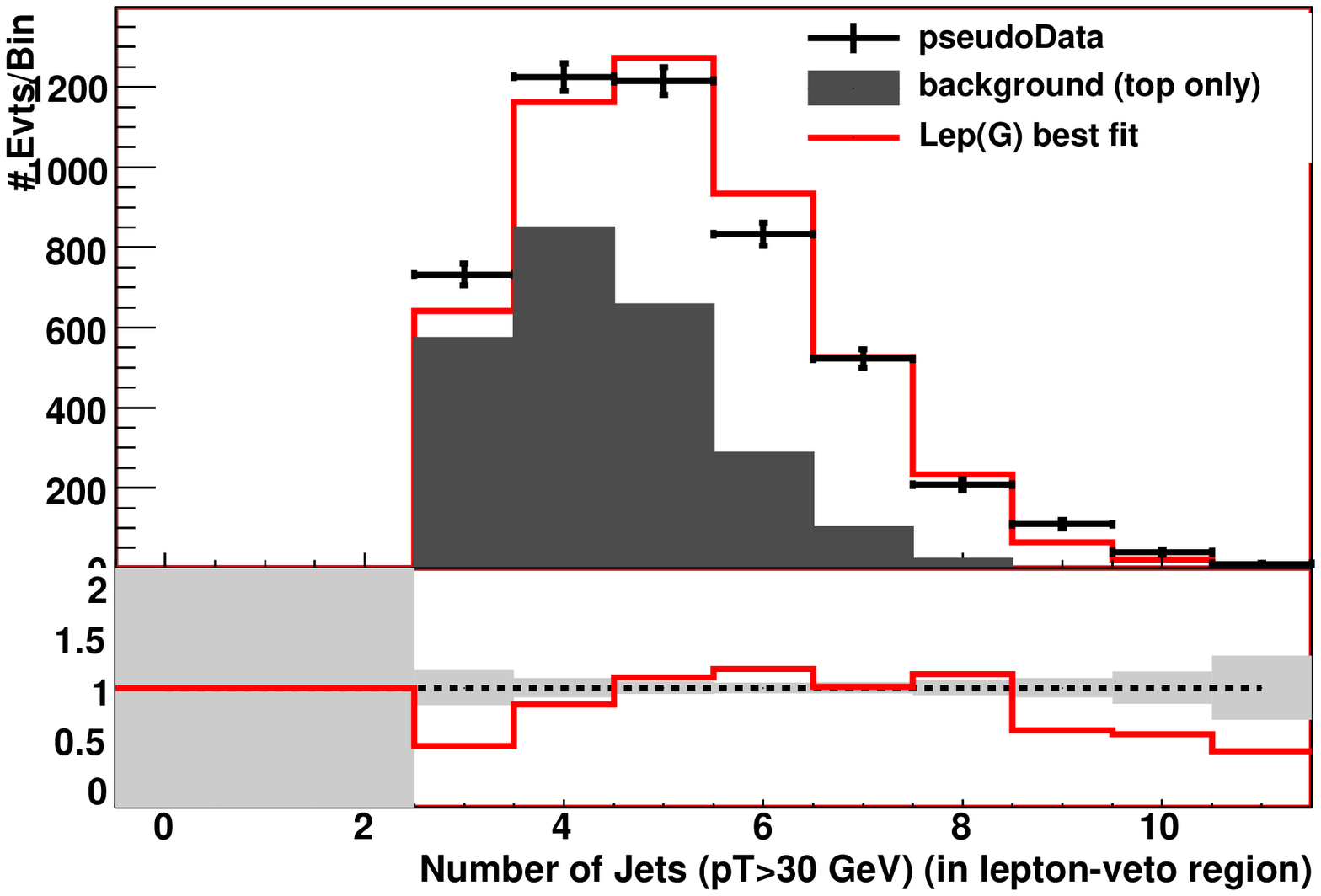}
\includegraphics[width=3in]{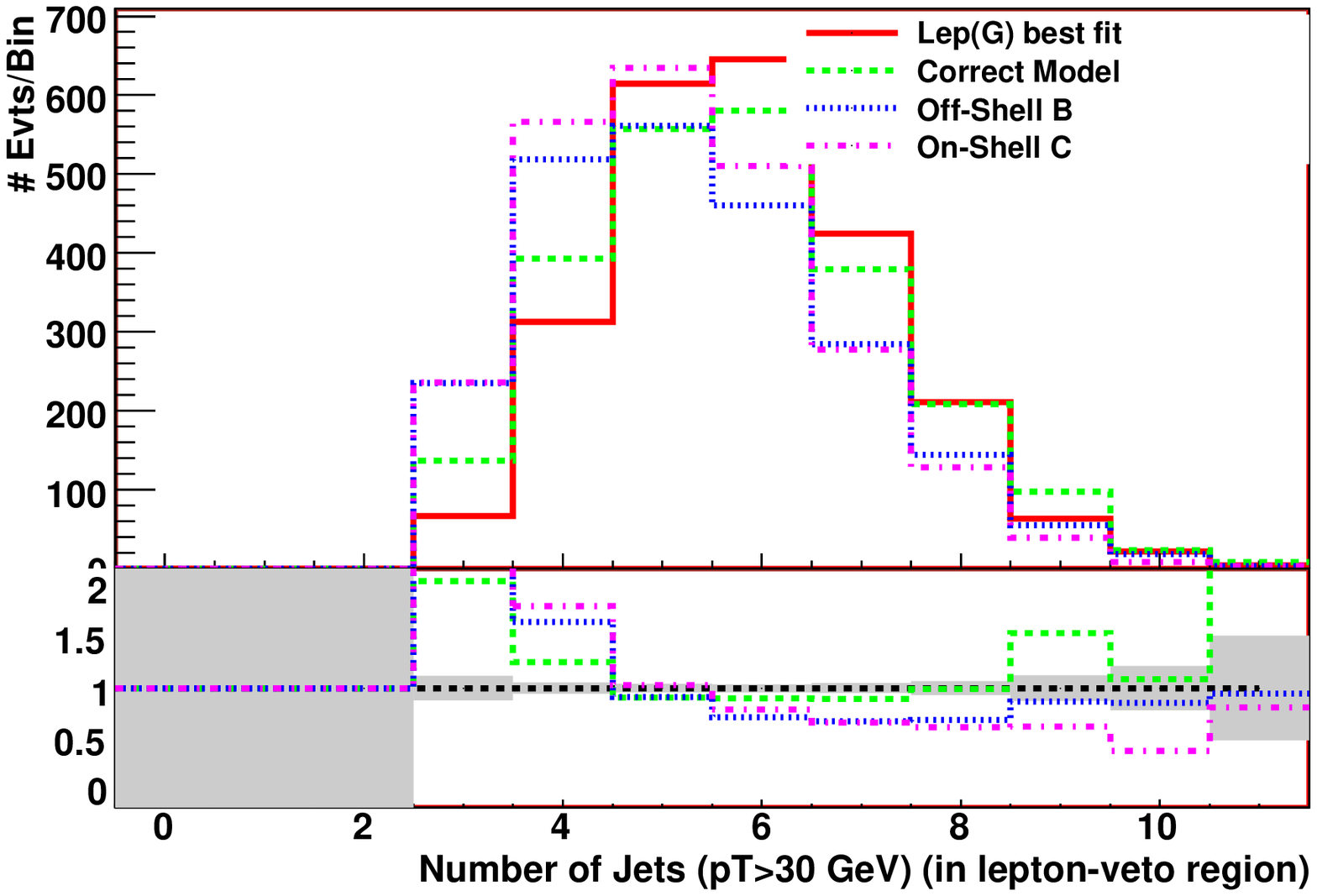}

\includegraphics[width=3in]{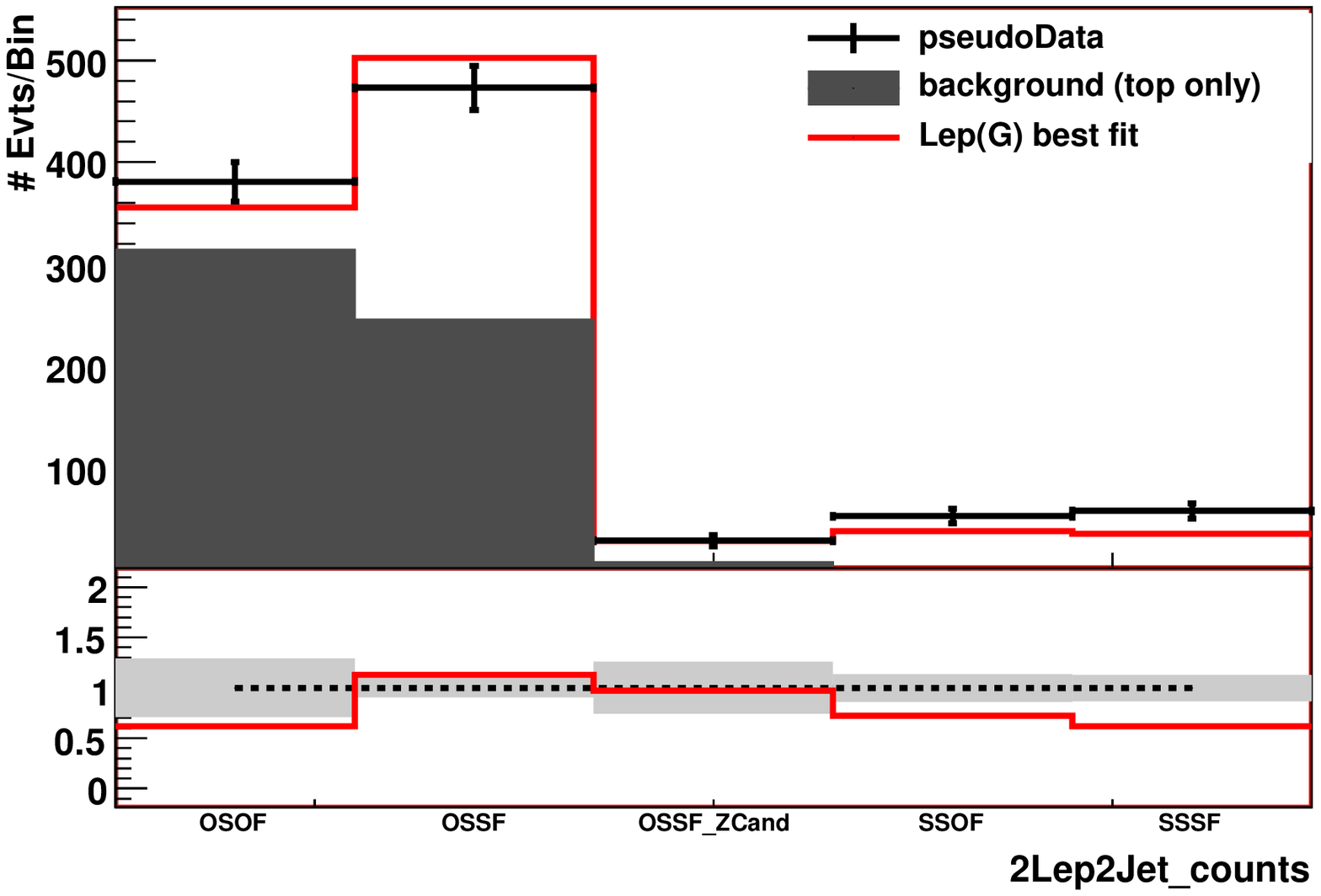}
\includegraphics[width=3in]{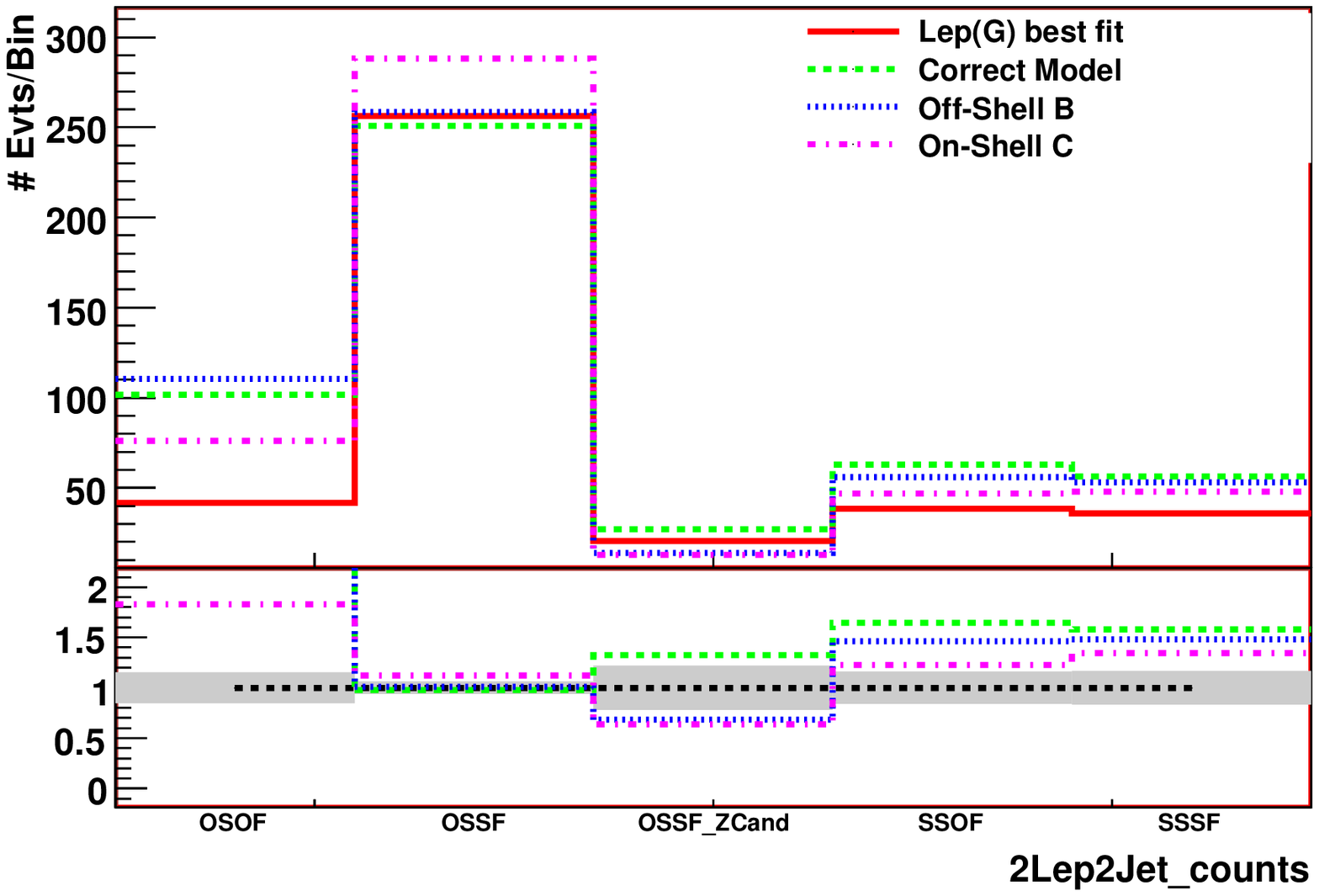}

\includegraphics[width=3in]{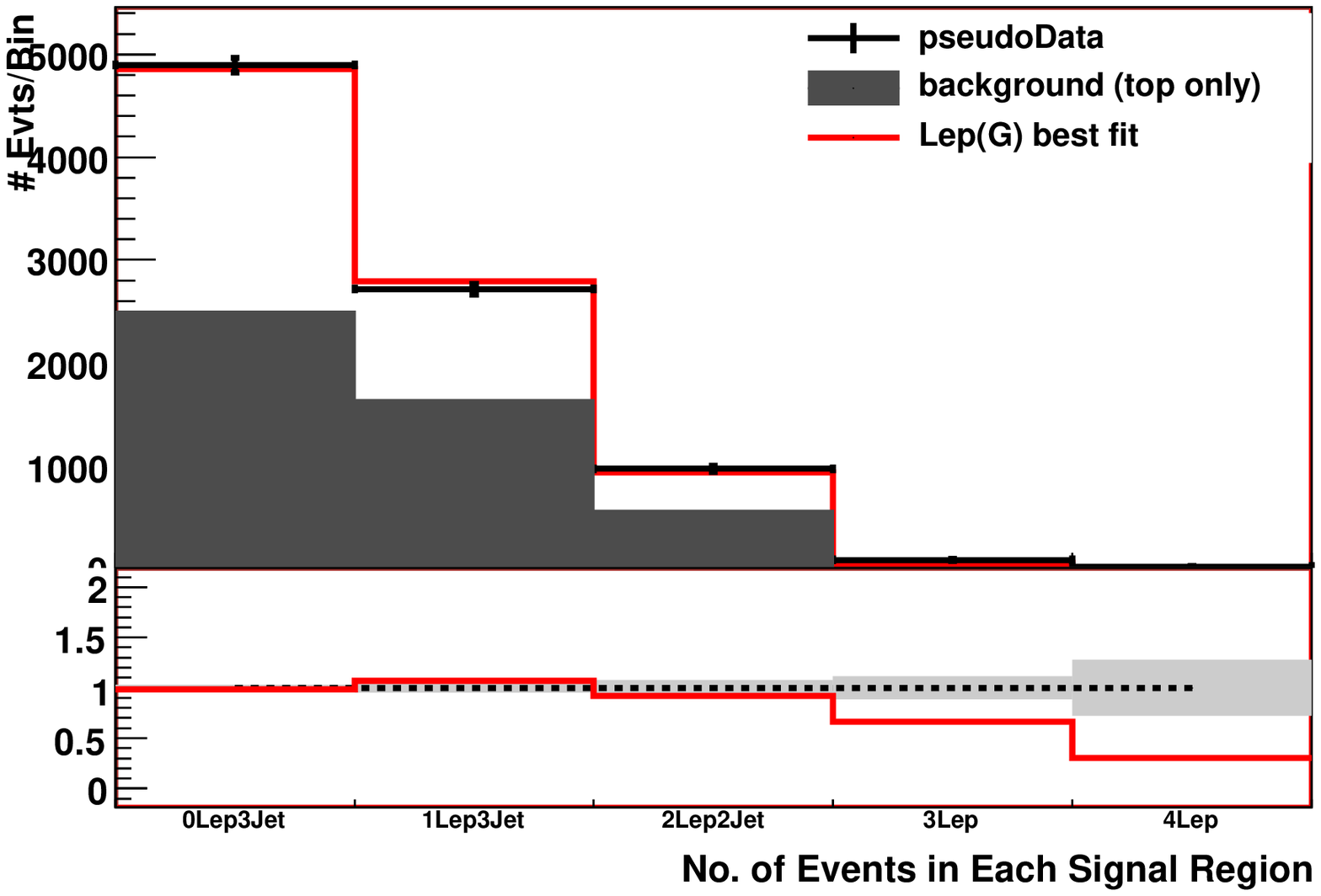}
\includegraphics[width=3in]{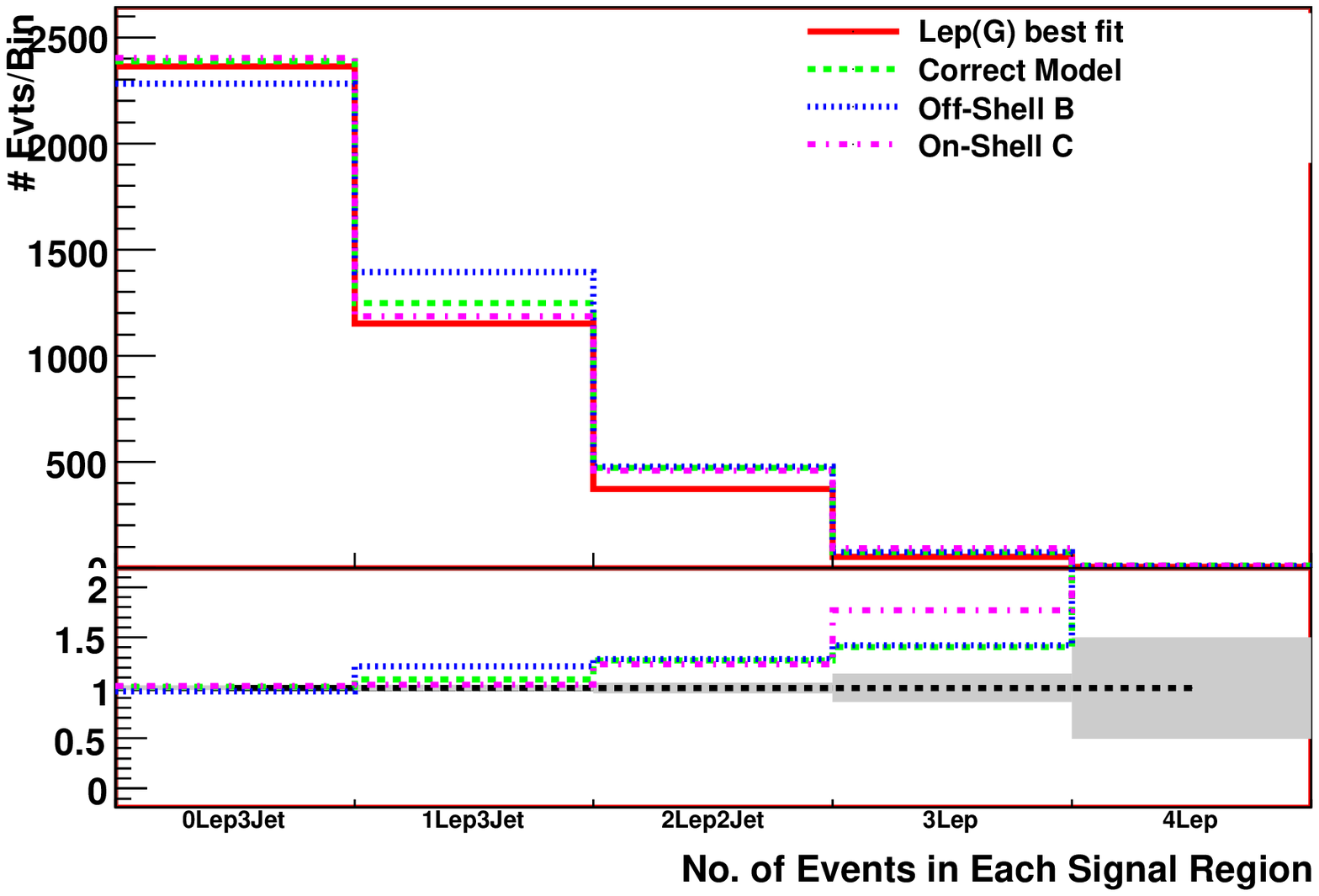}

\includegraphics[width=3in]{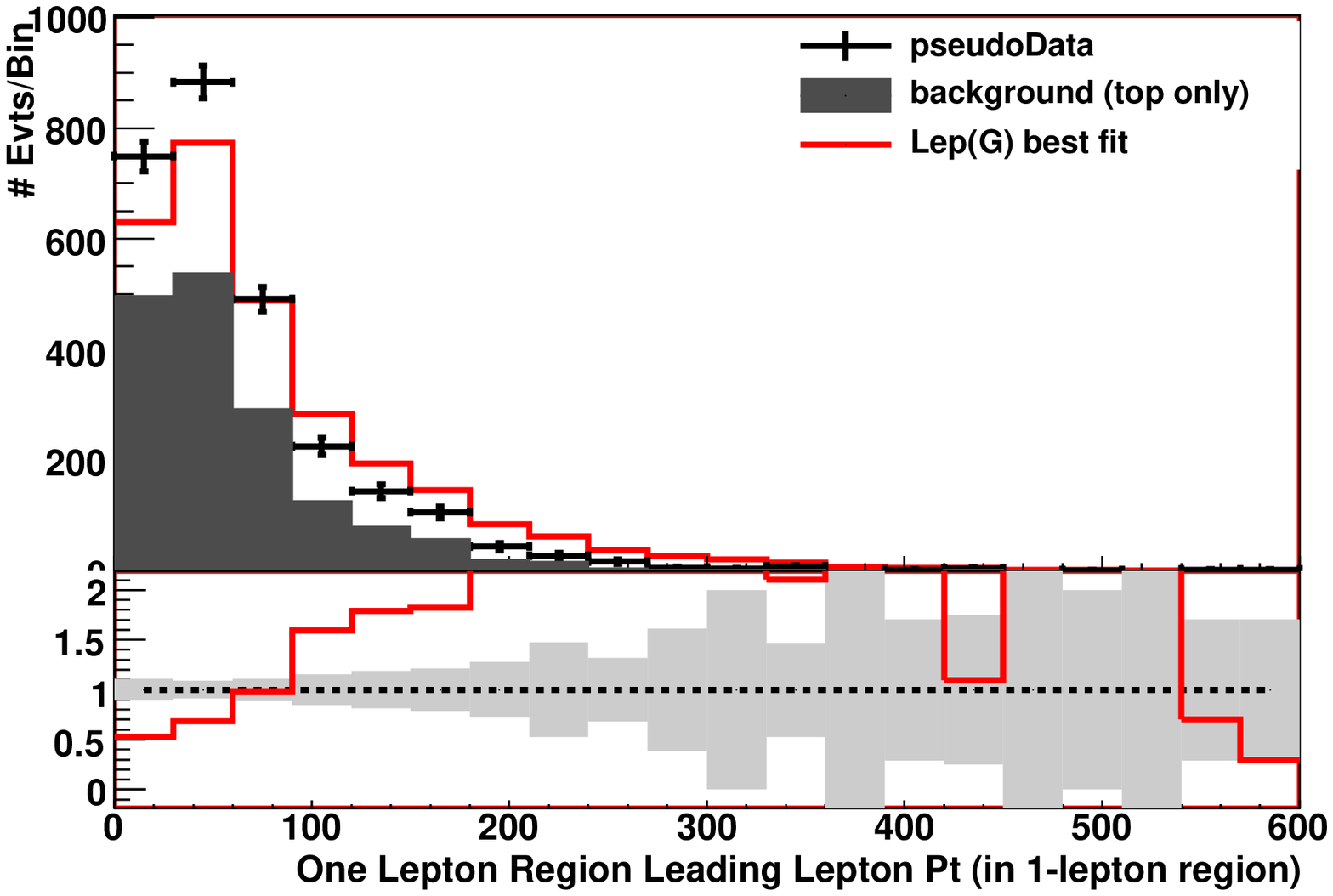}
\includegraphics[width=3in]{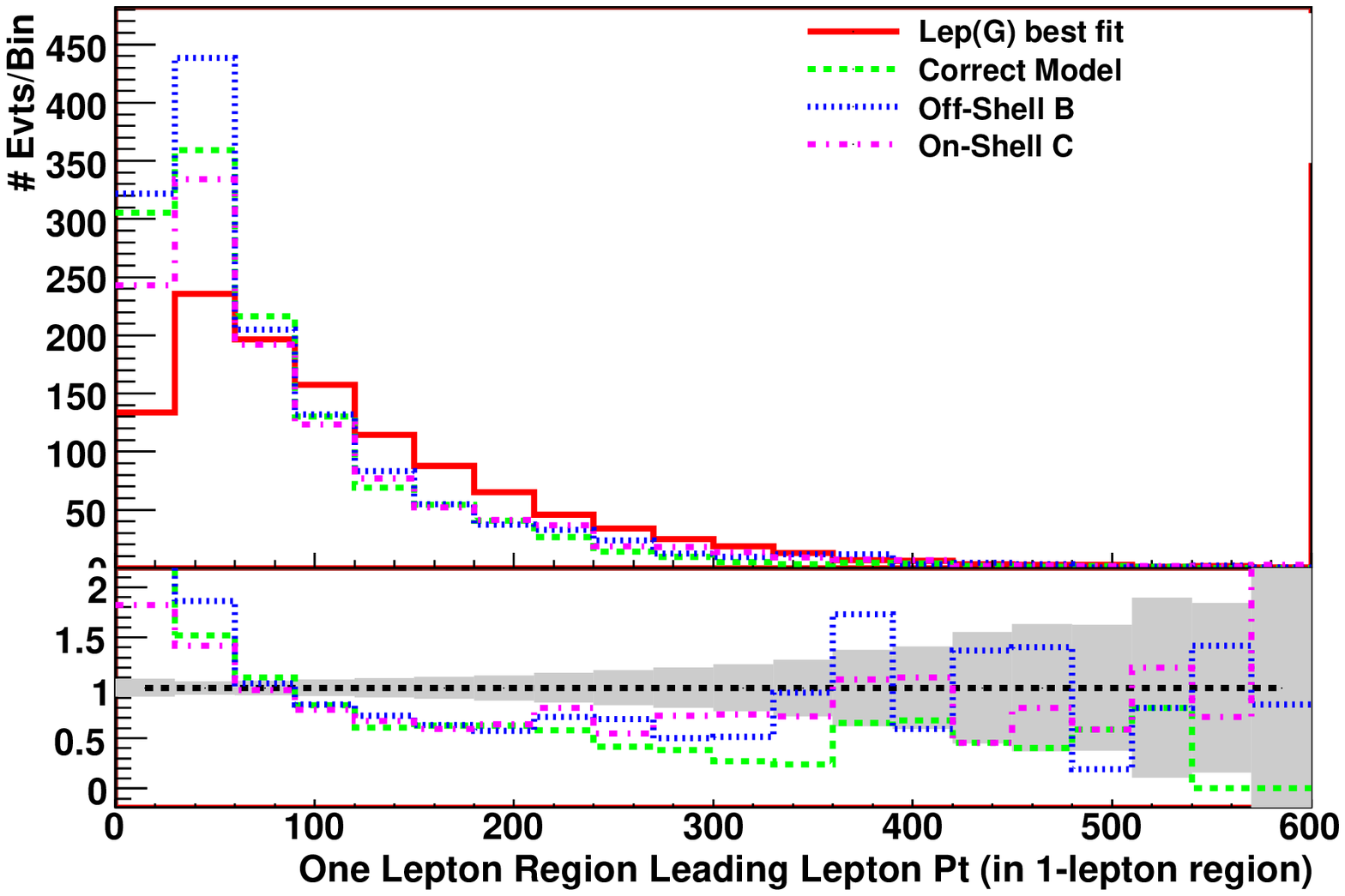}

\caption{
Left: Comparisons of Data, corresponding to $500$pb$^-1$ with $t\bar{t}$ backgrounds superimposed, to Simplified Model Lep(G) with
  parameters $M_G=700$ GeV, $M_I=440$ GeV, $M_{LSP}=100$ GeV, $\sigma=11.5$pb, $B_{ll}=6.3\%$, $B_W=87.2\%$, $B_Z=6.5\%$. Right.
  Comparison of the same simplified model to SUSY models off-shell B, and on-shell C, as well as the correct model. Error bars have been suppressed on the model comparisons, but they can be taken as statistical. From top to bottom, the distributions shown are: number of jets ($p_T > 30$ GeV,0l region), di-lepton counts (2l region), overall lepton counts, lepton $p_T$ (1l region).}
  \label{fig:ex2_and_models_LCM}	
\eef

\bef
\vspace{-0.7in}
\includegraphics[width=3in]{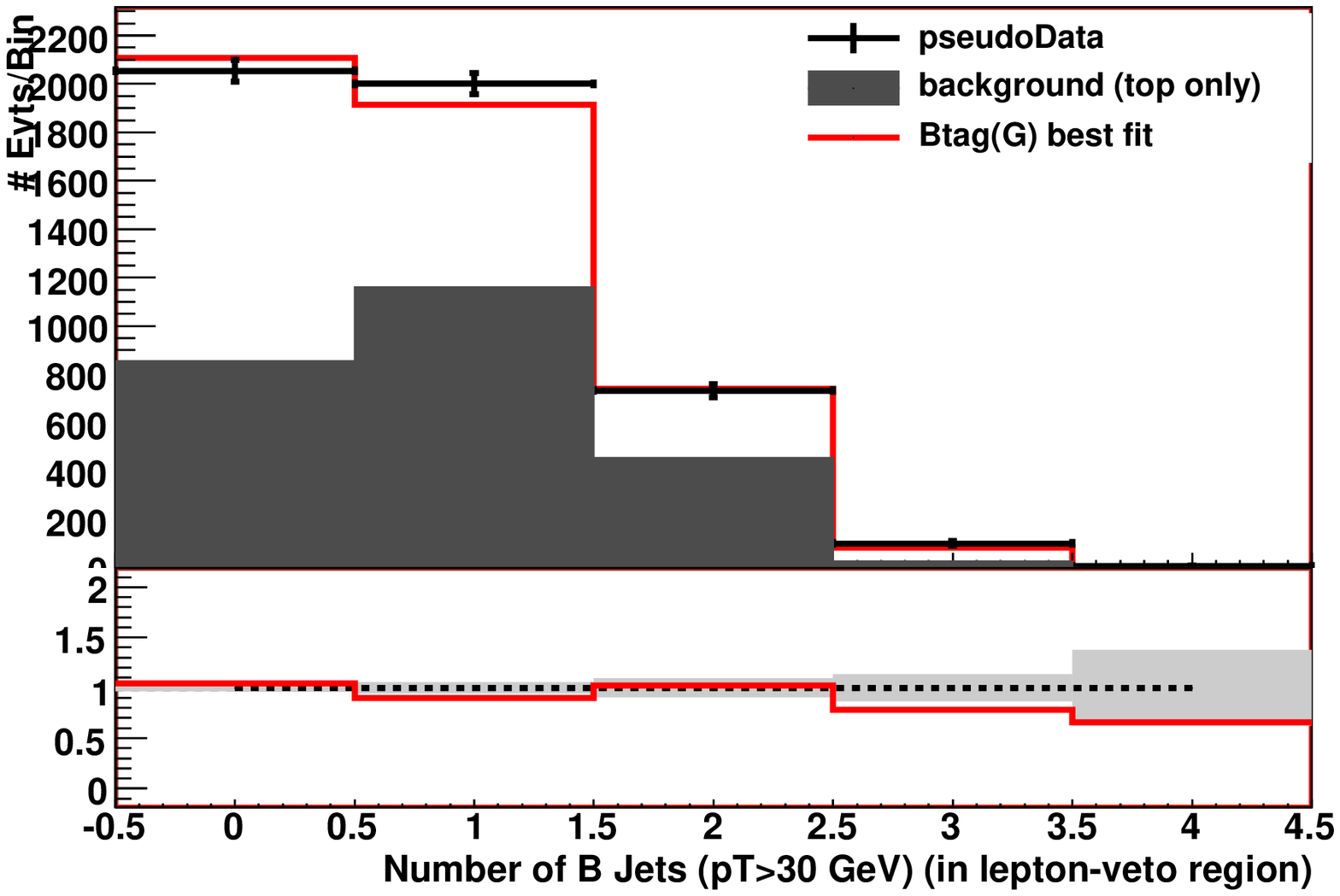}
\includegraphics[width=3in]{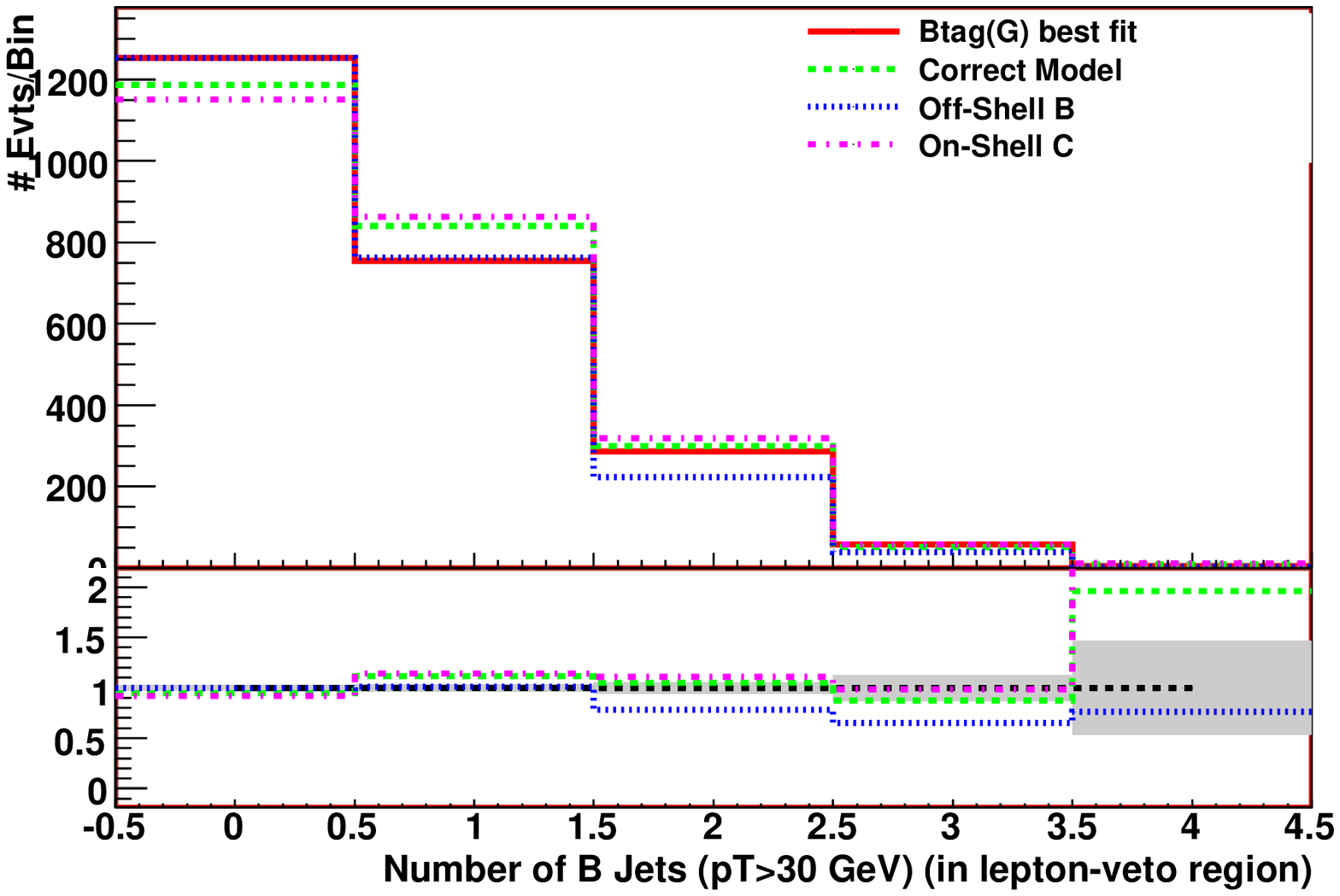}

\includegraphics[width=3in]{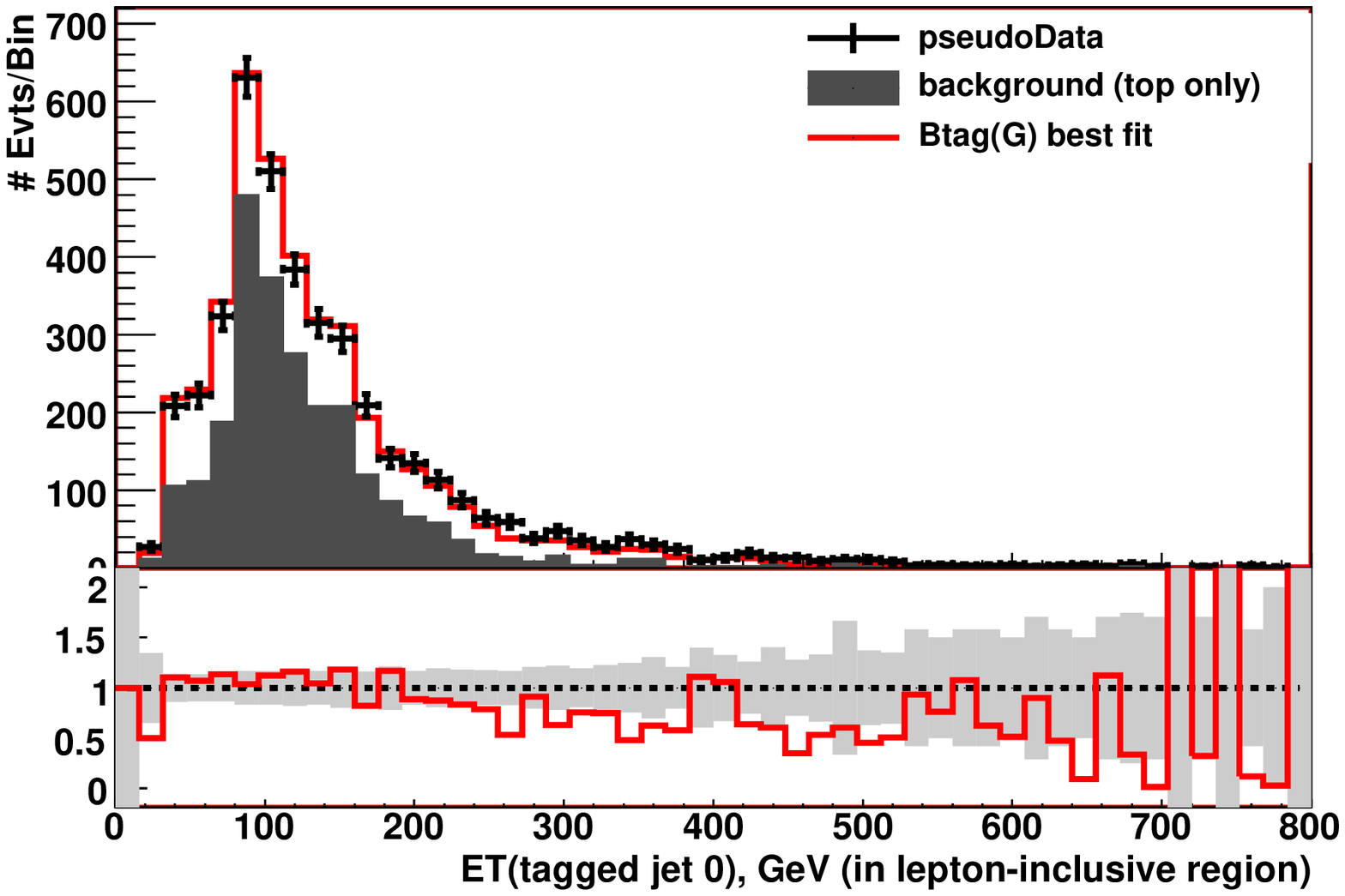}                 
\includegraphics[width=3in]{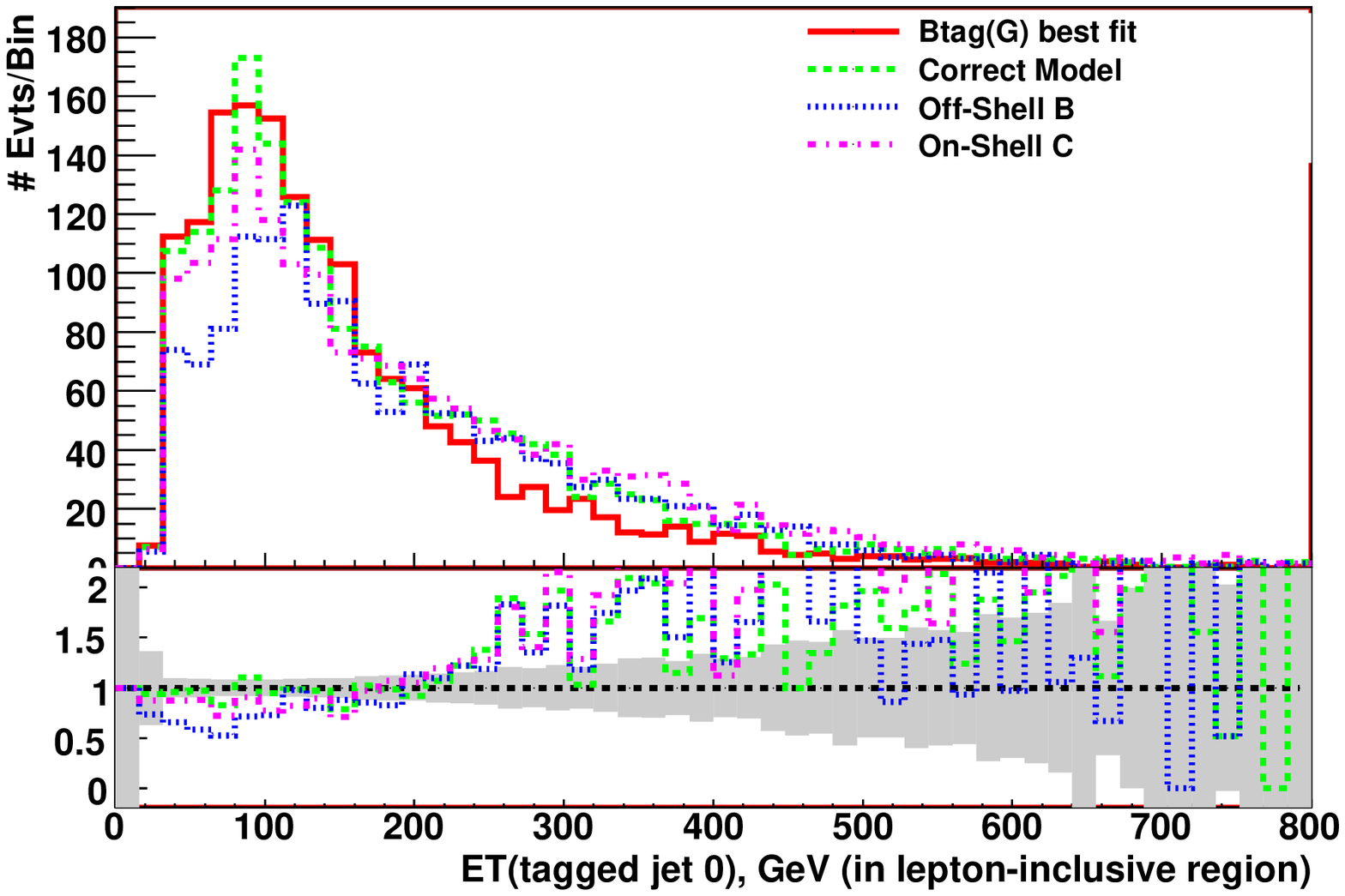}                 

\includegraphics[width=3in]{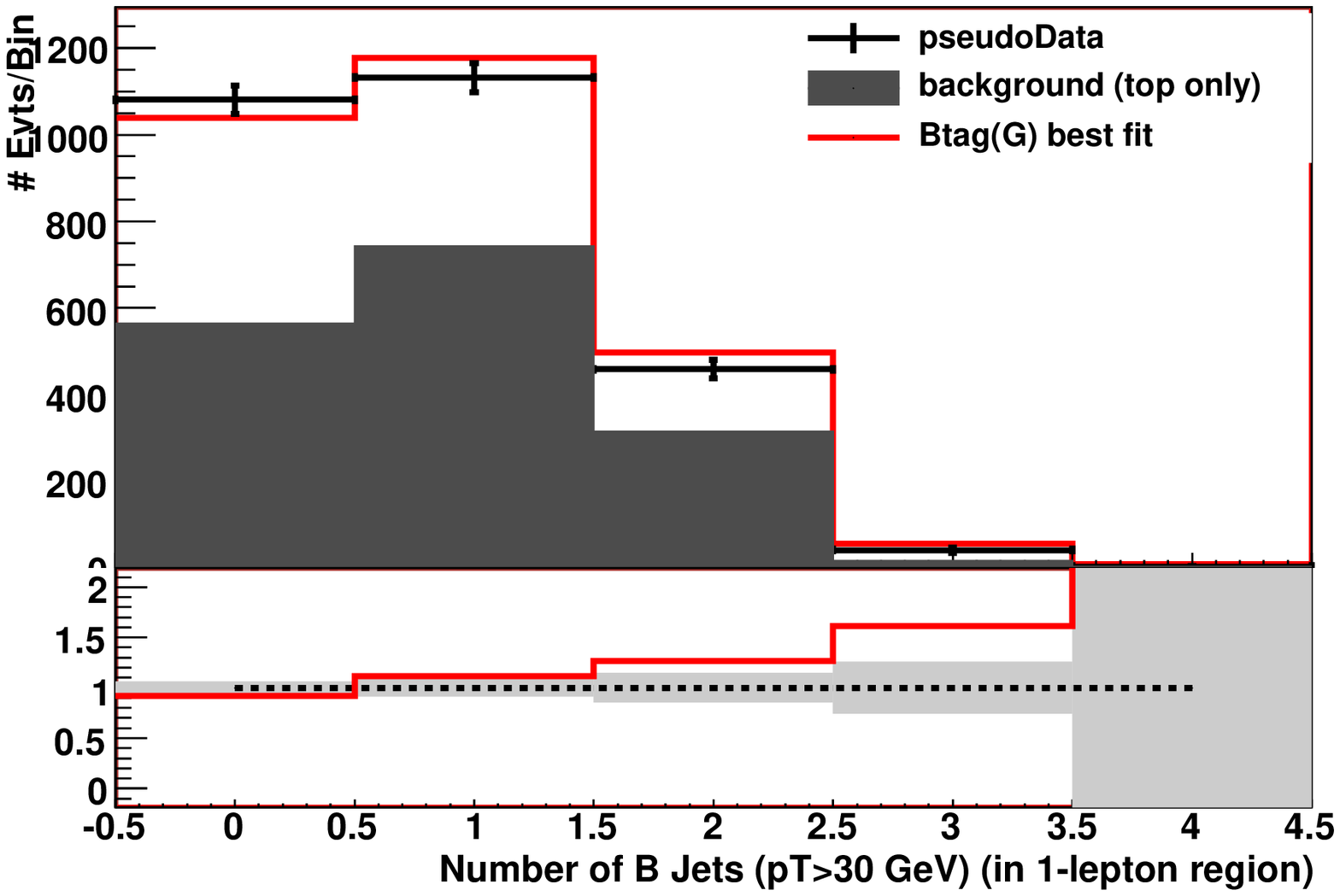}
\includegraphics[width=3in]{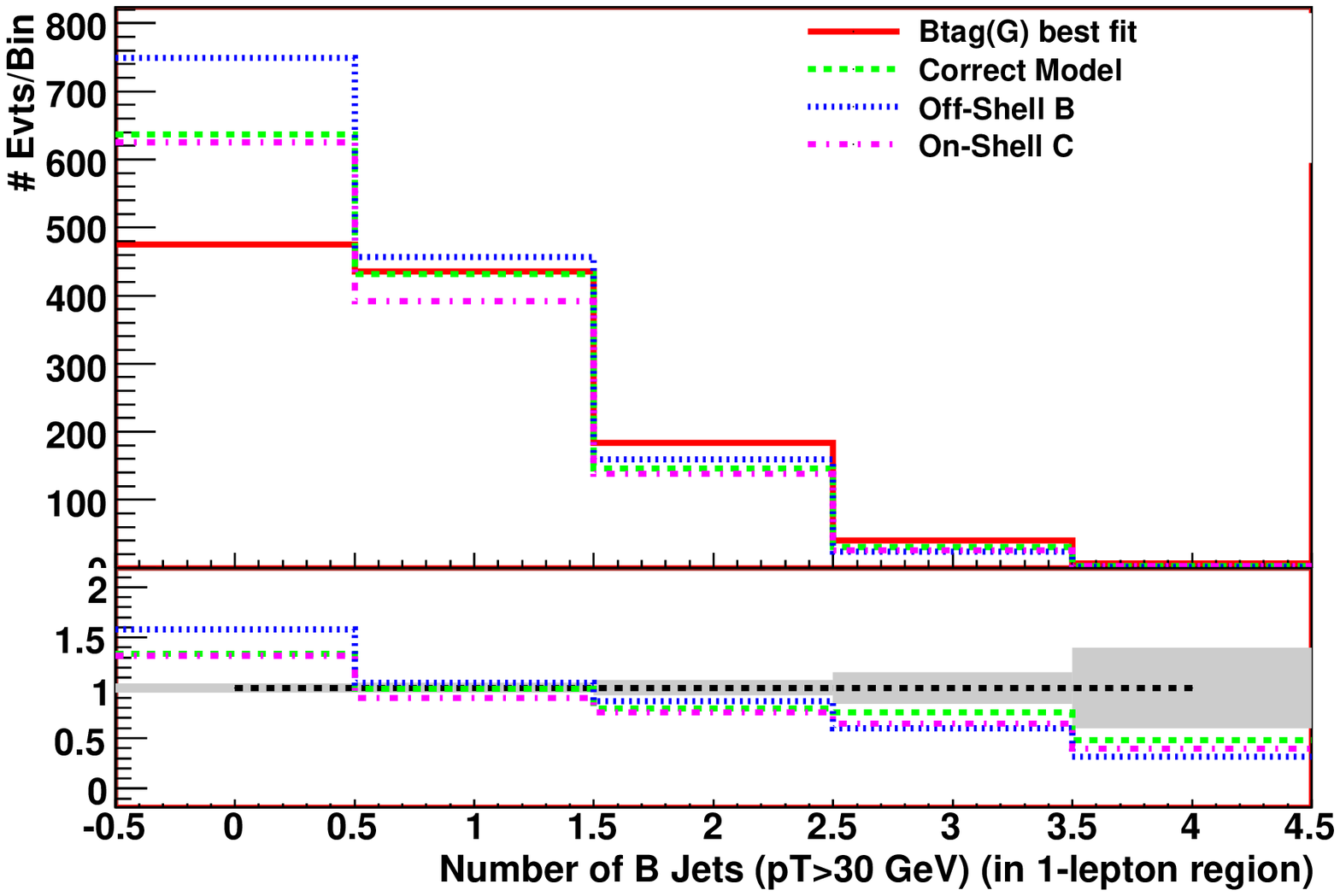}

\includegraphics[width=3in]{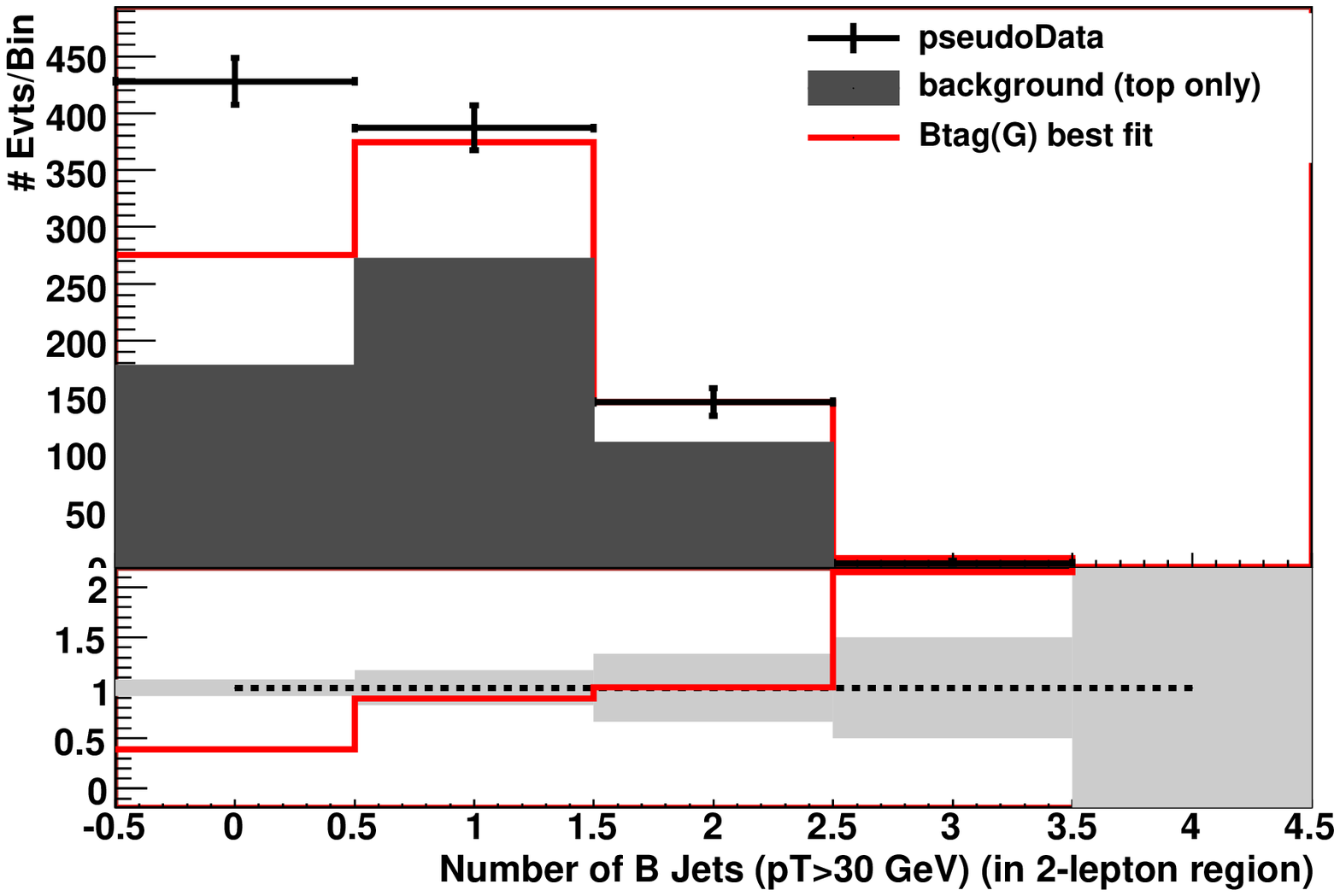}
\includegraphics[width=3in]{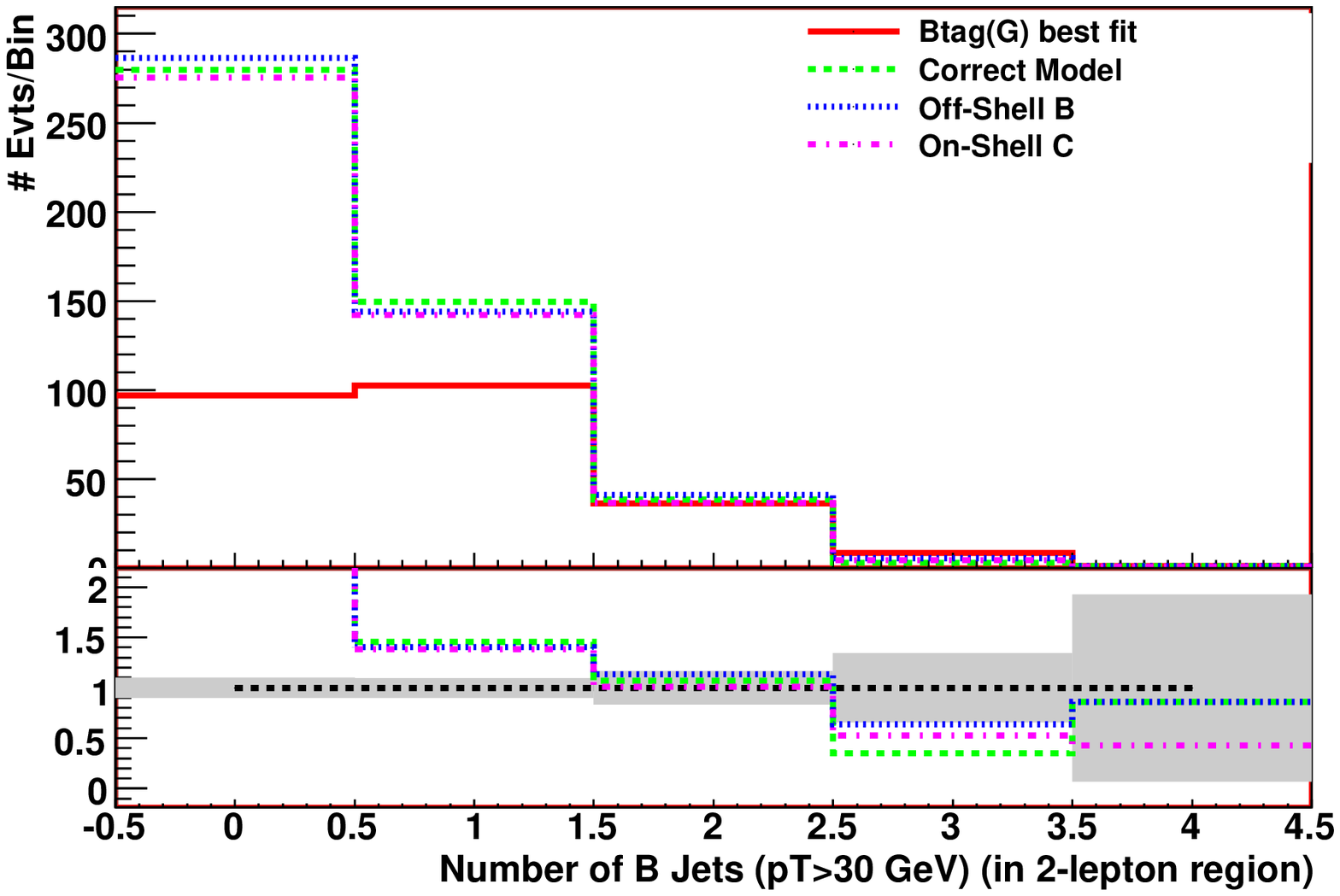}

\caption{Left: Comparisons of Data, corresponding to $500$pb$^-1$ with $t\bar{t}$ backgrounds superimposed, to Simplified Model Btag(G) (exclusive lepton fit) with parameters $M_G=700$ GeV, $M_{LSP}=100$ GeV, $\sigma=11.8$pb, $B_{uu}=35.4\%$, $B_{bb}=2.8\%$, $B_{tt}=61.8\%$. Right.
  Comparison of the same simplified model to SUSY models off-shell B, and on-shell C, as well as the correct model. Error bars have been suppressed on the model comparisons, but they can be taken as statistical. From top to bottom, the distributions shown are: number of $b$-jets and $p_T$ of hardest $b$-jet   (lepton-inclusive region), number of $b$-jets in 1-lepton and 2-lepton regions.} 
\label{fig:ex2_and_models_HFM}
\eef

\bef
\includegraphics[height=3.5in]{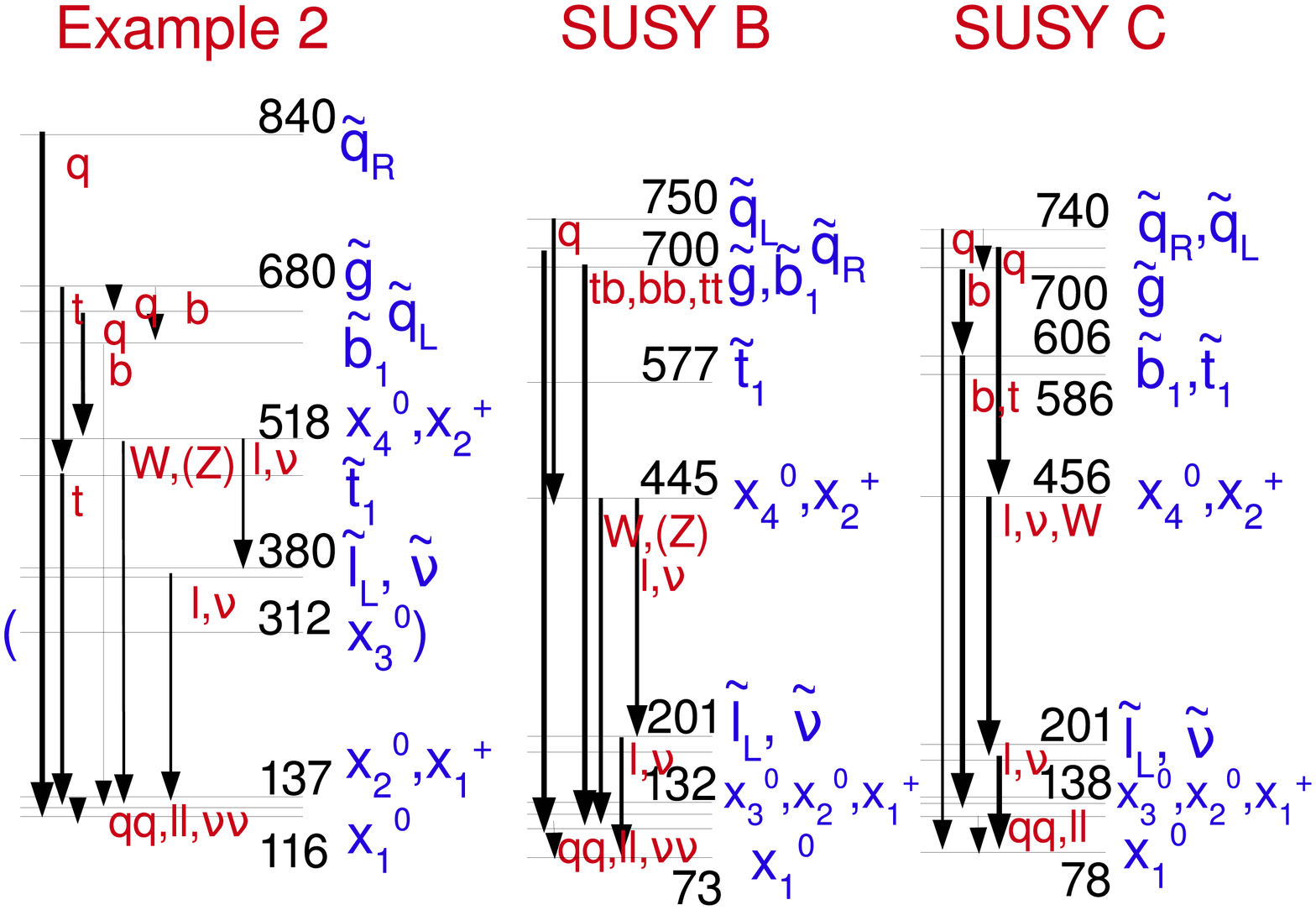}
\caption{Left: Spectrum cartoon for the model used in Example 2 (parameters
  in Appendix \ref{app:susyPythiaCardExample2}). 
Right: Spectra for SUSY models offB and onC used in the comparison in sec.~\ref{sssec:susymodels}
(parameters in Appendix \ref{app:susyPythiaConjecturesExample2})}
\label{fig:ex2_spectrum} 
\eef

Figures \ref{fig:ex2_and_models_LCM} and \ref{fig:ex2_and_models_HFM}
show comparisons of the three models to the gluon-partner-initiated
leptonic decay (Lep(G)) and b-tag (Btag(G)) models respectively.
In each case, we have included both the ``experimental'' comparison of
the simplified model to data, and the ``theoretical'' comparison of
the simplified model to different points in parameter space.  

Figures \ref{fig:ex2_and_models_LCM} and \ref{fig:ex2_and_models_HFM}
are meant to reinforce four general points. First, the simplified
models allow a description of the data independent of the background
and detector effects. It should be noted however that the topic of
quantifying systematic errors arising from detector-modeling errors
merits further study.

Second, provided the basic jet and lepton kinematics are well-modeled,
we expect that the simplified model fits can be simulated in a crude
detector simulator (with approximately similar features as the
experimental environment, such as cone size, and overall geometry),
and then used as a target for vetting models that any particular
theorist has in mind. Where the simplified model fully describes the
data --- the $H_T$ distribution and lepton and jet counts in the case
of Lep(G), lepton-inclusive $b$-tag counts and $b$ kinematics in
Btag(G) --- it can be used as a target for full models. We are
\emph{not} saying that strict exclusions can be derived from
comparisons to the fits, but certainly the approximate consistent
regions of parameter space can be identified, and others broadly ruled
out.

Third, sources of tension in the fits, such as the soft lepton
deficits in this example, can be used quite readily in the
comparisons. As can be seen in figures \ref{fig:ex2_and_models_LCM}
and \ref{fig:ex2_and_models_HFM}, qualitative differences from the
simplified models can be seen to agree with those in the models
considered.  For instance, all of the models have softer leptons than
in the Lep(G) model (because they have several light states with small
splittings), as does the signal, and similar enhancements of events
with leptons but no $b$-jets over the Btag(G) best-fit (because they
have sources of leptons associated with light-flavor quarks).

Finally, it is very easy for a broad range of very different models to
match the simplified models, and hence the data. In fact, the three
models shown in figures \ref{fig:ex2_and_models_LCM} and
\ref{fig:ex2_and_models_HFM} have qualitatively different SUSY
production and decay modes. Nonetheless, all look similar to the data
at low statistics.  

These points underscore why characterizing the data independent of
background and detector effects is valuable. The experimentally
difficult task of fittings and calibrating backgrounds is done in a
framework where parameters are well constrained by gross properties of
the data.  The characteristics of the data, in this language, allows
theorists to very efficiently study a broad a range of candidate
theories (without having to burden experimental
collaborations until there is a well-motivated candidate
model). Naturally, the next step involves refining searches to
discriminate among the well-motivated candidates that emerge from this
process.

\section{Summary and Future Directions}\label{sec:conclusionFuture}
In this paper, we have proposed a framework for characterizing early
data excesses, in which detector effects and backgrounds can be
sharply unfolded, facilitating the comparison of theoretical models to
experimental data.  We consider a scenario in which the LHC
experiments have found solid excesses in a number of different
channels involving jets, leptons and missing transverse energy, with
rates consistent with the production of heavy, strongly interacting
particles.  We have defined four simplified models as a framework for
characterizing such excesses, assuming a ``SUSY-like'' structure.

The simplified models have a deliberately simple structure, so that
they can describe the most important features of the data with a
minimal number of parameters.  Specifically, each model includes only
one pair-produced particle, with decay modes than can produce either
leptons or $b$-tagged jets in the final state.  These models form a
``basis'' of representative phenomenology for SUSY-like physics,
providing a framework for qualitative study of jet structure, and
quantitative description of leptonic and heavy-flavor decay modes.  It
is striking how well these simplified models reproduce features of
models with very complex heavy-particle spectra.  Deviations of a
signal from the structure predicted by the simplified models may
motivate extensions of one of the models in a similar spirit (the
appropriate refinements will depend on what is observed in the data).
Nevertheless, we expect the four simplified models to include good
fits to SUSY-like physics at the LHC in early data.  Taken together,
such fits provide a quantitative description of the most important
features of the new-physics signal that is useful to theorists and
experimentalists alike.

Fits of the simplified models to data can be used as targets for
testing arbitrary models with SUSY-like phenomenology. A reasonable
hypothesis for the new physics is one that is consistent with the
simplified model, except that where the data differs from the
simplified model, the hypothesis differs in the same direction.
However, the latter comparison can be performed with a simplified
detector simulator.  In this sense, the simplified models are a
representation of the data that can be studied outside the
experiments, in which Standard Model backgrounds and details of
detector simulation have been properly incorporated.

We have motivated the use of simplified models in hypothesis-testing
from a theoretical standpoint, but it is complicated if mis-modeling
of object kinematics and multiplicities (particularly jets) in the
simplified models biases trigger efficiencies, search region
acceptances, or identification efficiencies for other leptons and
$b$-jets in an event.  A detailed study is required to assess whether
these effects are typically small enough that a simplified model
characterization of data can be meaningfully compared to full models
using a simulator, such as PGS, that does not quantitatively model the
detector.  Reasonable agreement and a means of estimating the
systematic effects introduced by this procedure are crucial in order
to call a simplified model fit to data a ``detector-independent
characterization'' of that data.

Comparisons of new-physics signals to simplified models are
complementary to the traditional methods of fitting to more or less
constrained Lagrangian models (such as mSUGRA or the 20-parameter
MSSM).  These Lagrangian fits are useful, first, as demonstration that
the new physics is consistent with a given model. Moreover, a good fit
can be used --- just as we have used the simplified models in the
examples in this paper --- as a detector-independent description of
the data.  A very thorough comparison may even find all consistent
points within the studied model, such as all three MSSM points
identified in Sec.~\ref{sssec:susymodels}.  But, no matter how large a
parameter space is searched, physicists will always wish to consider
generalizations and other models as possible explanations of new
physics, and test their assumptions.  For this purpose, it is
preferable to isolate the known and distinguishing features with as
few parameters as possible, and quantify how well masses and rates are
constrained by the data --- not how well they are constrained subject
to the assumptions of the MSSM.  Simplified models are a natural
framework for describing these constraints in early data.

\section*{Acknowledgments}
We are indebted to Nima Arkani-Hamed, Joseph Incandela, Sue Ann Koay,
Michael Peskin, Albert de Roeck, and Roberto Rossin for discussion of
early ideas that shaped the direction of this work.  We are
particularly grateful to Michael Peskin for extensive feedback on this
paper, and for his help in refining the set of simplified models, and
to Nima Arkani-Hamed for several useful discussions.

\appendix
\section{Implementations of Simplified Models Using Pythia or MadGraph} \label{app:Implementations}

For this paper, the simplified models were implemented in Pythia
6.404~\cite{Sjostrand:2006za}, using
Marmoset~\cite{ArkaniHamed:2007fw} to generate event topologies and
perform branching ratio and cross section fits. This implementation
uses the on-shell effective theory (OSET) approximation: flat
production matrix elements (i.e., production according to phase space)
and 2- and 3-body decay according to phase space. This gives
descriptions of the kinematics of production and decay of massive
particles at the LHC which are accurate to far better precision than
necessary for comparison with the kind of inclusive properties of
early data used here. The OSET definition files for the simplified
models are included in the standard Marmoset distribution. Note that
the intermediate color singlet state is always modeled as neutral,
while the LSP is modeled using a neutral and a charged particle, with
a mass splitting of 1 GeV. This is done in order to get charge
symmetric single-lepton decays.

An OSET implementation is clearly not enough for studies of e.g.~spin
correlations in data, which is in general only feasible with very
significant data samples. It is however still possible to use the
philosophy of the simplified models to do this type of studies. The
simplest way to do this is to fix the spin of the particles in the
simplified models to either be identical to the spin of their Standard
Model partner, or to the opposite spin (as in the MSSM). The
simplified models can then be implemented in
MadGraph/MadEvent~\cite{Alwall:2007st} or some other matrix element
generator.

\textbf{Production:} For pure QCD
production, including the interference between $s$- and $t$-channel
production, the most model-independent implementation includes QCD
couplings of the produced particles to gluons, with a multiplicative
factor that can be used to fit the cross section.

\textbf{Decay:} Cascade decays can be implemented either using one
intermediate particle with several decay modes, or using several
intermediate particles with identical mass, each coupling only to one
decay mode. The latter implementation makes the fixing of branching
ratios easier; the branching ratios are directly given by the relative
couplings of the QCD state to the intermediate states, except for the
direct decay into the LSP. The decay matrix elements for 2-body decays
are fixed by the spins of the participating particles and the coupling
constant. 3-body decays are most easily implemented using an off-shell
heavy particle. The mass of this particle (quark partners in the Lep(G)
and Btag(G), and lepton partners in the off-shell $\ell\ell$ or
$\ell\nu$ decays in Lep(Q/G)) is arbitrary, and can be set high enough not to
be seen in the spectrum.

It is also straightforward to describe the simplified models using
effective or renormalizable Lagrangians.

\section{Details of the Examples}\label{app:ExampleDetails}

In this appendix, we provide additional supporting information for the analysis of the examples in sections 4 and 6. We summarize specifications of the signal regions, basic observables used in fitting, and model parameters in the form of Pythia input information. 

\subsection{Definitions of Signal Regions and Analysis Objects}\label{app:signalRegions}

All Monte Carlo was generated at parton-level with Pythia 6.404 \cite{Sjostrand:2006za}, and passed to PGS \cite{PGS} for detector simulation and object reconstruction. We used the PGS cone jet algorithm with a cone size of $0.7$. All other object identification parameters were taken as default. We used a private C++ based analysis code to perform the studies discussed in sections 4 and 6. In table \ref{table:signalregions}, we summarize the primary cuts that define the signal regions in our examples. As mentioned in the text, only events passing these cuts were used in fitting the simplified models to the example data. 

\begin{table}
\begin{center}
\begin{tabular}{|l|l|}
\hline
\textbf{Signal Region} & \textbf{Requirement} \\
\hline
Lepton Inclusive & $E_T^{\rm miss}>100$ GeV \\
 & $N_{jet} \geq 3$, $p_T(j_{1,2,3}) > 75$ GeV \\
 & $H_T \equiv \sum_{i=1}^{4} p_T(j_i) + \sum_{lep}p_T(lep) + E_T^{\rm miss} > 350$ GeV \\
 & $\frac{E_T}{H_T} > 0.2$ \\
\hline
Lepton Veto & Number $e/\mu = 0$, $E_T^{\rm miss}>100$ GeV \\
 & $N_{jet} \geq 3$, $p_T(j_{1,2,3}) > 75$ GeV \\
 & $H_T > 350$ GeV, $\frac{E_T}{H_T} > 0.2$ \\
\hline
Single Lepton & Number $e/\mu = 1$, $E_T^{\rm miss}>100$ GeV \\
 & $N_{jet} \geq 3$, $p_T(j_{1,2,3}) > 75$ GeV \\
 & $H_T > 350$ GeV \\
\hline
Two Lepton & Number $e/\mu = 2$, $E_T^{\rm miss}>80$ GeV \\
 & $N_{jet} \geq 2$, $p_T(j_{1,2}) > 75$ GeV \\
 & $H_T > 350$ GeV \\
\hline
Three Lepton & Number $e/\mu = 3$, $E_T^{\rm miss}>80$ GeV \\
 & $H_T > 350$ GeV \\
\hline
Four Lepton & Number $e/\mu \geq 4$, $E_T^{\rm miss}>30$ GeV \\
\hline
\textbf{All Regions} & $\delta\phi_i \equiv \delta\phi(j_i, E_T^{\rm miss}) < 0.3$ rad ($i=1, 3$) \\
 & $\delta\phi_2 < 20^\circ$, $R_1 \equiv \sqrt{\delta\phi_2^2 + (\pi - \delta\phi_1)^2} < 0.5$ \\
 & $R_2 \equiv \sqrt{\delta\phi_1^2 + (\pi - \delta\phi_2)^2} < 0.5$ \\
\hline
\end{tabular}
\caption{Primary cuts that define the signal regions used in the analysis of the examples of sections 4 and 6. There are four exclusive lepton regions, and one lepton inclusive region. Cuts vary significantly among the different regions. Detector simulation and object reconstruction was done with PGS \cite{PGS}. Private analysis code was used for building signatures and fitting the simplified models to the example data.}\label{table:signalregions}
\end{center}
\end{table}

\subsection{Count Observables and Fitting}\label{app:fitData}

\begin{table}
\begin{center}
\begin{tabular}{|l|c|c|c|c|r}
\hline
Region / Fit Type & Lepton Fits & Lepton Inclusive B Fits & Lepton Exclusive B Fits \\
\hline
Lepton Inclusive & -- & Number of B-tags & -- \\
			& & \emph{$H_T$}, \emph{B-jet $p_T(b_{1,2,3})$} & \\
			& & \emph{Number of Jets}, \emph{$p_T(jet_{1,2,3})$} & \\
\hline
Single Lepton & Number of $e / \mu$, \emph{$H_T$}, \emph{$p_T(lep)$} & -- & Number of B-tags \\
			& \emph{Number of Jets}, \emph{$p_T(jet_{1,2,3})$} & & \emph{$H_T$}, \emph{B-jet $p_T(b_{1,2,3})$} \\
			& & & \emph{$p_T(lep)$} \\
\hline
Two Lepton & Number of $e / \mu$ & -- & Number of B-tags \\
			& OSSF, OSOF, Z Candidates & & \emph{$H_T$}, \emph{B-jet $p_T(b_{1,2,3})$} \\
			& SSSF, SSOF, \emph{$H_T$}, \emph{$p_T(lep_{1,2})$} & & \emph{$p_T(lep)$} \\
			& \emph{Number of Jets}, \emph{$p_T(jet_{1,2,3})$} & & \\
			& \emph{OSSF and OSOF invariant mass} & & \\
\hline
Three Lepton & Number of $e / \mu$ & -- & Number of B-tags \\
			& \emph{$H_T$}, \emph{$p_T(lep_{1,2,3})$} & & \emph{$H_T$}, \emph{B-jet $p_T(b_{1,2,3})$} \\
			& \emph{Number of Jets}, \emph{$p_T(jet_{1,2,3})$} & & \emph{$p_T(lep)$} \\
\hline
Four Lepton & Number of $e / \mu$ & -- & Number of B-tags \\
			& \emph{$H_T$}, \emph{$p_T(lep_{1,2,3,4})$} & & \emph{$H_T$}, \emph{B-jet $p_T(b_{1,2,3})$} \\
			& \emph{Number of Jets}, \emph{$p_T(jet_{1,2,3})$} & & \emph{$p_T(lep)$} \\
\hline
\end{tabular}
\caption{Signatures used for the fits and diagnostics discussed in sections 4 and 6. Signatures listed in plain text are counts used for quantitative fitting. Those listed in italics are the primary signatures used for mass estimates and to diagnose the quality of fit.}\label{table:signatures}
\end{center}
\end{table}

In table \ref{table:signatures}, we summarize the counts and kinematic signatures used in the analyses of sections 4 and 6. Count variables include; Numbers of electrons and muons, number of opposite sign same flavor lepton (OSSF) events, number of opposite sign opposite flavor lepton (OSOF) events , number of same sign same flavor lepton (SSSF) events, number of same sign opposite flavor lepton (SSOF) events, number of B-tagged jets with $p_T\geq 30$ GeV, numbers of jets with $p_T\geq 30,75,150$ GeV, and number of OSSF lepton events that reconstruct a Z to within $4$ GeV. Only a subset of these were actually used for quantitative fitting (see table \ref{table:signatures}). 

To perform parameter fits, we used fitting tools available with the Marmoset package \cite{ArkaniHamed:2007fw} as well as private analysis code. A $\chi^2$ metric was defined using the count variables listed in table \ref{table:signatures}. The Marmoset package simplex fitter was then used for the minimization. As we've emphasized in this paper, we have not optimized our fitting methods, and only use these tools to illustrate how to derive information from fitting the simplified models to data. Errors were artificially enlarged, as the primary source of error in our analysis is systematic, and we have not properly quantified them in this paper. 

\subsection{Pythia Parameters}

All examples were generated using PGS \cite{PGS} and Pythia 6.411\cite{Sjostrand:2006za}.  
The Pythia \emph{non-default} parameters are given below.

\subsubsection{Blind Example 1}\label{app:susyPythiaCardExample1}

{\tiny
\begin{minipage}{3in}
\begin{verbatim}
IMSS(1)=1     ! SUSY spectrum, specified by hand
MSEL=39       ! All SUSY production processes

RMSS(1)=101   !bino soft mass M1
RMSS(2)=301   !wino soft mass M2
RMSS(3)=500   !gluino soft mass M3
RMSS(4)=1140  !mu parameter
RMSS(5)=5      !tan(beta)

RMSS(6)=201    ! slepton-L mass
RMSS(7) = 201  ! slepton-R mass
\end{verbatim}
\end{minipage}
\begin{minipage}{3in}
\begin{verbatim}
RMSS(13)=201   ! stau L mass
RMSS(14)=201   ! stau R mass

RMSS(8)=1801    ! left-squark mass
RMSS(9)=1802    ! dR-squark mass
RMSS(22)=1803   ! uR-squark mass

RMSS(10)=1801   ! 3L squark mass
RMSS(11)=1802   ! sbottomR mass 
RMSS(12)=1803   ! stopR mass
\end{verbatim}
\end{minipage}
}

\subsubsection{Blind Example 2}\label{app:susyPythiaCardExample2}

{\tiny
\begin{minipage}{3in}
\begin{verbatim}
IMSS(1)=1
IMSS(9) = 1   ! use separate L/R squark masses
IMSS(3) = 1   ! 1: RMSS(3) is pole mass (0=def : use RGE)

MSEL=40

RMSS(1)=305   ! bino
RMSS(2)=505    !wino
RMSS(3)=680   ! gluino
RMSS(4)=130   ! higgsino
RMSS(5)=40  !tan beta

RMSS(6)=380  ! slepton-L
RMSS(7)=380  ! slepton-R 
\end{verbatim}
\end{minipage}
\begin{minipage}{3in}
\begin{verbatim}
RMSS(13)=380 ! stau L
RMSS(14)=380  ! stau R

RMSS(8)=650  ! left-squark
RMSS(9)=840 ! dR-squark
RMSS(22)=840 ! uR squark

RMSS(10)=620 ! 3L squark 
RMSS(11)=685 ! sbottomR 
RMSS(12)=450 ! stopR  

RMSS(15)=0.2 !A_b
RMSS(16)=0.1 !A_t
RMSS(19)=600 ! MA  
\end{verbatim}
\end{minipage}
}

\subsubsection{SUSY Conjectures for Blind Example 2}\label{app:susyPythiaConjecturesExample2}

The spectrum for the guess ``SUSY B'' (middle spectrum shown in Figure
\ref{fig:ex2_spectrum}):

\vspace{24pt}

{\tiny
\begin{minipage}{3in}
\begin{verbatim}
IMSS(1)=1     ! user-specified MSSM   
IMSS(3)=1     ! use gluino pole mass  
IMSS(9) = 0   ! don't use separate L/R squark masses 
MSEL=40 
  
RMSS(1)=100    !M1 (~bino mass) 
RMSS(2)=440    !M2 (~wino mass) 
RMSS(3)=700    !gluino pole mass
RMSS(4)=110    !mu (~higgsino mass)
RMSS(5)=40     !tan(beta) 

RMSS(6)=201   ! slepton-L
RMSS(7)=501   ! slepton-R
\end{verbatim}
\end{minipage}
\begin{minipage}{3in}
\begin{verbatim}
RMSS(13)=201  ! stau L
RMSS(14)=501  ! stau R

RMSS(8)=750   ! left-squark
RMSS(9)=720   ! dR-squark 
RMSS(22)=720  ! uR squark

RMSS(10)=705  ! 3L squark 
RMSS(11)=705  ! sbottomR  
RMSS(12)=575  ! stopR  
\end{verbatim}
\end{minipage}
}

\vspace{24pt}
The spectrum for the guess ``SUSY C'' (right-hand spectrum shown in Figure
\ref{fig:ex2_spectrum}):
\vspace{24pt}

{\tiny
\begin{minipage}{3in}
\begin{verbatim}
IMSS(1)=1     ! user-specified MSSM 
IMSS(3)=1     ! use gluino pole mass  
IMSS(9) = 0   ! don't use separate L/R squark masses
MSEL=40
 
RMSS(1)=101   !M1 (~bino mass)
RMSS(2)=441   !M2 (~wino mass)
RMSS(3)=700   !gluino pole mass  
RMSS(4)=121   !mu (~higgsino mass)  
RMSS(5)=60    !tan(beta)

RMSS(6)=201   ! slepton-L  
RMSS(7)=201   ! slepton-R  
\end{verbatim}
\end{minipage}
\begin{minipage}{3in}
\begin{verbatim}
RMSS(13)=201  ! stau L  
RMSS(14)=201  ! stau R  

RMSS(8)=722    ! left-squark
RMSS(9)=742    ! dR-squark (artificially decoupled: should be 718)
RMSS(22)=742   ! uR squark

RMSS(10)=620   ! 3L squark   (should be 385)
RMSS(11)=620   ! sbottomR    (should be 262)
RMSS(12)=620   ! stopR       (should be 560)
\end{verbatim}
\end{minipage}
}

\subsubsection{L/R Splitting Example from Sec.~\ref{sec:LCM_LR}}\label{app:Example51}

{\tiny
\begin{minipage}{3in}
\begin{verbatim}
! Pythia card for pseudodata

IMSS(1)=1
IMSS(3)=1
IMSS(9)=1      ! use separate uR/dR squark masses

MSEL=40

RMSS(1)=120    ! bino
RMSS(2)=503    !wino
RMSS(4)=1011.1 ! higgsino 
RMSS(10)=1050  ! 3L squark 
RMSS(11)=960   ! sbottomR 
RMSS(12)=940   ! stopR  

RMSS(5)=20     !tan beta
\end{verbatim}
\end{minipage}
\begin{minipage}{3in}
\begin{verbatim}
RMSS(6)=390    ! slepton-L
RMSS(7)=395    ! slepton-R 

RMSS(13)=620   ! stau L
RMSS(14)=610   ! stau R

RMSS(15)=0     !A_b
RMSS(16)=0     !A_t
RMSS(19)=800   ! MA  

RMSS(8)=700    ! left-squark
RMSS(9)=710    ! dR-squark
RMSS(22)=710   ! uR squark
RMSS(3)=2005   ! gluino
\end{verbatim}
\end{minipage}
}




\bibliography{SimplifiedModels}

\IfFileExists{\jobname.bbl}{}
 {\typeout{}
  \typeout{******************************************}
  \typeout{** Please run "bibtex \jobname" to obtain}
  \typeout{** the bibliography and then re-run LaTeX}
  \typeout{** twice to fix the references!}
  \typeout{******************************************}
  \typeout{}
 }

\end{document}